%% file: paper20220609.tex
\documentclass[prd,aps,twocolumn,a4paper,showkeys,nofootinbib,showpacs]{revtex4-1}

\usepackage[colorlinks=true,backref=true,linkcolor=blue,linktocpage=true,anchorcolor=black,citecolor=blue,filecolor=black,menucolor=black,pagecolor=black,urlcolor=blue]{hyperref}
\usepackage{graphicx,psfrag}
\usepackage{mathrsfs}
\usepackage{amsmath,amsfonts,amssymb,textcomp}
\usepackage{multirow}
\usepackage{comment}
\usepackage{bm}
\usepackage{paralist}
\usepackage{ulem}
\usepackage{xcolor}
\usepackage{enumitem}
\usepackage{bm}

\newcommand{\be}{\begin{equation}}
\newcommand{\ee}{\end{equation}}
\newcommand{\bea}{\begin{eqnarray}}
\newcommand{\eea}{\end{eqnarray}}
\newcommand{\bel}{\begin{align}}
\newcommand{\eel}{\end{align}}

\newcommand{\ord}{\mathcal{O}}
\newcommand{\f}{\frac}

\newcommand{\el}{\ell}
\newcommand{\nn}{\nonumber}
\newcommand{\mbf}[1]{\mathbf{#1}}

\newcommand{\Lhat}{\!\hat{\,\mathbf{L}}_\text{N}}

\newcommand{\Sa}{\mbf{S}_1}
\newcommand{\Sb}{\mbf{S}_2}
\newcommand{\Lhatdot}{\dot{\hat{\,\mathbf{L}}}_\text{N}}

\newcommand{\LN}{\mbf{L}_\text{N}}

\newcommand{\Li}{L_0}

\newcommand{\J}{\mbf{J}}
\renewcommand{\S}{\mbf{S}}
\renewcommand{\L}{\mbf{L}}

\def\i{{\rm i}}

\def\Msun{M_{\odot}}

\def\GMc2{G M_{\odot} c^{-2}}

\def\O{\mathcal{O}}

\def\F{{\cal F}}

\def\F{{\cal F}}

\def\O{{\cal O}}

\def\Msun{M_\odot}
\def\bajes{\texttt{bajes}}

\def\TEOBResumS{\texttt{TEOBResumS}}

\def\TEOB{\TEOBResumS}
\def\TEOBPv0{\texttt{TEOBResumPv0}}

\def\PhenomPv3HM{\texttt{PhenomPv3HM}}

\def\NRsurP{{\texttt{NRSur7dq4}}}

\DeclareSymbolFontAlphabet{\mathrsfs}{rsfs}
\DeclareMathAlphabet{\mathcal}{OMS}{cmsy}{m}{n}
\DeclareSymbolFontAlphabet{\mathrsfs}{rsfs}
\DeclareMathAlphabet\mathbfcal{OMS}{cmsy}{b}{n}

\usepackage{color}
\definecolor{cyan}{rgb}{0,0.9,0.9}
\definecolor{orange}{rgb}{0.9,0.5,0}
\definecolor{magenta}{rgb}{1,0,1}
\definecolor{purple}{rgb}{0.8,0.4,0.8}
\definecolor{gray}{rgb}{0.8242,0.8242,0.8242}
\definecolor{dodgerblue}{rgb}{0.12, 0.56, 1.0}

\begin{document}

\title{Effective-one-body waveforms for precessing coalescing compact binaries with post-newtonian twist
}

\author{Rossella \surname{Gamba}$^{1}$}
\author{Sarp \surname{Ak\c{c}ay}$^{2}$}
\author{Sebastiano \surname{Bernuzzi}${}^{1}$}
\author{Jake \surname{Williams}${}^{2}$}

\affiliation{${}^1$Theoretisch-Physikalisches Institut, Friedrich-Schiller-Universit{\"a}t Jena, 07743, Jena, Germany}
\affiliation{${}^2$University College Dublin, Dublin, Ireland}

\normalem 

\begin{abstract}
Spin precession is a generic feature of compact binary coalescences, which leaves clear imprints in the gravitational waveforms.
Building on previous work, we present an efficient time domain
inspiral-merger-ringdown effective-one-body model (EOB) for precessing binary black holes, which incorporates subdominant modes beyond $\ell=2$, and  the \textit{first} EOB frequency domain approximant for precessing binary neutron stars.
We validate our model against 99 ``short'' numerical relativity precessing waveforms, where we find median mismatches of $5\times 10^{-3}$, $7 \times 10^{-3}$ at inclinations of $0$, $\pi/3$, and 21 ``long'' waveforms with median mismatches of $4 \times 10^{-3}$ and $5 \times 10^{-3}$ at the same inclinations.
Further comparisons against the state-of-the-art \texttt{NRSur7dq4} waveform model yield median mismatches of $4\times 10^{-3}, 1.8 \times 10^{-2}$ at inclinations of $0, \pi/3$ for 5000 precessing configurations with the precession parameter $\chi_p$ up to 0.8 and mass ratios up to 4.
To demonstrate the computational efficiency of our model we apply it to parameter estimation, and re-analyze the gravitational-wave events GW150914, GW190412, and GW170817.

\end{abstract}


\maketitle

\section{Introduction}
\label{Sec:introduction}
From the very first direct detection of a coalescing binary black hole (BBH) system,
gravitational wave astronomy has provided the scientific community with
important (and at times unexpected) indications about black hole (BH) and neutron star (NS) physics \cite{Abbott:2016blz, Abbott:2018wiz,Abbott:2018exr}.
The LIGO-Virgo collaborations have detected (as of the first half of the third 
observing run) $\sim 50$ BBH systems
\cite{Abbott:2016nmj, Abbott:2017gyy, Abbott:2017oio, Abbott:2017vtc, LIGOScientific:2018mvr, 
Venumadhav:2019tad,Venumadhav:2019lyq, Nitz:2018imz, Nitz:2019hdf, Abbott:2020khf, LIGOScientific:2020stg, Abbott:2020tfl}, 
two binary neutron star mergers (BNSs) \cite{TheLIGOScientific:2017qsa,Abbott:2020uma}, 
and two BHNS systems \cite{LIGOScientific:2021qlt} with 
a large number of significant triggers whose parameters have not been published yet.
Analyses so far indicate that -- out of the detected BBH events -- 
two systems have at least one spinning BH component, and that for eleven 
binaries the effective spin-parameter $\chi_{\rm eff}$ [see Eq.~\eqref{eq:chi_eff}] is 
nonzero with more than $95\%$ credibility.
Two events, GW190521~\cite{Abbott:2020tfl,Abbott:2020mjq} and GW190412 \cite{LIGOScientific:2020stg},
individually exhibit marginal evidence for the phenomenon of precession \footnote{Note, however, that alternative interpretations have been put forward for GW190521, see e.g. \cite{Bustillo:2020syj, Gayathri:2020coq, Gamba:2021gap}.} \cite{LIGOScientific:2020ibl,LIGOScientific:2020kqk}, 
whereby the orbital plane and the black holes' spins precess about the total angular momentum of the system.
This affects the resulting gravitational waveforms by modulating the amplitude and contributing to the phase
in a time-dependent manner \cite{Apostolatos:1994mx}.
Wider population studies of all the currently observed signals further 
indicate evidence for nonvanishing spin-orbit misalignment among the population of merging BBH events,
with the spin-precession parameter $\chi_p$ [see Eq.~\eqref{eq:chi_p}] non-null at more than $99\%$ credibility.

Source properties are extracted from the data with matched filtering techniques that employ
 waveform templates, i.e., theoretical models of 
the gravitational waves (GWs) emitted by the two coalescing bodies. 
Waveform templates -- also called \textit{approximants} -- should incorporate a large amount of
physical information: the physics that can be extracted from the data is (at best)
that which is contained within the model itself. Further, such approximants should also
be faithful to numerical relativity (NR) simulations at high frequencies and computationally efficient, 
in order to be employed in parameter estimation (PE) which can require the generation of up to $\ord(10^5-10^6)$ waveforms.

The last decade has seen a flurry of activity in the development of various waveform approximants. 
Several different approximant ``families'' have now matured into their fourth generation versions that incorporate higher multipolar modes, spin precession,
and robust fits to NR data for the merger-ringdown stages. 
These families are clearly distinguished by the underlying approaches that are employed to build the waveforms.

The Effective-One-Body (EOB) semi-analytic family of waveforms for BBH \cite{Buonanno:1998gg, Buonanno:2000ef,Damour:2000we,Damour:2001tu,Buonanno:2005xu,Damour:2015isa} and BNS \cite{Damour:2009wj, Damour:2012yf,Bernuzzi:2012ci} systems maps the general-relativistic two-body problem into the effective problem of describing the evolution of a test mass orbiting around a deformed Kerr metric. Such a system is entirely defined by a Hamiltonian, which describes the conservative motion, and a prescription for the dissipative dynamics, namely the waveform and radiation reaction. 
By relying on systematic resummation of Post-Netwonian (PN) results, EOB models are faithful also in the high-speed, strong-field regime. 
They can be extendend beyond the merger stage of the binary with information from NR simulations.
Two main avatars of this family exist: the \texttt{SEOBNR} \cite{Bohe:2016gbl,Babak:2016tgq,Cotesta:2018fcv,Hinderer:2016eia,Steinhoff:2016rfi,Lackey:2018zvw,Matas:2020wab,Ossokine:2020kjp} and the \TEOB{} \cite{Damour:2014yha, Bernuzzi:2014owa, Nagar:2017jdw,Nagar:2018zoe,Akcay:2018yyh,Nagar:2020pcj} approximants. 
They differ from one another in choices of resummation within the conservative Hamiltonian, the amount of PN and NR information employed and the radiative sector. A detailed investigation of the differences between the two conservative Hamiltonians is presented in Ref.~\cite{Rettegno:2019tzh}. Both models include higher-order modes, tidal effects and precession (albeit $\TEOB{}$ description of the latter previously being limited to the inspiral), with $\TEOB{}$
also being able to generate waveforms along generic orbits \cite{Chiaramello:2020ehz, Nagar:2020xsk, Albanesi:2021rby, Nagar:2021gss, Nagar:2021xnh}.

The most commonly used approximant family in PE are the
phenomenological models \cite{Ajith:2007qp, Ajith:2007kx, Ajith:2009bn, Santamaria:2010yb, Husa:2015iqa, Khan:2015jqa,Pratten:2020fqn,Estelles:2020osj,Estelles:2020twz, Estelles:2021gvs,Hamilton:2021pkf} which combine PN waveforms with fits to hybrid EOB-NR waveforms to obtain computationally cheap, yet accurate, waveforms
that can cover the binary evolution from the early inspiral regime (for which no NR simulation is available) up to merger. 
These models include higher-order modes \cite{London:2017bcn,Garcia-Quiros:2020qpx,Khan:2019kot}, precession \cite{Hannam:2013oca,Schmidt:2014iyl,Khan:2018fmp,Pratten:2020ceb} and tidal effects, which can
be incorporated via time or frequency domain \texttt{NRTidal} models \cite{Dietrich:2017aum, Dietrich:2019kaq}.

Finally, NR surrogates \cite{Blackman:2014maa, Blackman:2017dfb, Varma:2018mmi, Varma:2019csw, Williams:2019vub}
directly interpolate large sets of NR simulations, and are able to provide fast and extremely accurate waveforms
within their parameter space of validity.

In this paper we focus on the description of precessing compact binaries. We work within the EOB
approach \cite{Buonanno:1998gg, Buonanno:2000ef}, and improve on the model \TEOB{} first presented in Ref.~\cite{Akcay:2020qrj}.
In particular, we extend that precessing waveform model to
(i) incorporate higher modes in the waveform for both BBHs and BNSs;
(ii) incorporate the ringdown description for BBHs, to obtain a complete inspiral-merger-ringdown (IMR) approximant, 
(iii) provide an alternative, fast, frequency-domain approximant for BNS inspiral-merger and long BBH inspiral events based on the nonprecessing approach of Ref.~\cite{Gamba:2020wgg}.
\TEOB{} IMR precessing model for BBHs is validated by directly computing (mis)matches against a significant number of NR {\tt SXS} \cite{SXS:catalog} waveforms and against the waveform model $\tt{NRSur7dq4}$ \cite{Varma:2019csw}.
We additionally indirectly test its performance against the state-of -the-art $\tt{IMRPhenomXPHM}$ model \cite{Pratten:2020ceb}, by comparing EOB/NR and Phenom/NR mismatches.
Finally, we perform full PE to further compare the model to other existing approximants, and estimate the source parameters of the binary black hole merger events GW150914 \cite{Abbott:2016blz}, GW190412 \cite{LIGOScientific:2020stg}, and the first binary neutron star inspiral-merger event GW170817 \cite{TheLIGOScientific:2017qsa}.

This article is organized as follows: in Sec.~\ref{Sec:model}, we review our model and introduce the improvements we have made.
Sec.~\ref{Sec:validation} details the validation of our model against {\tt SXS} and \NRsurP{} waveforms.
Sec.~\ref{Sec:PE} presents the results of our PE studies.
We summarize our results and discuss future directions in Sec.~\ref{Sec:conclusions}.

We work with geometrized units where $G=c=1$. 
$m_1$ and $m_2$ denote the masses of the primary and secondary components of the compact binary system.
Accordingly, we define the mass ratio as $q\equiv m_1/m_2 \ge 1$, the symmetric mass ratio $\nu = q/(1+q)^2$, the total mass as $M=m_1+m_2$ and the mass fraction $X_i$ as $X_i = m_i/M$ with $i=1,2$.
As is standard,we employ mass-rescaled units within the EOB and the precession dynamics.
However, when we make comparisons using interferometer sensitivity which varies with GW frequency $f$, thus with  $M$, we restore the physical dimensions of $f$ and $M$, which we express in Hertz (Hz) and in solar masses ($M_\odot$), respectively.
$\Sa$, $\Sb$ denote the dimensionful spin vectors of the binary components with their respective dimensionless
spin parameters given by $\bm{\chi}_i = \mbf{S}_i/m_i^2$ with $\chi_i\equiv |\bm{\chi}_i|$. 
The spin dependence in our baseline EOB model is expressed via the spin variables 

\begin{align}
\hat{\S}   =& M^{-2}(\Sa + \Sb)= (X_1^2 \bm{\chi}_1 +  X_2^2 \bm{\chi}_2)  \, ,\\
\hat{\S}_* =& M^{-2}\Bigl(\frac{X_2}{X_1}\Sa + \frac{X_1}{X_2} \Sb \Bigr) = X_1 X_2 (\bm{\chi}_1 + \bm{\chi}_2)\, .
\end{align}

Let us also recall two standard definitions in the literature that pertain to the mass-weighted projections
of the spins parallel and perpendicular to the Newtonian orbital angular momentum of the system $\LN$.
For $q\ge 1$, the parallel scalar is given by \cite{Ajith:2011ec, Damour:2001tu, Racine:2008qv}
\be
\begin{split}
\chi_\text{eff}\equiv & (X_1 \bm{\chi}_1 + X_2 \bm{\chi}_2)\cdot\Lhat \\
= & (\hat{\S} + \hat{\S}_*)\cdot \Lhat \\ 
= & \, \tilde{a}_0 \label{eq:chi_eff}
\end{split}
\ee
where $\Lhat\equiv \LN/|\LN|$, and $\tilde{a}_0$ is one of the spin-parameters which often enter EOB models. 
This is a conserved quantity of the orbit-averaged precession equations over the precession timescale \cite{Racine:2008qv}. 
The perpendicular scalar, first introduced in Ref.~\cite{Schmidt:2014iyl}, is defined as
\be
\chi_p \equiv {m_1^{-2}}\max\left\{|\mbf{S}_{1,\perp}|,
 q\f{4+3q}{3+4q}|\mbf{S}_{2,\perp}|\right\} \label{eq:chi_p},
\ee
where $\mbf{S}_{i,\perp}\equiv \mbf{S}_i - (\Lhat\cdot\mbf{S}_i)\Lhat$ are the components of $\mbf{S}_i$ perpendicular to $\Lhat$ for $i=1,2$.

\section{Method}
\label{Sec:model}

In this section we describe in some detail the technical improvements that have been 
implemented in the model with respect to its first version in Ref.~\cite{Akcay:2020qrj}.
After succintly recalling the conventions that we employ and describing the spin-aligned baseline model, 
we will highlight the main advancements introduced, 
which pertain to the coupling of the spin-evolution equations to the
EOB orbital dynamics, the extension of the waveform to merger-ringdown, and the inclusion of higher ($\ell > 2 $) modes.

\subsection{Reference frames and co-precessing waveform model}

When the spins of the binary components $\Sa,\Sb$ are not aligned with the orbital angular momentum $\L$ of the system, all three vectors precess around
the total angular momentum vector $\J=\L+\Sa+\Sb$ \cite{Apostolatos:1994mx}. Accordingly, the orbital plane of the binary is not fixed throughout its evolution, but rather precesses, inducing non-trivial modulations in the gravitational waves detected by an inertial observer.
In this scenario, the dominant emission of gravitational radiation happens along the direction
perpendicular to the orbital plane, i.e., along the Newtonian orbital angular momentum $\LN$ \cite{Schmidt:2010it, Schmidt:2012rh, Boyle:2011gg, OShaughnessy:2011pmr}.
It is then possible to identify a special ``co-precessing'' non-inertial frame, which follows the evolution of $\LN$. In this frame, the modulations of amplitude and phase due to precession effectively disappear
and the waveform can be well approximated by that emitted from an aligned spin system.
Then, given the evolution of the co-precessing frame (i.e., of the vectors $\Sa, \Sb$ and $\LN$) and a co-precessing waveform, it is possible to rotate the latter into the inertial source frame and obtain the associated precessing waveform  \cite{Schmidt:2010it, Schmidt:2012rh}. This technique is usually referred to as the ``twist'', due to the time-dependent rotation which relates the two frames. 

To perform the twist, one can in principle use either the frame set by the Newtonian
orbital angular momentum $\LN$ or $\L$. 
Since, by definition, $\LN$ remains orthogonal to the orbital plane,
we employ this frame for the twist. Ref.~\cite{Pan:2013rra} has shown that the differences in the $\L$-frame vs. $\LN$-frame twisted waveforms as compared with precessing NR waveforms are marginal.
Accordingly, we define our inertial source frame such that its $z$ axis is aligned with the initial Newtonian orbital angular momentum $\LN(0)$. 
Then, following usual conventions, we choose the line of sight vector $\hat{\bm{N}}$ to have spherical angles $(\iota, \pi/2-\phi_{\rm ref})$ and define the initial spin components, $\Sa(0), \Sb(0)$, in this so-called $\Li$ frame. 
We then track the evolution of $\LN$ with respect to this frame via its spherical angles $\alpha$ and $\beta$, 
defined using the Cartesian components of the unit vector $\Lhat$
\begin{align}
\alpha =& \arctan(\Lhat{}_y/\Lhat{}_x), \label{eq:alpha_eq1}\\
\beta  =& \arccos(\Lhat{}_z) \label{eq:beta_eq1}. 
\end{align}
A third angle $\gamma$, which identifies the co-precessing frame univocally with respect to the $\Li$ frame \cite{Boyle:2011gg}, is given by 
\begin{equation}
\dot{\gamma} = \dot{\alpha}\cos\beta \label{eq:gamma_eq1},
\end{equation}
where the overdot denotes differentiation with respect to time.
With $\alpha, \beta$, and $\gamma$, the twisted (precessing) multipolar waveform $h^T_{\ell m}$ is obtained via an Euler rotation
\begin{equation}
h^T_{\ell m} = \sum_{m'=-m}^{m} h^A_{\ell m'} D^{(\ell)}_{m', m}(-\gamma, -\beta, -\alpha)\label{eq:twist},
\end{equation}
where $h^A_{\ell m}$ are the spin-aligned waveforms in the co-precessing frame and $D^{(\ell)}_{m', m}(-\gamma, -\beta, -\alpha)$ are the Wigner D matrices, defined as
\begin{equation}
D^{(\ell)}_{m', m}(\alpha, \beta, \gamma) = e^{- i m' \alpha} e^{-i m \gamma} d^{\ell}_{m', m}(\beta) \label{eq:WigD_eq}
\end{equation}
with
\begin{align*}
d^{\ell}_{m', m}(\beta) =& \sum_{k_i}^{k_f} (-1)^{k-m+m'} \\ 
\times& \frac{\sqrt{(\ell+m)!(\ell-m)!(\ell+m')!(\ell-m')!}}{k!(\ell+m-k)!(\ell-k-m')!(k-m+m')!} \\
\times& \Bigl[\cos\frac{\beta}{2}\Bigr]^{2\ell-2k+m-m'}\Bigl[\sin\frac{\beta}{2}\Bigr]^{2k-m+m'}.
\end{align*}
Finally, the plus and cross GW polarizations $h_{+}$ and $h_{\times}$ are obtained from the twisted modes as 
\begin{equation}
h_{+} - i h_{\times} = \sum_{\ell, m}h^T_{\ell m}\,  {}_{-2}Y^{\ell m}(\iota, \pi/2-\phi_{\rm ref}) \, \label{eq:hpc_eq},
\end{equation}
where ${}_{-2}Y^{\ell m}(\iota,\phi)$ are the standard spin weight $s= -2$ spherical harmonics.

Our baseline model for the spin-aligned co-precessing waveforms is $\TEOB{}$ {\rm v2} \cite{Nagar:2019wds,Nagar:2020pcj}, a multipolar EOB model
for quasi-circular BBH and BNS coalescences. Its non-spinning
orbital sector contains analytical PN information suitably resummed via Pad\'e approximants in both the Hamiltonian $H_{\rm EOB}$ and radiation reaction  
$\mathcal{F}_{\varphi}$. It is informed by NR via an effective parameter entering the EOB circular-orbit potential, $A$, at the relative 5PN order, $a_6$.
Spin-orbit and spin-spin effects are included within the model Hamiltonian via the centrifugal radius $r_c$ \cite{Damour:2014sva} (or its inverse $u_c = r_c^{-1}$) and 
the gyro-gravitomagnetic coefficients $G_S$ and $G_{S*}$ \cite{Damour:2014sva} given in the DJS gauge \cite{Damour:2008qf}, and heceforth considered ar next-to-next-to-leading order (NNLO) \cite{Nagar:2018zoe}.
The EOB dynamics further contains a spin-orbit parameter, $c_3$, that is tuned to NR. Multipolar waveforms up to $\ell=m=5$ additionally contain spin-dependent terms, 
which are factorized and analytically resummed to improve their robustness in the strong-field regime \cite{Nagar:2020pcj}.
Tidal effects are included in both the multipolar waveform and the EOB conservative dynamics
\cite{Nagar:2018zoe, Nagar:2018plt, Akcay:2018yyh}. The tidal sector of \TEOB{} contains contributions from the multipolar $\ell= 2,3,4$ gravitoelectric and 
$\ell = 2$ gravitomagnetic tidal coefficients, and the former are resummed via a gravitational self-force inspired expression~\cite{Bini:2012gu, Bini:2014zxa, Akcay:2018yyh}. 
The equation of state-dependent quadrupole-monopole terms are included at NNLO via $r_c$~\cite{Nagar:2011fx, Nagar:2018plt}.

\subsection{Spin dynamics}

\subsubsection{Time evolution of the orbital frequency}
To obtain the time and frequency evolution of the Euler angles $\alpha, \beta$ and $\gamma$
we follow Ref.~\cite{Akcay:2020qrj} and solve the PN spin-evolution equations for $\Sa, \Sb$, and $\Lhat$
at the next-to-next-to-next-to-next to leading order (N4LO).
To avoid clutter, we present the evolution equations up to the next-to-leading order, which can be cast into classical precession equations
\begin{subequations}
\begin{align}
 \dot{\mbf{S}}_i &= \bm{\Omega}_i \times \mbf{S}{}_i,  \label{eq:Sidot_NLO}\\
 \Lhatdot &= \bm{\Omega}_\text{NLO} \times \Lhat \label{eq:LNdot_NLO}, 
 \end{align}
\end{subequations}
for $i=1,2 $. The precession frequencies are given by
\begin{subequations}
\begin{align}
 \label{eq:Omega_i}
 \begin{split}
\bm{\Omega}_1 =& v^5  \Bigl[\nu\left(2+\f{3}{2}\frac{X_2}{X_1}\right)\Lhat
 +\f{v}{2}\Bigl\{\mbf{S}_2 \\
 & -3 \Bigl[\Bigl(\frac{X_2}{X_1} \mbf{S}_1+\mbf{S}_2\Bigr)\cdot\Lhat\Bigr]\Lhat\Bigr\}\Bigr] \, ,
 \end{split}\\
\bm{\Omega}_\text{NLO}=&-\f{v}{\nu} \left(\bm{\Omega}_1+\bm{\Omega}_2  \right) \label{eq:Omega_NLO},
 \end{align}
\end{subequations}
where $v$ is the relative speed of the binary components and $\bm{\Omega}_2$ is obtained from $\bm{\Omega}_1$ with $ 1 \leftrightarrow 2$. The N$4$LO expression for this system of nine coupled ODEs is explicitly given in equations (4a)-(4b) and (7) of Ref.~\cite{Akcay:2020qrj},
together with an expression for the evolution of the orbital frequency under radiation reaction for which we employed a \texttt{TaylorT4}-resummed PN expression in our previous work: $\dot\omega^{\rm PN}$ \cite{Buonanno:2002fy, Buonanno:2009zt}. Specifically, we employed the expressions from Ref.~\cite{Chatziioannou:2013dza} up to 3.5PN, which 
are integrated together with the spin ODEs in order to evolve the system under radiation reaction.
Here we study three alternative reformulations for $\dot\omega$. 

First, we consider a hybrid PN-EOB expression denoted by $\dot\omega^{\rm HY1}$. 
To obtain it, we begin by calculating the magnitude of the PN-corrected orbital angular momentum $L^\text{PN}(\omega)$ up to relative 4PN order (see App. G of Ref.~\cite{Pratten:2020ceb}) and 
the orbital separation $r^\text{PN}(L^\text{PN})$ up to 4PN [Eqs.~(12), (17), (18) of Ref.~\cite{Nagar:2018plt}]. The resulting quantities are then employed to evaluate the EOB radiation reaction $\mathcal{F}_{\varphi}\equiv dL/dt$ along circular orbits for
which the EOB radial momentum $p_{r *}$ is set to zero. The expression for $\dot\omega^{\rm HY1}$
is thus obtained as 
\begin{equation}
\label{eq:omgdot_hybrid}
\dot\omega^{\rm HY1} = \frac{\mathcal{F}_{\varphi}(r^\text{PN}, L^\text{PN})}{dL^\text{PN}/d\omega}  \, .
\end{equation}

Alternatively, we can compute another hybrid PN-EOB expression, denoted by $\dot\omega^{\rm HY2}$, where we obtain $r(\omega)$ by numerically inverting the Hamilton equation for $\omega$, evaluated at $p_{r*}=0$:
\begin{equation}
\label{eq:omega_EOB}
\omega = \frac{\partial H_{\rm EOB}}{\partial p_{\varphi}} = \frac{1}{\nu H_{\rm EOB} H_{\rm eff}^{\rm orb}}\Bigl[A p_{\varphi}u_c^2 +H_{\rm eff}^{\rm orb}(G_S \hat{S} + G_{S_*} \hat{S}_*)\Bigr] \, ,
\end{equation} 
where $H_{\rm eff}^{\rm orb}$ is the orbital effective EOB hamiltionian.
The magnitude of the orbital angular momentum $L(r)$ is then computed as the EOB circular-orbit angular momentum by analytically solving the equation
\begin{equation}
\partial_r H_{\rm EOB} = 0 \, .\\
\end{equation}
Once $r(\omega)$ and $L(r)$ are known, we can compute $dL/d\omega = (dL/dr)(d\omega/dr)^{-1}$. The first piece, $(dL/dr)$, can be immediately obtained by differentiating the analytical solution found above; the second piece, $(d\omega/dr)$, is computed from Eq.~\eqref{eq:omega_EOB}. Then, we obtain $\dot\omega^{\rm HY2}$ as in Eq.~\eqref{eq:omgdot_hybrid}. 

Finally, we also consider ``aligned'' $\omega(t)$ relation as given by the integrating the EOB dynamics \textit{before}
computing the evolution of the spins. In this case, instead of solving an ODE system of 9 + 1 equations, we 
compute the cubic spline of $\omega(t)$ and use it to drive the spins evolution. 
Notably, the time axis of the EOB dynamics may in principle have a different origin with respect 
to the time axis of the spin dynamics, for which $t=0$ always corresponds to the reference (initial)
frequency $\omega_{\rm ref}$ 
at which the spin components are specified. Therefore, by solving 
$\omega(t_{\rm EOB}) = \omega_{\rm ref}$ for $t_{\rm EOB}$
we compute the timeshift $\Delta t = t_{\rm EOB}$ necessary to align the two time axes. 
Then, the frequency at each timestep of the spin dynamics is simply given by 
$\omega_i = \omega(t_i + \Delta t)$.

\begin{figure}[h!]
\includegraphics[scale=0.495]{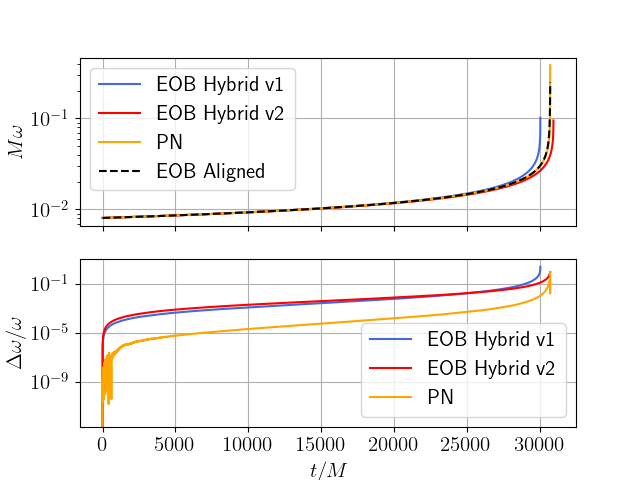}
\caption{
\label{fig:fluxes} Example plot of the frequency evolution $\omega(t)$ of the spin dynamics, obtained by integrating pure PN (orange) or hybrid PN-EOB (blue and red) $\dot\omega(\omega)$ ODEs, or by employing the $\omega(t)$ relation as given by the aligned-spin EOB Hamilton equations (dubbed as ``EOB aligned'', black dashed inline). 
The bottom panel displays the relative difference of the hybrid and PN methods with respect to the EOB aligned method. For the system considered, $\omega(t)^{\rm PN}$ remains the closest to the EOB frequency evolution at all times.}
\end{figure}

Figure \ref{fig:fluxes} shows the different $\omega(t)$ relations obtained with the four methods 
described above for a system with $(q, \chi_{\rm eff}, \chi_{\rm p}) = (1.56,-0.06, 0.28)$ at a 
reference frequency of $M\omega=0.0025$.
The pure PN implementation $\omega^{\rm PN}$ is the closest 
to the EOB ``aligned'' $\omega$ evolution. The hybrid PN-EOB evolutions, instead, appear to either overestimate ($\rm HY1$) or underestimate ($\rm HY2$) dissipation effects by GW emission.
Further, the hybrid evolution stops at lower 
values of $M \omega$ than the PN or ``aligned EOB'' one, because of the denominator of Eq.~\ref{eq:omgdot_hybrid} becoming zero. 
The difference between the two hybrid versions can be qualitatively understood by considering a simple spinless, equal mass case. 
During the inspiral ($M \omega <0.06$), at a fixed $M \omega$ value $L^{\rm HY1}\geq L^{\rm HY2}$ and $|dL/d\omega|^{\rm HY1} \leq |dL/d\omega|^{\rm HY2}$. 
Therefore, by applying Eq.~\eqref{eq:omgdot_hybrid}, the first hybrid version will emit GWs faster -- and thus, have a steeper orbital frequency evolution -- than the second hybrid, 
whose phase in turn will evolve slower than the ``aligned'' EOB one due to the (strong) assumption that the coalescence is along circular orbits and $p_{r*}=0$.
Although all options are available in the publicly released \TEOB{} code, we find that the fully PN expression for $\dot\omega$ consistently gives the better performance 
in terms of accuracy and speed when
computing mismatches against NR waveforms (see Sec.~\ref{Sec:validation}). Therefore,
all results obtained in this paper are obtained with $\dot\omega = \dot\omega^{\rm PN}$. 
We leave to future work the exploration of different purely PN formulations of $\dot\omega$.

\subsubsection{Backward in time integration and $\alpha$ initial conditions}
\label{subsubsec:back}
We further discuss two technical points: the initial condition for the Euler angle $\alpha$ and the behavior of the Euler angles when
integrating backward in time. Neither is discussed in depth in the literature, as the former 
has no real implication on waveforms (since the in plane spin components are determined only up to a rotation, the single $x-y$ projections can vary: this is equivalent to choosing different $\alpha(0)$ initial conditions)
and the latter stems from our hybrid PN-EOB implementation of the dynamics.

Regarding the first point, in our source frame, the initial conditions at $t=t_0$ (set to 0 without loss of generality) for the spin components and the angular momentum are straightforward to obtain. However, since $\Lhat(0)$ is parallel to the $z$ axis at $t=0$, $\alpha(t)$ is apparently undefined
at $t=0$. 
Nonetheless, an expression for $\alpha(0)$ can be obtained using the direction of the initial ``torque'' at $t=0$ which only has $x,y$ components given by
$\Lhatdot{}_x(0), \Lhatdot{}_y(0)$ yielding
\be
\alpha(0) = \arctan\left(\f{\Lhatdot{}_y(0)}{\Lhatdot{}_x(0)} \right) \label{eq:alpha_0_eq1} .
\ee
Explicit expressions, as well as a comparison between $\alpha$ obtained at different PN orders, can be found in
App. \ref{app:alphaIC}. 
As far as we can tell, this is the only physically motivated initial condition for $\alpha(0)$ and although for simple precession
$\alpha(0)$ is often in the fourth quadrant of the $x$-$y$ plane, it only equals the commonly-used $-\pi/2$ for special cases.
On the other hand, there does not seem to be a physically motivated initial condition for the third Euler angle $\gamma$ defined by $\dot{\gamma}=\dot{\alpha}\cos\beta$. Therefore, one has the freedom to set $\gamma(0)=0$ or $\pm \pi/2$ or even $\alpha(0)$.

\begin{figure}[t]
\includegraphics[scale=0.45]{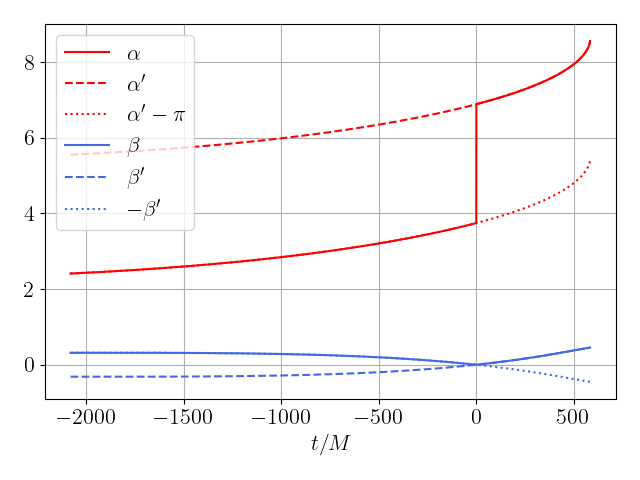}
\caption{\label{fig:backint}
Backward time evolution of the Euler angles of Eqs.~\eqref{eq:alpha_eq1}, \eqref{eq:beta_eq1}. At $t=0$, $\alpha$ (straight red line) shows a jump of $\pi$, while $\beta$ (straight blue line) has a cusp. To correct this behavior and avoid interpolation issues, we compute $\alpha'$ and $\beta'$, which are then corrected to their true values after interpolation.}
\end{figure}

As for the second point, we observe that since the spin evolution is solved independently from the EOB dynamics, 
it may happen that for some binaries the initial orbital EOB frequency $\omega_0^\text{EOB}$ is smaller than the initial spin-evolution frequency $\omega_0^{S}$ specified by the user via $\omega_0^S= \pi f_0^{\rm Hz} M$, where $f_0^{\rm Hz}$ is input initial GW frequency in Hertz.
Then, in order to twist the \textit{entire} EOB waveform, it is necessary to integrate the spin dynamics
backwards in time, at least to below $\omega_0^\text{EOB}$. Since all the directly evolved quantities vary continuously when going from $t>0$ to $t<0$, this procedure may appear straightforward at a first glance.
However, as can be seen from Fig. \ref{fig:backint}, $\alpha(t)$ can exhibit a jump by $\pi$ due to the sign-change of both $\Lhat{}_x$ and $\Lhat{}_y$ when $\Lhat$ passes through the origin at $t=0$. Since a number of numerical interpolations of $\alpha(t), \beta(t)$ are required to compute the twist, when integrating backwards we compute
\begin{align}
\alpha'_{t<0} =& \arctan(\Lhat{}_y/\Lhat{}_x) + \pi, \, \\
\beta'_{t<0}  =& -\arccos(\Lhat{}_z). 
\end{align}
Note that this is analogous to changing into a different but equivalent source frame, in which $\beta\rightarrow -\beta$, $\alpha\rightarrow \alpha + \pi$ and $\Lhat = (\Lhat{}_x,\Lhat{}_y,\Lhat{}_z)$ is mapped into itself.
After removing the kinks this way then interpolating, we restore the original frame via 
$\alpha_{t<0} = \alpha'_{t<0}-\pi$, $\beta_{t<0} = -\beta'_{t<0}$, then perform the standard waveform twist.

\subsection{Coupling of the PN spin evolution to the EOB dynamics}
When analyzing long signals, neglecting the evolution of 
the spins in the aligned-spin dynamics can lead to non-negligible errors.
In principle, one should evolve the full EOB equations, coming from a
genereral 
Hamiltonian where the orbital plane is not fixed. 
Similarly, the waveform and radiation reaction of the model, too, 
would need to be extended to incorporate the effect of the planar components of the spins. 
This general approach would increase the already significant computational cost related to the solution of the Hamilton equations. Luckily, it was found \cite{Pan:2013tva} that good agreement with NR
waveforms can be achieved by simply replacing
in the waveform and radiation reaction the fixed values of $\chi_{i}$ with the time-dependent projections of the spin vectors onto the orbital angular momentum, i.e., $\Lhat(t)\cdot \bm\chi_i(t)$.

\begin{figure}[h!]
\includegraphics[width=0.495\textwidth]{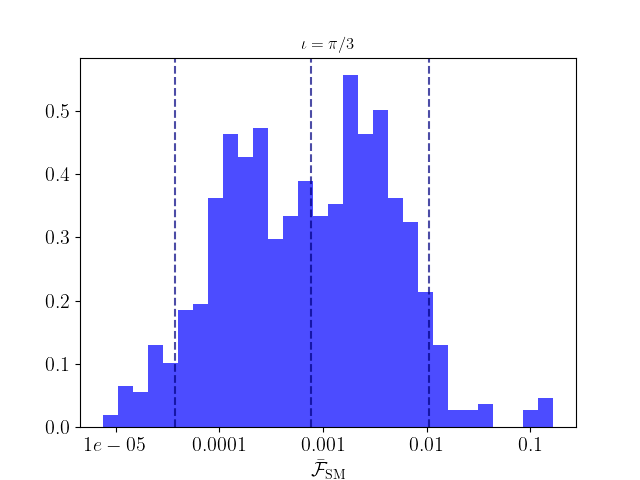}
\caption{\label{fig:projection} Comparison between waveforms obtained with and without projected
  spin dynamics for systems with the same intrinsic parameters as those examined in Sec.~\ref{subsec:eobnr} (see also App.~\ref{app:nrdata}).
  For waveforms with large in-plane spin components, the effect of the spin projection is non-negligible
  and can lead to mismatches larger than $1\%$.}
\end{figure}

In our model, we (optionally) employ the spin dynamics to compute the projections of the spins onto $\Lhat$ either
in the time or frequency domain. 
We proceed as follows:
(i) the PN spin-dynamics is independently evolved with the N4LO description of the precession equations
with $\dot\omega^\text{PN}$ 
detailed above; (ii) we interpolate the spin and angular momentum components as functions of the ``spin'' orbital frequency $\omega^{S}$; 
(iii) at each step of the EOB evolution, we compute the EOB orbital frequency and evaluate 
$\bm{\chi}_{i,z}(\omega_\text{EOB})\equiv \Lhat(\omega_\text{EOB})\cdot \bm\chi_i(\omega_\text{EOB})$
via the splines calculated above; (iv) finally, these quantities are inserted into the appropriate places in the EOB dynamics.
This generic procedure is applied both when numerically evolving the ODE system and 
when applying the postadiabatic approximation (PA) of Ref.~\cite{Nagar:2018gnk}.

Figure \ref{fig:projection} displays the mismatches (see Sec. \ref{sec:faithfulness}) 
obtained between \TEOB{} waveforms with an inclination of $\iota=\pi/3$, 
evolved either with or without spin projection, for a set of waveforms with $q \in [1,6]$, $M\in[50, 225]\Msun$, $\chi_{\rm p} \in [0., 0.8]$ and $\chi_{\rm eff} \in [-0.45, 0.65]$. Notably, although a large portion
of the mismatches lie below the $10^{-3}$ threshold, the effect of the spin projection
can be relevant for binaries with large in-plane spin components, i.e., $|\mbf{S}_{i\perp}|$, 
for which the parallel components of the spins to the orbital angular momentum varies more.

\subsection{BBH Merger-Ringdown}
\label{Subsec:RD}

To model the final state of the BBH one can employ the fits of Ref.~\cite{Jimenez-Forteza:2016oae} with minor modifications
to account for the non-null planar components of the BHs' spins.
Following Ref.~\cite{Pratten:2020fqn}, we define the remnant spin as:
\begin{equation}
\label{eq:spinfit_prec}
\chi_f = \sqrt{(\chi_{f||})^2 + (S_{\perp}/M_f)^2},
\end{equation}
where $\chi_{f||}$ and $M_f$ are estimated from the fits of Ref.~\cite{Jimenez-Forteza:2016oae} using the parallel component of the spins
to the 
orbital angular momentum at merger, and $\mathbf{S}_{\perp}$ is given by
\begin{equation}
\mathbf{S}_{\perp} = \Sa(\omega_{\rm mrg}) - \mathbf{S}_{1 ||}(\omega_{\rm mrg}) + (1 \leftrightarrow 2) \, .
\end{equation}

\begin{figure*}
\includegraphics[width=0.99\textwidth]{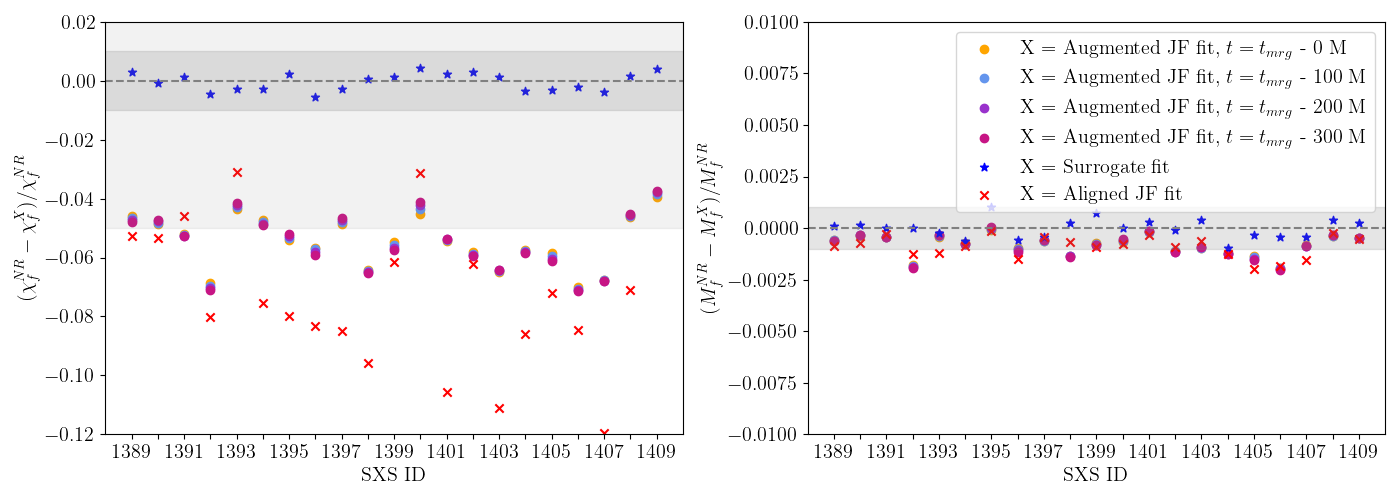}
\caption{\label{fig:final_state}
Relative differences in the dimensionless spin $\chi_f$ of the final black hole (left) and its mass $M_f$ (right)
between NR simulations and fits of Ref.~\cite{Varma:2019csw} (blue stars) and the various Jimenez-Forteza (JF) fits of Ref.~\cite{Jimenez-Forteza:2016oae}.
The latter are evaluated with the initial $z$ component spins (red crosses) or 
corrected to account for the precession by employing the spins at a reference time before merger like in Eq.~\eqref{eq:spinfit_prec}. While the remnant mass is always estimated at the order of $10^{-3}$ and the two methods give comparable results, the surrogate fit for the remnant spin is up to an order of magnitude more precise. In the left panel, light (dark) gray bands highligt the $5\%$ ($1\%$) relative error interval; in the right panel the $1$\textperthousand~ one.}
\end{figure*}

Figure~\ref{fig:final_state} displays the accuracy of the fits when compared to a handful of $\tt SXS$ NR simulations. We also compare the output of the fit above to the values obtained with the ``simple'' aligned-spin fit, and with the fits provided by the surrogate model of Refs.~\cite{Varma:2018aht, Varma:2019csw}. 
Notably, while the mass of the remnant is approximated (by all approaches) at the level of $10^{-3}$, the difference $|\chi_f^{\rm NR} - \chi_f^{\rm surr}|$ is up to ten times smaller with
respect to the other approaches.
This result is not surprising, and is in line with the discussion presented in Ref.~\cite{Varma:2019csw}. 
Therefore, although all the results presented in this paper will employ Eq.~\eqref{eq:spinfit_prec}, we also implemented the option to
take $\chi_f$ and $M_f$ as input parameters. This way, by externally computing the remnant properties with the surrogate model (using the {\tt surfinBH} package), we can easily obtain a precise description of the final BH. 

For a complete model of the ringdown phase, it is necessary to also extend the Euler angles $\alpha, \beta, \gamma$ beyond the merger.
The precession of the orbital momentum effectively stops at the merger, and the direction of the spin of the final black hole can be thought of as constant, and well-enough approximated by the direction of the angular momentum at merger.
Therefore, one option is to simply prolong the angles by fixing them to their value at merger.
Alternatively, it was observed that the evolution of the $\alpha$ angle can be approximately described through the difference of the $\ell = 2$ fundamental quasinormal modes (QNMs) \cite{OShaughnessy:2012iol, Ossokine:2020kjp}
\begin{equation}
\alpha(t) = 
  \begin{cases}
    \alpha(t_{\rm mrg}) + (\omega_{22} - \omega_{21})(t - t_{\rm mrg}), &\bm\chi_f \cdot \Lhat{}_f > 0, \\
	\alpha(t_{\rm mrg}) + (\omega_{2-1} - \omega_{2-2})(t - t_{\rm mrg}), & \bm\chi_f \cdot \Lhat{}_f < 0,
   \end{cases}
\end{equation}
where $ \bm{\chi}_f =\bm{\chi}_1(\omega_\text{mrg})+\bm{\chi}_2(\omega_\text{mrg}) $ and $\omega_{\ell m}$ are the fundamental quasinormal-modes $\omega_{\ell m 0}$ for $\ell = 2$ and $m=2,1,-1,-2$ \cite{Berti:2005ys}. 
One can then fix $\beta$ to its value at the merger. $\gamma$ is subsequently computed by integrating its evolution equation \eqref{eq:gamma_eq1}.
Both options for the post-merger evolution of $\alpha$ are curently available in \TEOB{} public code, and users can choose between one or the other. The default behavior is given by the quasinormal-modes extension, which gives marginally better results when computing mismatches between EOB and NR waveforms (see Sec. \ref{Sec:validation}).

\begin{figure*}
   \includegraphics[width=0.495\textwidth]{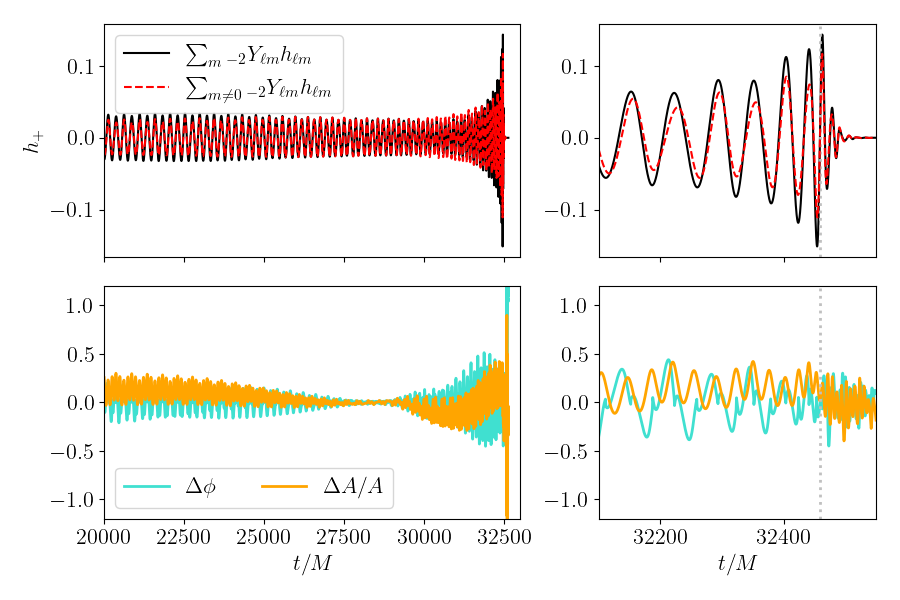}
\includegraphics[width=0.495\textwidth]{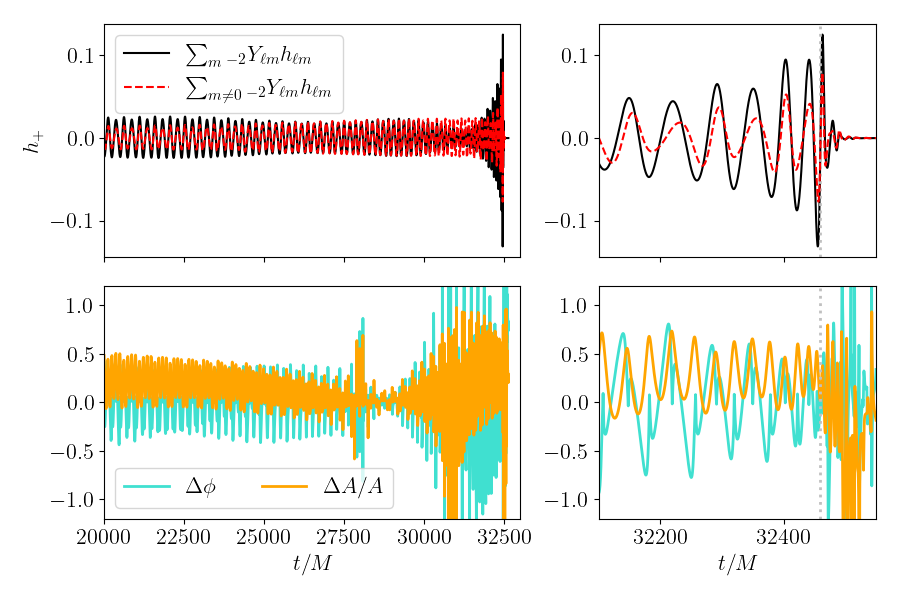}
\caption{\label{fig:hl0}
Plus polarization $h_+$ of the NR simulation $\tt{SXS:BBH:1409}$ having $(q, \chi_{\rm eff}, \chi_{p}) = (4, -0.16, 0.41)$, obtained with all modes with $\ell \leq 5$ either including $m=0$ modes (black) or not (red). Two different binary inclinations are considered: $\iota=\pi/3, \pi/2$. The largest impact of the $h^T_{\ell,0}$ modes is in the amplitude of the waveforms close to merger, which can be underestimated up to $50\%$ for edge on binaries.}
\end{figure*}

\begin{figure}
\includegraphics[width=0.495\textwidth]{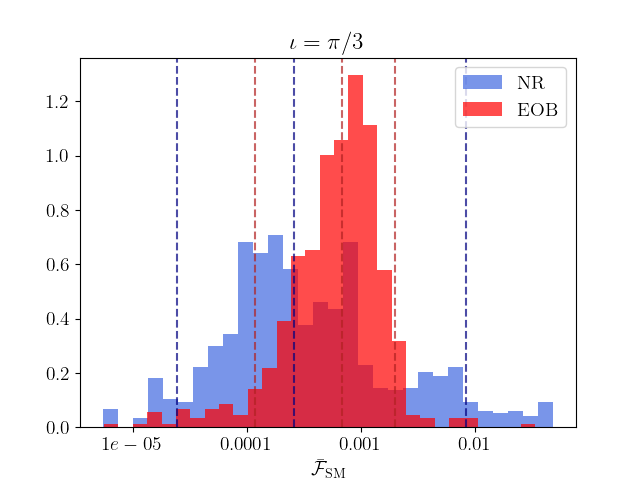}
\caption{\label{fig:sxs_m0} 
Distribution of $\bar{\mathcal{F}}_{\rm SM}$ between SXS waveforms constructed with all modes up to $\ell=8$ and the 
same modes except $m=0$ ones (blue) or EOB waveforms constructed with all modes up to $\ell=5$ and the 
same modes except $m=0$ ones. We consider systems with the same parameters as those examined in 
Sec.~\ref{subsec:eobnr}. 
For a fixed inclination of $\iota=\pi/3$, we find that both sets span the interval
$\bar{\mathcal{F}}_{\rm SM} \in [10^{-5}, 5 \times 10^{-2}]$.
}
\end{figure}

\begin{figure}
\includegraphics[scale=0.495]{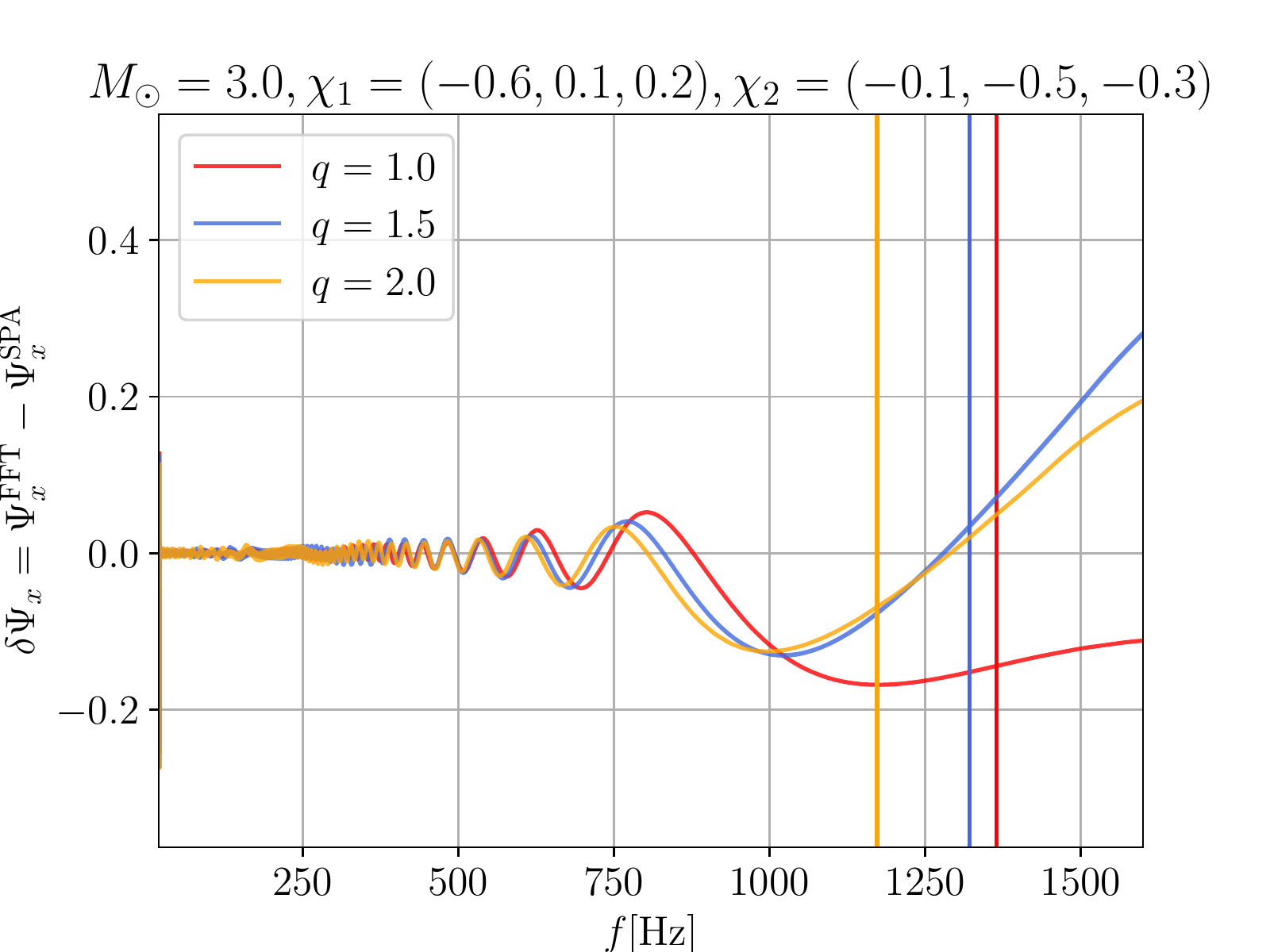}
\caption{\label{fig:twisted_SPA}
Phase differences between the frequency domain cross polarizations $h_{\times}$ obtained either by twisting SPA-transformed modes or directly via FFT.
The three fiducial BNS systems considered have varying mass ratios $q=\{1, 1.5, 2\}$ 
and fixed spins, total masses and tidal parameters. 
Vertical colored lines denote the merger frequency. At merger, the largest phase difference amounts to $\approx -0.2$ radians.}
\end{figure}

\subsection{Higher modes and $m=0$}

The higher modes are obtained by twisting the $\ell>2$ spin-aligned modes of 
\TEOB{} {\tt v2}. Precessing \TEOB{} computes
all modes with $\ell \leq 5$ including the twisted $m=0$ modes.
Note that, similarly to most of the currently available approximants,
we compute the co-precessing $m<0$ modes by means of symmetry with the $m>0$ modes. This approximation, which
is valid in absence of precession of the orbital plane, does not hold when 
describing precessing systems close to merger \cite{Ossokine:2020kjp, Ramos-Buades:2020noq}. Nonetheless,
it was found that the effect of employing this approximation is subdominant to other sources of error \cite{Ossokine:2020kjp}.

The contribution of $m=0$ modes, negligible when dealing with spin-aligned waveforms, 
can become relevant for precessing binaries.

For example, Fig.~\ref{fig:hl0} shows how the precessing $h^T_{\ell,0}$ modes of the $\tt{SXS:BBH:1409}$ 
NR simulation contribute to the total waveform polarization $h_+$ for two binaries with inclinations $\iota=\pi/3$ and $\iota=\pi/2$.
While the amplitude of $h^T_{\ell,0}$ modes can become as large as that of $h^T_{2,2}$ close
to merger, the mode sum of $h^T_{\el m}$ with spin-weighted spherical harmonics 
decreases the overall importance of the $m=0$ modes when computing the polarizations.
Nonetheless, the contribution to the amplitude for large-inclination binaries is non-negligible, whereas the
phase difference between the $h_+$'s obtained with and without the $m=0$ modes oscillates around zero during the inspiral and remains below $\sim1$ rad at the merger. 
To systematically quantify the importance of precessing $m=0$ modes, we compute the sky-maximized 
and SNR-weighted unfaithfulness $\bar{\F}_{\rm SM}$ between our entire {\tt SXS}
validation set of waveforms (see Sec.~\ref{sec:faithfulness} and \ref{subsec:eobnr}), constructed by either setting the precessing $m=0$ modes to zero or by considering them
in the construction of the polarizations. 
Fig.~\ref{fig:sxs_m0} shows the distribution of the mismatches computed for binaries with $\iota=\pi/3$.  
We find that while most of the mismatches obtained are $\O(10^{-4})$, for some high mass ratio systems
$\bar{\mathcal{F}}_{\rm SM} > 3\%$.
Repeating this analysis with EOB waveforms yields  
qualitatively similar results, with few mismatches surpassing the $3\%$ threshold.
Thus, accurate modelling of the precessing $m=0$ modes is important for very asymmetrical binaries. 
We note however that current modelling of the twisted $m=0$ modes is not complete as the default \TEOB{} 
treatment of the spin-aligned $m=0$ modes sets them to zero thus overlooking their contributions in the twist formula \eqref{eq:twist}.
We plan to study the  effects employing nonzero spin-aligned $m=0$ modes in the future.

\subsection{BNS Frequency-domain waveforms}\label{subsec:SPA}

Spin-aligned EOB models can be straightforwardly extended to the 
frequency domain by applying a stationary phase approximation (SPA)
to the multipolar modes $h_{\ell m}(t)$. The frequency domain, spin-aligned modes $\tilde{h}_{\ell m}(f)$
can then be twisted and combined into plus and cross polarization as \cite{Pratten:2020ceb}:
\begin{subequations}
\begin{align}
h_{+} =& \frac{1}{2}\sum_{\ell\geq 2}\sum_{m'>0}e^{i m'\gamma} \tilde{h}_{\ell m'} \label{eq:hp_twistFD} \\
&\times\sum_{m=-\ell}^{\ell} \Bigl [ e^{-i m \alpha}d^{\ell}_{m m'} {}_{-2}Y^{\ell m} 
        + (-1)^{\ell}e^{i m\alpha}d^{\ell}_{m -m'} {}_{-2}Y^{\ell m\ast}\Bigr] \,,\nn\\
h_{\times} =& \frac{1}{2}\sum_{\ell\geq 2}\sum_{m'>0}e^{i m'\gamma} \tilde{h}_{\ell m'} \label{eq:hx_twistFD}\\ 
&\times \sum_{m=-\ell}^{\ell} \Bigl [ e^{-i m \alpha}d^{\ell}_{m m'} {}_{-2}Y^{\ell m} 
        - (-1)^{\ell}e^{i m\alpha}d^{\ell}_{m -m'} {}_{-2}Y^{\ell m\ast}\Bigr] \,\nn .
\end{align}
\end{subequations}
The sign differences in our expressions with respect to those presented in 
Ref.~\cite{Pratten:2020ceb} come from the EOB convention that the phase of the time domain multipoles $h_{\ell m}$ with
$m > 0$ is positive. Hence, $\tilde{h}_{\ell m}(f) = 0$ for $m > 0$ and $f < 0$.
The Euler angles $\alpha, \beta, \gamma$ are all evaluated at the SPA frequencies $2 \pi f/m$.

Figure~\ref{fig:twisted_SPA} displays the phase difference in the frequency domain
of the cross polarization $h_\times$ computed between the FFT of precessing \TEOB{} time domain signals 
and the SPA-based model described above. We consider three nominal BNS systems with fixed spins 
$\bm{\chi}_1 = (-0.6,0.1,0.2)$, $\bm{\chi}_2=(-0.1,-0.5,-0.3)$ inspiralling from an initial frequency $f_0= 20$ Hz, tidal polarizability parameters $\Lambda_1 = \Lambda_2 = 400$, total mass $M = 3 \Msun$ and mass ratios of 1, 1.5, and 2.
For all three cases considered, we find that the phase difference at the merger (represented by the vertical lines) lies below 0.2 rad.
The conclusions of Ref.~\cite{Gamba:2020ljo} regarding the validity of the SPA up to merger can be applied also to precessing BNS systems. At the same time, the SPA-based model is less computationally expensive than its TD counterpart thanks to the non-uniform time grid which
is employed for the inspiral. Moreover, and more importantly, it opens to the possibility of generating waveforms directly over a non-uniform frequency grid, optimized for PE, allowing the application of techniques such as relative binning \cite{Zackay:2018qdy} or multibanding \cite{Vinciguerra:2017ngf}.

\section{Validation}
\label{Sec:validation}

In this section we compare our EOB model to (i) the set of  $99$ precessing ${\tt SXS}$ simulations also employed
in Ref.~\cite{Pratten:2020fqn}, supplemented with the longer precessing simulations ${\tt SXS:BBH:1389}$ to ${\tt SXS:BBH:1409}$,
and (ii) 5000 ${\tt NRSur7dq4}$ (henceforth {\tt NRsur}) waveforms, spanning $q \in [1, 4]$ and 
$|\chi_i| \in [0.1, 0.8]$ yielding a range of $-0.8 \le \chi_\text{eff} \le 0.8$ and $ 0.0 \le \chi_p\le 0.8$.
We compute the sky-averaged faithfulness (see Sec.~IV of Ref.~\cite{Harry:2016ijz}) for all considered templates. Then, for a selected number
of systems, we align the time-domain polarizations and compute the cumulative phase difference of the waveform $h = h_+ - i h_\times$.
Overall, we find that the maximum mismatch between \TEOB{} and ${\tt SXS}$ is obtained for very asymmetric,
highly spinning binaries. The same statement holds for ${\tt NRsur}-\TEOB{}$ mismatches.

\subsection{Faithfulness}
\label{sec:faithfulness}

The Wiener product between two time domain waveform templates $a(t), b(t)$ is defined as
\begin{equation*}
(a, b) = 4 \text{Re} \int \frac{\tilde{a}^{*}(f) \tilde{b}(f)}{S_n(f)}\, \mathrm{d}f,
\end{equation*} 
where $S_n(f)$ is the power spectral density (PSD) of the detector and $\tilde{a}, \tilde{b}$
denote the Fourier transform of the waveforms.
The agreement between a target model $s$ and a generic template $h$ is usually quantified through the 
faithfulness (or match) $\mathcal{F}$, defined as the normalized inner product between $s$ and $h$, 
maximized over the reference time and phase $t_0, \varphi_0$:
\begin{equation}
\mathcal{F} = \max_{t_0, \varphi_0} \frac{(s, h)}{\sqrt{(s,s)(h,h)}} \label{eq:standard_match}.
\end{equation}
However, when the template waveform incorporates higher modes or if the system is precessing, 
this definition is not completely independent of the extrinsic parameters of the binary. 
In general, the target and template waveforms are obtained from the plus and cross polarizations as:
\be
\begin{split}
k_i &= F_{+}(\theta^i, \phi^i,\psi^i)\; k_{+}(\iota^i, \varphi_0^i, t_0^i, \Theta^i) \\
    &+  F_{\times}(\theta^i, \phi^i,\psi^i)\; k_{\times}(\iota^i, \varphi_0^i, t_0^i, \Theta^i) \label{eq:ki_eq1},
\end{split}
\ee
where $i = s, h$ and $\theta, \phi, \psi, \iota, \Theta$ are, respectively, the right ascension, declination, polarization, inclination, and intrinsic parameters (masses, spins, tidal parameters etc.) of the binary system.
Equation \eqref{eq:ki_eq1} can be rearranged into 
\be
\begin{split}
k_i =  \mathcal{A}(\theta^i,\phi^i)[&\cos\kappa (\theta^i,\phi^i,\psi^i) k_{+}(\iota^i, \varphi_0^i, t_0^i, \Theta^i)  \\
      &+ \sin\kappa(\theta^i,\phi^i,\psi^i) k_{\times}^i(\iota^i, \varphi_0^i, t_0^i, \Theta^i) ] \, \label{eq:ki_eq2},
\end{split}
\ee
where $\kappa$ denotes the effective polarizability and

\begin{align}
e^{i \kappa(\theta,\phi,\psi)} &= \left[F_{+}(\theta,\phi,\psi) + i F_{\times}(\theta,\phi,\psi)\right]/\mathcal{A}(\theta,\phi) \, , \label{eq:kappa}\\
\mathcal{A}(\theta,\phi) &= \sqrt{F^2_{+}(\theta,\phi,\psi) + F^2_{\times}(\theta,\phi,\psi)} \label{eq:calA} \, .
\end{align}

When only $(2,\pm 2)$ modes are considered,
it can be shown that Eq.~\eqref{eq:standard_match}
depends on extrinsic parameters only through
overall amplitude and phase factors.
On the other hand, when higher modes are considered, the dependence on the extrinsic quantities
is nontrivial.

We define the (template) sky-maximized (SM) faithfulness between the target strain and the waveform template as
\begin{equation}
\mathcal{F}_\text{SM}  = \max_{t_0^h, \varphi_0^h, \kappa^h} \frac{(s, h)}{\sqrt{(s, s)(h, h)}},
\end{equation}
where we dropped the explicit depencence on intrinsic and extrinsic parameters in the right hand side.
Accordingly, the unfaithfulness is given by $ \bar{\mathcal{F}}_{SM} = 1 - {\mathcal{F}}_{SM}$.
We follow the procedure outlined in Sec.~IV of Ref.~\cite{Harry:2016ijz}.
The maximization over $\kappa$ is performed analytically, while $t_0$ is maximized via the inverse FFT.
The maximization over the reference phase $\varphi_0$ is performed numerically through a dual annealing algorithm, similar to what is done in Ref.~\cite{Pratten:2020ceb}.
Finally, we mention that for precessing systems one additional degree of freedom remains: the freedom
to perform a rigid rotation of the in-plane spin components about the initial $\hat{z}$ axis,
which is equivalent to choosing different initial conditions for the $\alpha$ (and
$\gamma$) Euler angles. We further maximise $\mathcal{F}_\text{SM}$ over such a rotation by once more
relying on a dual annealing algorithm.
We note that this procedure differs from the one employed in Ref.~\cite{Ossokine:2020kjp}, where instead the initial (reference)
frequency is varied, and the initial in-plane spin components are kept fixed to their nominal target value.

Once $\mathcal{F}_\text{SM}$ (or, equivalently,  $\bar{\mathcal{F}}_\text{SM}$) is computed as described, we normalize it over the SNR of the signal and further average over the sky angles of the target waveform, in order to completely marginalize over any dependence of the mismatch on the sky position and obtain values which depend exclusively
on the intrinsic parameters of the source. We consider $N_{\varphi}$ values of $\varphi_0^s \in [0, 2\pi)$ and $N_{\kappa}$ values of $\kappa^s \in [0, 2\pi)$, and present the average value over $N_\varphi \times N_\kappa$ values.

\begin{figure}[t]
\label{fig:sxs_99}
\includegraphics[width=0.42\textwidth]{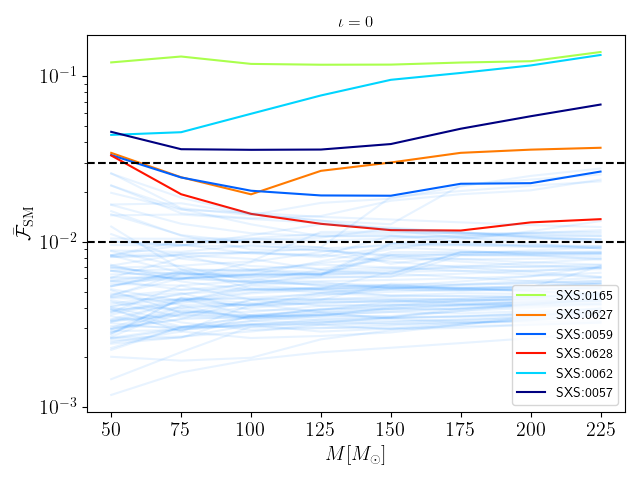}
\includegraphics[width=0.42\textwidth]{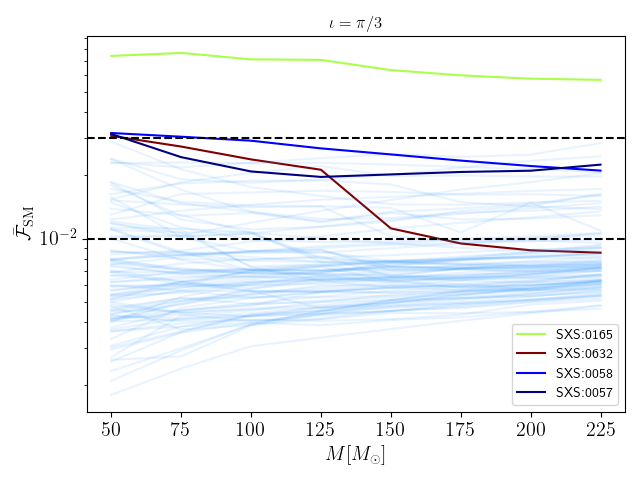}
\includegraphics[width=0.42\textwidth]{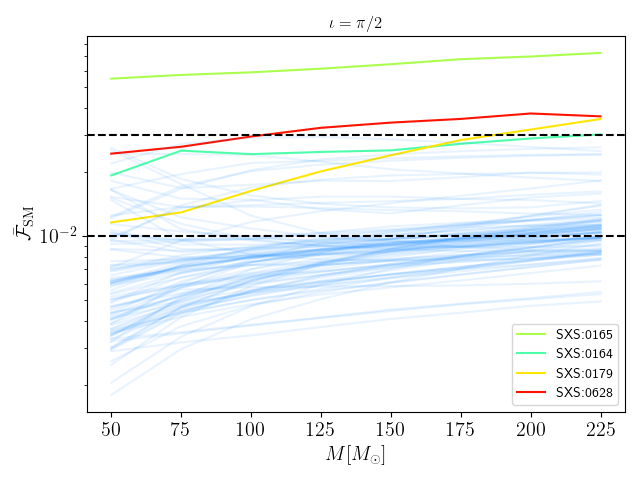}
\caption{\label{fig:sxs_short_eobnrmm}
NR/EOB mismatch, $\bar{\mathcal{F}}_{\rm SM}$, for the 99 {\tt SXS} short precessing simulations computed with strain mode content
$\ell \leq 4 + (5, \pm 5)$, plotted as a function of the total mass of the system and computed with a fixed inclination of the binary of $\iota = \{0, \pi/3, \pi/2\}$ (top, middle and bottom panels, respectively). A total of ten configurations have $\bar{\mathcal{F}}_{\rm SM}$ which reaches up to $3\%$. The dashed horizontal lines in each panel mark the $3\%$ and $1\%$ thresholds.
}
\end{figure}

\begin{figure*}[t]
\label{fig:sxs_99}
\includegraphics[width=0.495\textwidth]{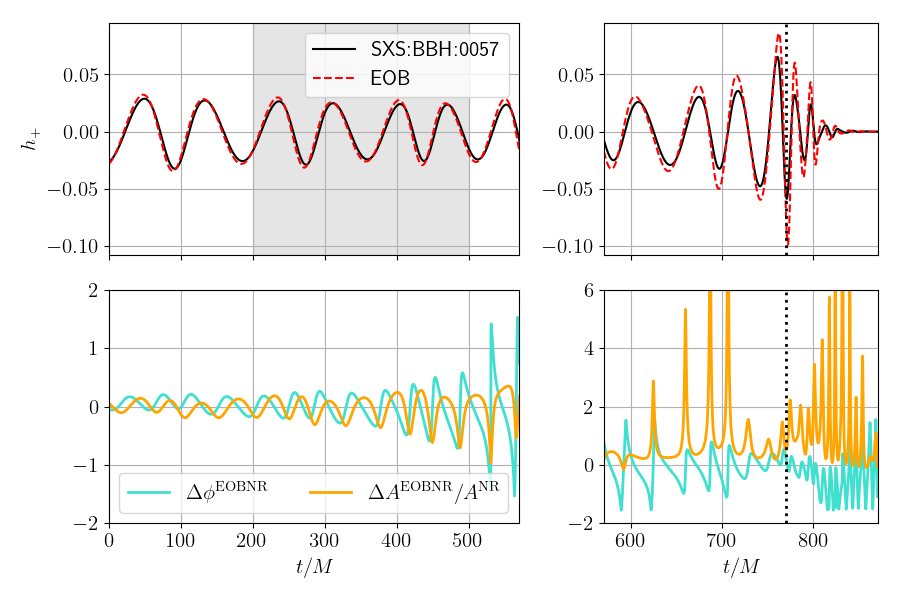}
\includegraphics[width=0.495\textwidth]{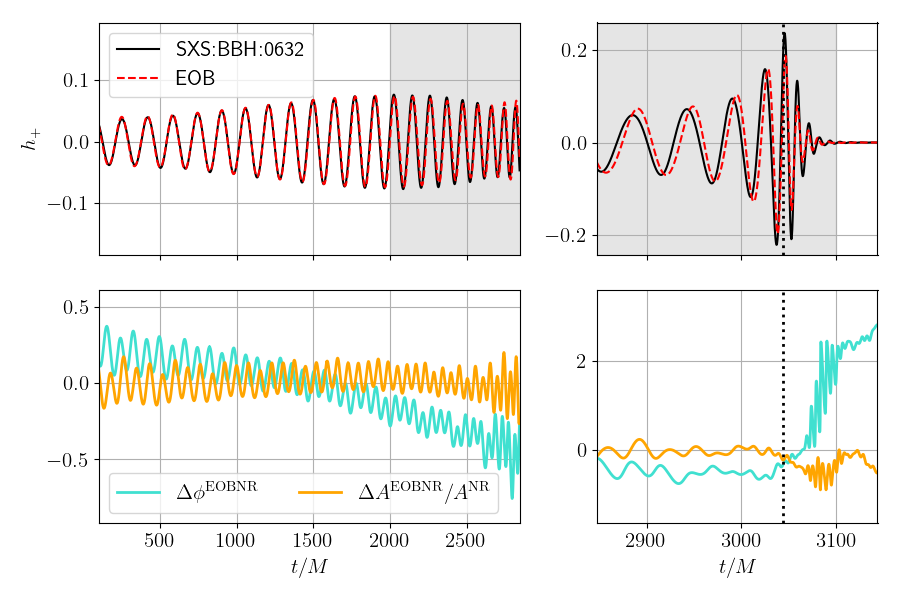}
\caption{\label{fig:sxs_phasing_short}
  Visual comparison between the $h_+$ NR waveforms (black, solid curves) computed 
  from {\tt SXS:BBH:0057} (left panel) and {\tt SXS:BBH:0632} (right panel)
  and the \TEOB{} waveforms obtained 
 with the same intrinsic parameters (red, dashed), inclination $\iota = \pi/3$ and all modes with $\ell \leq 4$. 
 The phase difference $\Delta\phi^{\rm EOBNR} = \phi_{\rm NR} - \phi_{\rm EOB}$ 
 is shown in cyan, and the relative 
 amplitude error $\Delta A^{\rm EOBNR}/A^{\rm NR}$ in orange. Merger 
 is indicated by a black dotted line. The waveforms are aligned by minimizing 
 the phase difference in the time window highlighted in gray, see Eq.~(30) of 
 Ref.~\cite{Dietrich:2019kaq}. Both systems are characterized by very large 
 in-plane spins at their initial reference frequency, with
 $\tt SXS:BBH:0057$ also having $q>5$.
 While the phase difference oscillates during the inspiral and generally 
 remains below ~1 rad, 
 the dephasing and the amplitude relative differences increase at merger, 
 indicating that an improved description of the final moments of the coalescence 
 will be required.
 }
\end{figure*}

\begin{figure*}
\includegraphics[width=0.495\textwidth]{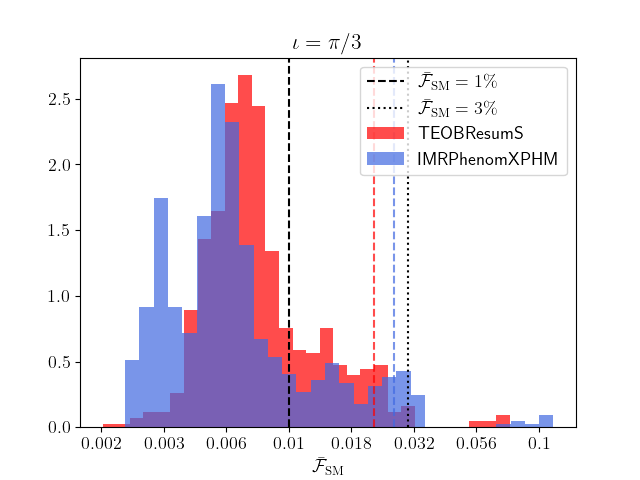}
\includegraphics[width=0.495\textwidth]{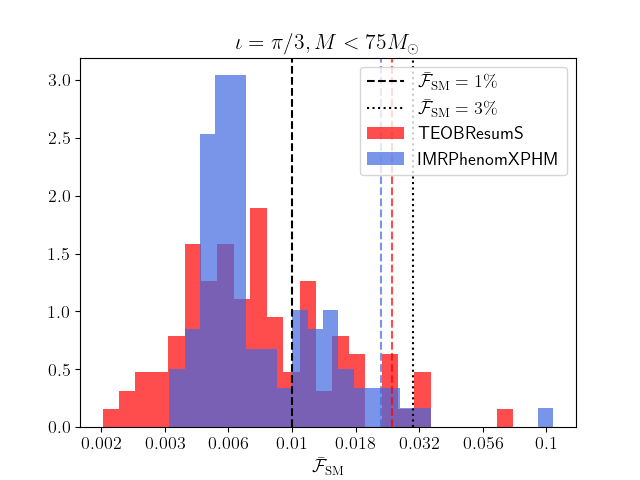}
\caption{\label{fig:sxs_short_eobphenmm}
Left panel: the distribution of NR/\TEOB{} and NR/{\tt IMRPhenomXPHM} mismatches for the 99 {\tt SXS} short precessing simulations of Fig.~\ref{fig:sxs_short_eobnrmm}, at a fixed binary inclination of $\iota = \pi/3$. The black dashed and the dotted black vertical lines mark the $1\%$ and $3\%$ thresholds, and dashed colored lines the $95^{\rm th}$ percentiles. We find that the performance of $\tt IMRPhenomXPHM$ is comparable to that of our EOB approximant, with $\bar{\mathcal{F}}_\text{SM}^{\rm EOB}$ falling in the range $0.002 - 0.06$ with median at $0.007$, and $\bar{\mathcal{F}}_{SM}^{\rm XPHM}$ falling within the interval $0.002 - 0.1$ and having a median of $0.005$.
Right panel: the same plot as above, with total masses restricted to below $75 \Msun$. Overall, \TEOB{} performs slightly better than $\tt IMRPhenomXPHM$ for lower masses, and slightly worse for higher ones.}
\end{figure*}

\subsection{BBH IMR EOB/NR comparison}
\label{subsec:eobnr}

To validate the performance of our model, we compare our waveforms with a set of selected {\tt SXS}
NR simulations. In particular, we focus on two different sets: 99 ``short'' 
waveforms, with $\chi_p \lesssim 0.84$, $\chi_{\rm eff} \in [-0.45, 0.65]$ and $q \lesssim 6$,
and 21 ``long'' simulations with $\chi_p \lesssim 0.49$, $\chi_{\rm eff} \in [-0.2, 0.3]$ and $q \lesssim 4$, 
spanning from $\sim 60$ to $\sim 146$ orbits. 
To translate the NR data from the NR frame into the source frame 
described in Sec.~\ref{Sec:introduction}, we make use of the public catalog tools available 
at \cite{sxs_tools} and described in, e.g., Ref.~\cite{Schmidt:2017btt}.
For all unfaithfulness computations, we consider total detector-frame masses $M \in [50, 225] \Msun$, employ the zero-detuned high-power PSD of Ref.~\cite{aLIGODesign_PSD} and average $\bar{\mathcal{F}}_{\rm SM}$ over a grid $\kappa^{\rm NR} = \{0, \pi/2, \pi, 3/2 \pi\}$ and $\varphi_0^{\rm NR} = \{0, 2\pi/5, 4\pi/5, 6\pi/5, 8\pi/5\}$. 
We perform our computations over the frequency range $[f_{\rm min},2048]$ Hz,
where $f_{\rm min}$ 
is the initial GW frequency of the NR waveform, expressed in physical units.

\subsubsection{``Short'' SXS simulations}
Figure~\ref{fig:sxs_short_eobnrmm} shows the sky-averaged $\bar{\mathcal{F}}_{\rm SM}$
as a function of the total binary mass for three different choices of the binary inclination, $\iota = \{ 0, \pi/3, \pi/2\}$. 
We find that when $\iota = 0$ ($\pi/3, \pi/2$), all but six (four) notable simulations
the EOB/NR unfaithfulness lies below the $3\%$ threshold for all values of masses considered,
and that $80\%$ ($76\%, 68\%$) of the averaged $\bar{\mathcal{F}}_{\rm SM}$ computed are smaller than $1\%$.
The configurations for which the EOB/NR faithfulness lies above the $3\%$ threshold are highly 
asymetrical $(q>5)$ or strongly precessing $(\chi_{\rm p} > 0.7)$ systems, with ${\tt SXS:BBH:0165}$ being the most challenging one, as it is a $(q, \chi_{\rm eff}, \chi_{\rm p}) = (6, -0.45, 0.77)$ coalescence.
In Fig.~\ref{fig:sxs_phasing_short} we consider two more of these systems ($\tt SXS:BBH:0057$ and $\tt SXS:BBH:0632$), and align the time-domain NR and EOB waveforms by minimizing their phase difference $\Delta\phi^{\rm EOBNR} = \phi^{\rm NR}-\phi^{\rm EOB}$ over a chosen time-window (see e.g.~\cite{Dietrich:2019kaq}). 
We find that the EOB waveform correctly 
captures the behavior of the NR waveform up to few orbits before merger, where differences in phase and amplitude start to grow.

For comparison, we also compute $\bar{\mathcal{F}}_{\rm SM}$ between the set of NR simulations here considered 
and the waveform approximant $\tt IMRPhenomXPHM$ \cite{Pratten:2020ceb}, with fixed inclination $\iota=\pi/3$.
Figure \ref{fig:sxs_short_eobphenmm} shows the results of this calculation. 
We find that 
$\bar{\mathcal{F}}_{SM}^{\rm EOB}$ varies between $\sim 0.002$ and $0.06$, with the distribution median peaking
at $0.007$; while $\bar{\mathcal{F}}_{SM}^{\rm XPHM}$ spans the interval $\sim 0.002$ to $0.1$, with a median of $0.005$.

Overall, the two approximants give consistent results, with \TEOB{} generally performing marginally worse at high masses, and marginally better for $M < 75 \Msun$.

\subsubsection{``Long'' SXS simulations}
\begin{figure}
\includegraphics[width=0.495\textwidth]{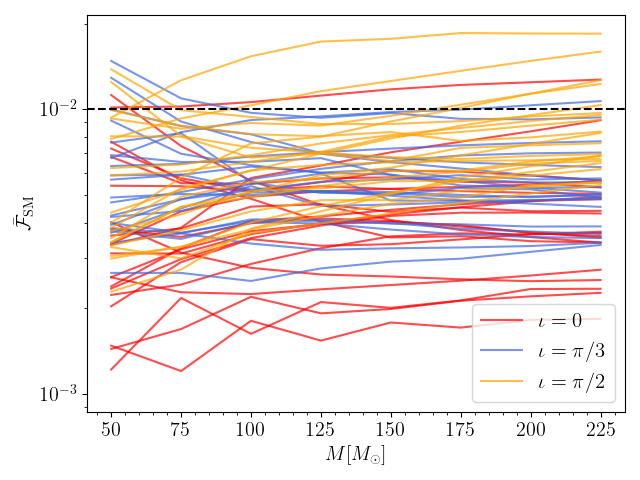}
\caption{\label{fig:higher_modes}
NR/EOB mismatch for the {\tt SXS} long precessing simulations 1389 to 1409, plotted as a function of the total mass of the system and computed with a fixed inclination of the binary of 
$\iota = 0, \pi/3, \pi/2$ (red, blue and orange lines, respectively). The dashed black horizontal line marks the $1\%$ threshold. No simulations have $\bar{\mathcal{F}}_{\rm SM} > 3\%$ for any of the considered inclinations.}
\end{figure}

\begin{figure*}
\includegraphics[width=0.495\textwidth]{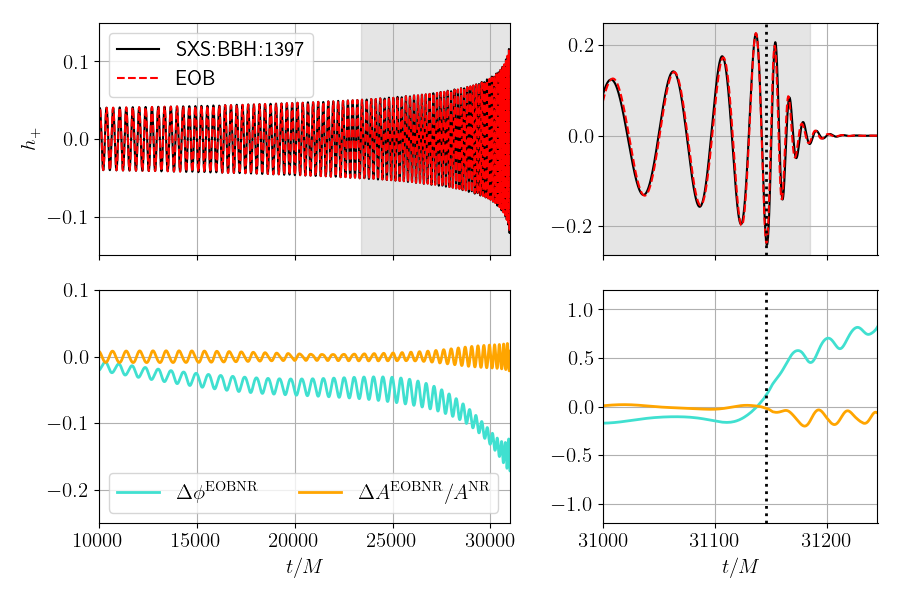}
\includegraphics[width=0.495\textwidth]{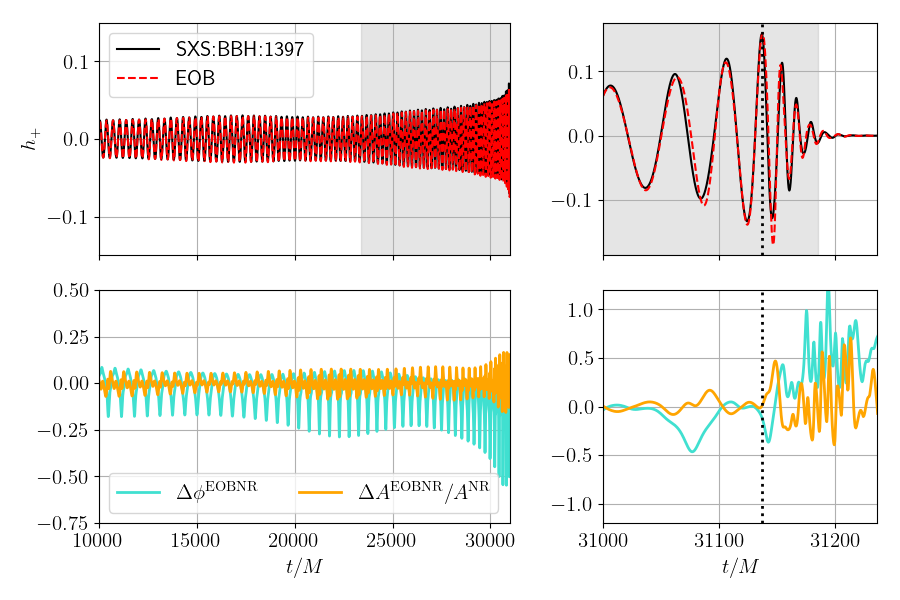}
 \caption{\label{fig:SXS1397}
 Visual comparison between the $h_+$ waveform computed from {\tt SXS:BBH:1397} 
 (black, solid curves) with $\ell \leq 4$ and the \TEOB{} waveform obtained 
 with the same intrinsic parameters (red, dashed) for two different inclinations, 
 $\iota = 0$ (left panel) and $\iota = \pi/3$ (right panel). The phase difference 
 $\Delta\phi^{\rm EOBNR}= \phi_{\rm NR} - \phi_{\rm EOB}$ is shown in cyan, and the relative 
 amplitude error $\Delta A^{\rm EOBNR}/A^{\rm NR}$ in orange. Merger 
 is indicated by a black dotted line. The waveforms are aligned by minimizing 
 the phase difference in the time window highlighted in gray.
 As the inclination increases, so do the importance of higher modes and the 
 amplitude modulations due to precession.
}
\end{figure*}

Figure~\ref{fig:higher_modes} once more shows the sky-averaged $\bar{\mathcal{F}}_{\rm SM}$
as a function of the total binary mass for two different choices of the binary inclination, $\iota = 0, \pi/3, \pi/2$. 
The mismatches behave similarly to what we described above in the sense that they generally degrade 
for increasing magnitude of in-plane spins and growing inclinations.
This well-known fact can be appreciated also from Fig.~\ref{fig:SXS1397}, where we align the
NR waveform ${\tt SXS:BBH:1397}$ and the corresponding EOB waveform. We compute the phase difference $\Delta\phi^{\rm EOBNR}$ 
between the two, and find that for $\iota=0$ it is constantly smaller than $0.1$ rad during the inspiral,
growing to $\sim 0.6$ rad after merger. For $\iota=\pi/3$ the phase difference displays larger 
oscillations, which are however always smaller than $0.5$ rad. The relative difference in the amplitude $\Delta A^{\rm EOBNR}/A^{\rm NR} = (A^{\rm NR}-A^{\rm EOB})/A^{NR}$, instead, degrades after merger for the $\iota = \pi/3$ case. Nonetheless, for the case considered the behavior of both the EOB phase and amplitude remain correct during the merger.
  
Overall we find that all the mismatches computed lie below $3\%$ for the inclinations 
considered, and $93\%$ ($98\%$, $87\%$)  below $1\%$ for $\iota=0$ ($\iota=\pi/3, \pi/2$).
 
\begin{figure}
\includegraphics[scale=0.49]{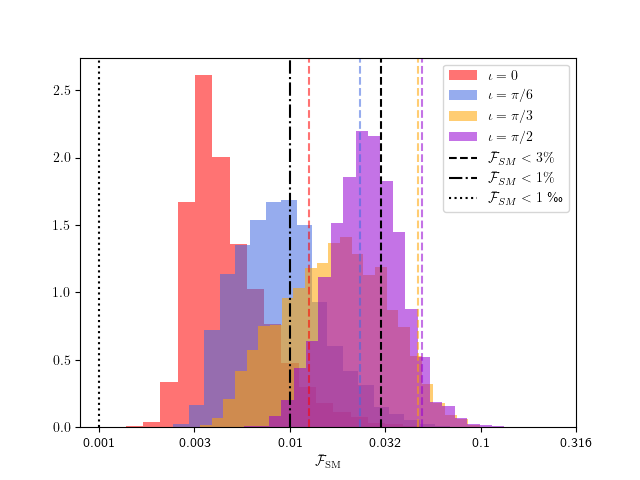}
\includegraphics[scale=0.49]{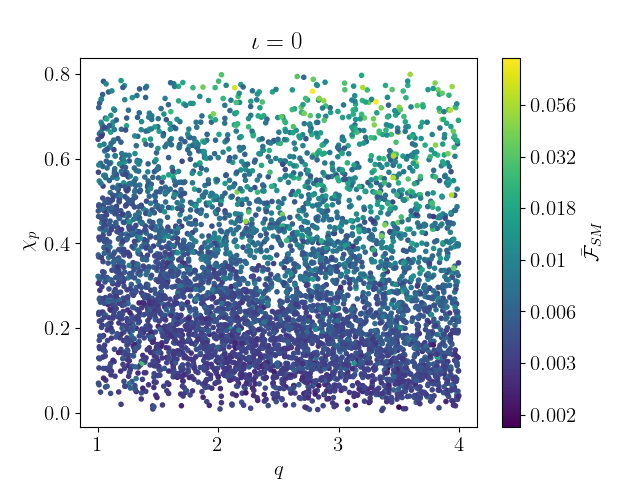}
\includegraphics[scale=0.49]{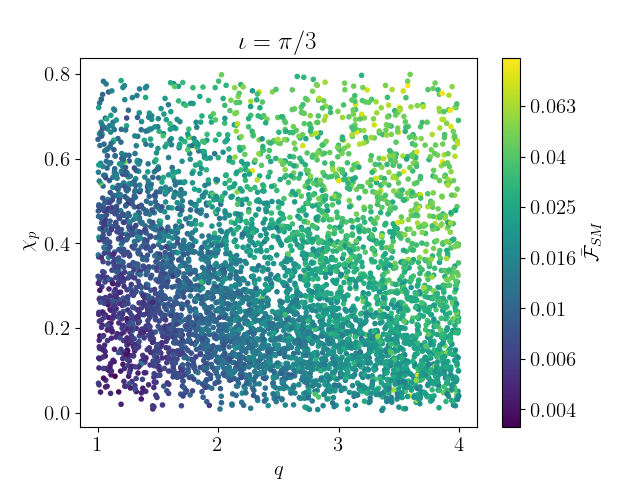}
\caption{\label{fig:nrsurr} \
Top panel: {\tt NRSur7dq4}/\TEOB{} sky maximized $\ell\leq 4$ mismatch for 5000 systems with $q\in[1, 4]$, spin magnitudes $\chi_i \in [0.1, 0.8]$ and random spin directions, computed from an initial frequency range of 20 to 37.5\,Hz with the aLIGO design PSD noise curve up to $1024$ Hz. 
The dotted, dot-dashed, and dashed vertical black lines mark unfaithfulness of $1$\textperthousand, $1\%$, and $3\%$, respectively.
The colored, dashed vertical lines mark the 95th percentiles for the four distributions.
Middle and bottom panels: the behavior of the mismatch over the $\{q,\chi_p\}$ parameter space for inclinations of 0 and $\pi/3$.
The higher unfaithfulness values are obtained for highly asymmetrical systems, with large in-plane spins (high $\chi_p$) and mass ratios of $q > 2$.}
\end{figure}

\begin{figure*}
\includegraphics[scale=0.49]{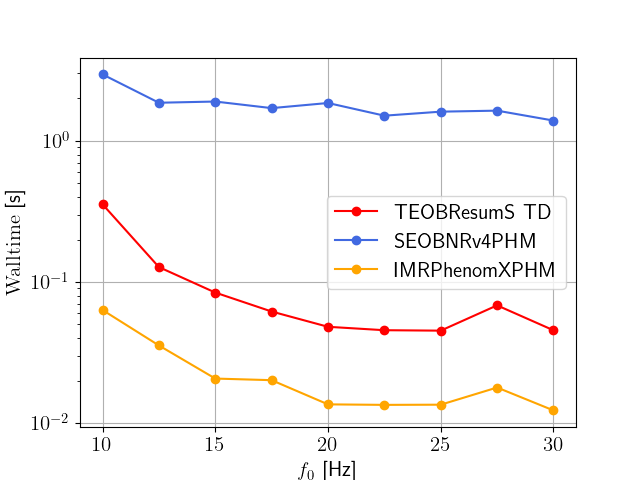}
\includegraphics[scale=0.49]{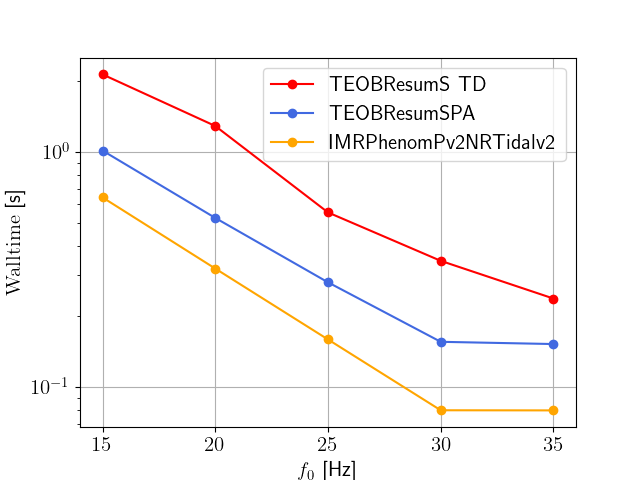}
\caption{\label{fig:timing}
Left panel: BBH evaluation time for a $q=1, M=60 \Msun$ precessing system containing the $(\ell, m) = (2,1), (2,2), (3,2),(3,3), (4,4)$ modes. 
Three different state of the art approximants are considered: \TEOB, ${\tt IMRPhenomXPHM}$ and ${\tt SEOBNRv4PHM}$. 
\TEOB~ is approximately three times slower than ${\tt IMRPhenomXPHM}$, and up to an order of magnitude faster than
${\tt SEOBNRv4PHM}$. Right panel: BNS evaluation time for a $q=1, M=3.5\Msun$ precessing system, 
whose waveform is constructed with the $(2,2)$ mode. TD denotes the standard time domain \TEOB{} model 
with SPA denoting the frequency domain version of Sec.~\ref{subsec:SPA}}
\end{figure*}

\begin{figure*}
\includegraphics[width=0.495\textwidth]{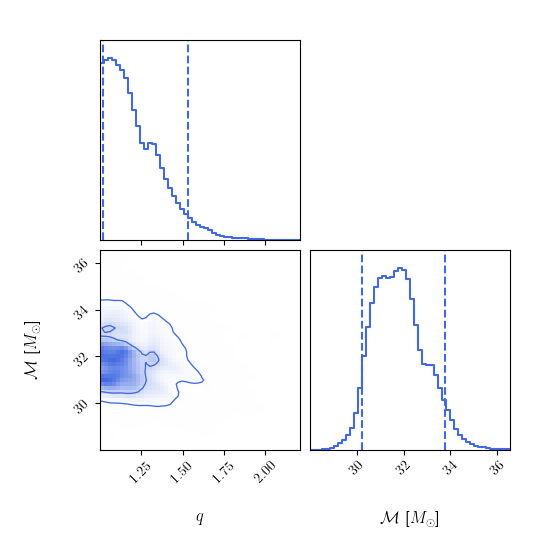} 
\includegraphics[width=0.495\textwidth]{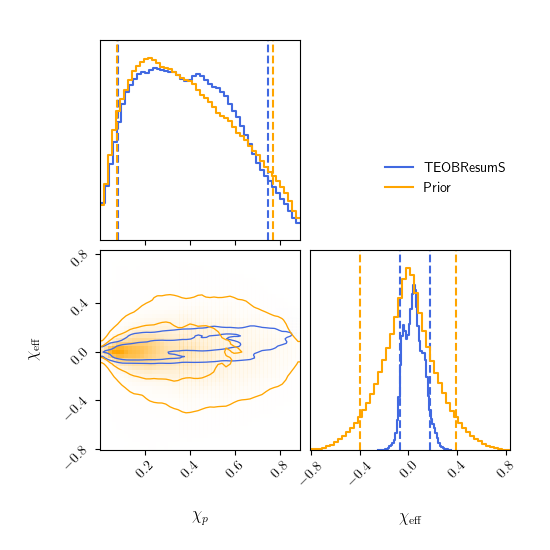}
\caption{\label{fig:GW150914}
Posteriors for chirp mass, mass ratio and spins obtained by analyzing the GW150914 data with \TEOB{}, as discussed in Sec.~\ref{Subsec:GW150914}. 
For comparison, we also plot the prior distributions for $\chi_{\rm eff}$ and $\chi_p$. As expected, the posterior distribution of $\chi_p$ is consistent with its prior.
The masses obtained in our analysis are consistent with the ones previously reported in Ref.~\cite{Breschi:2021wzr}, but the uncertainties on $q$ and $\mathcal{M}$ are larger. This is due to the addition of four degrees of freedom, namely the in-plane spin components.}
\end{figure*}
 
\begin{figure*}
\includegraphics[width=0.495\textwidth]{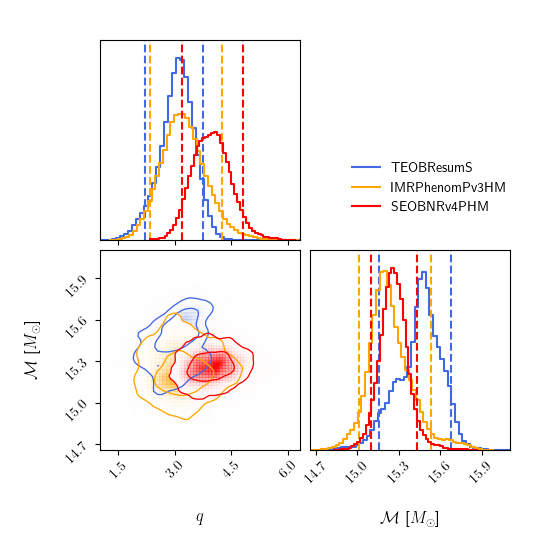}
\includegraphics[width=0.495\textwidth]{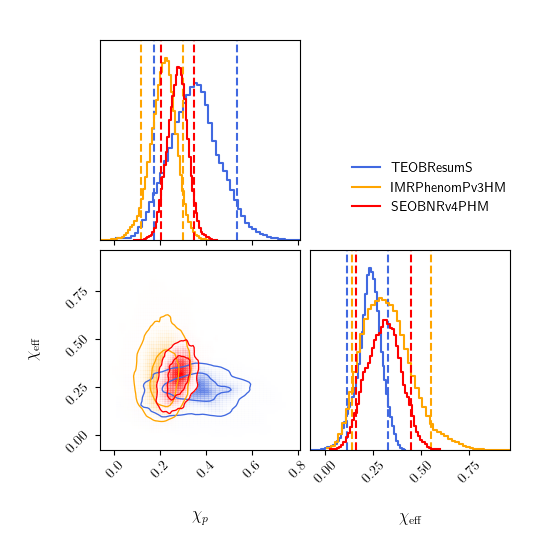}
\caption{\label{fig:GW190412}
Posteriors for chirp mass, mass ratio and spins obtained by analyzing the GW190412 data with \TEOB{}, as discussed in Sec.~\ref{Subsec:GW190412}. 
We compare our results to the public LVC posteriors obtained with the precessing models {\tt SEOBNRv4PHM} and {\tt IMRPhenomPv3HM}. 
The results of our analysis are broadly consistent with the ones obtained by LVC, although some model systematics are clearly present
between the three approximants.}
\end{figure*}

\begin{figure*}
\includegraphics[width=0.495\textwidth]{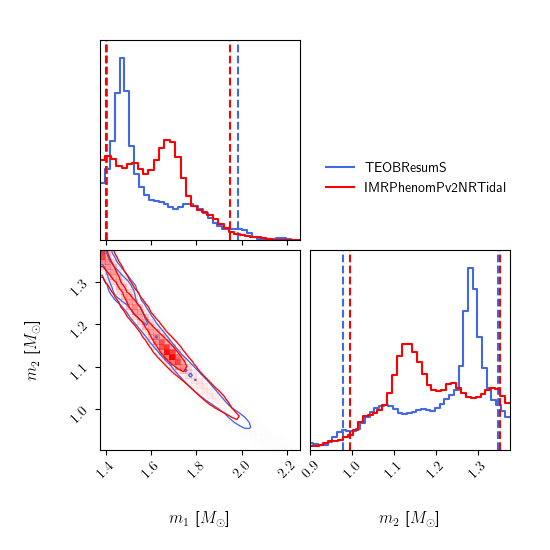}
\includegraphics[width=0.495\textwidth]{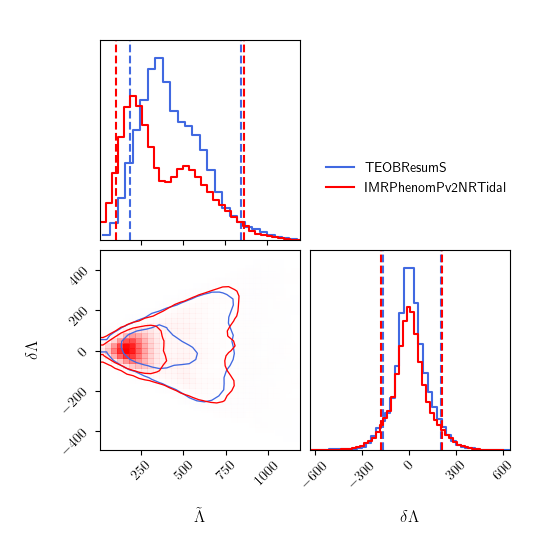}
\caption{\label{gw170817} Marginalized, two dimensional posteriors for detector frame masses (left) and tidal parameters (right) for GW170817, obtained with the precessing \TEOB{} model or the phenomenological ${\tt IMRPhenomPv2NRTidal}$ model, from the analysis of Ref.~\cite{LIGOScientific:2018mvr}. 
The $90\%$ intervals are compatible between the two models. We not that the ${\tt IMRPhenomPv2NRTidal}$ posteriors for $\tilde\Lambda$ display some bimodalities, which are due to the higher frequency cutoff employed for the analysis.}
\end{figure*}

\subsection{Comparison with \tt{NRSur7dq4}}
To extend the comparison to a larger number of binaries, we additionally computed $\bar{\mathcal{F}}_{\rm SM}$ between our model and the NR surrogate model $\tt{NRSur7dq4}$ using all modes with $\ell \leq 4$. 
We considered $5000$ systems with $q \in [1, 4]$, which is the calibration region of the surrogate, and spin magnitudes $\chi_{1,2} \in [0.1, 0.8]$ with uniformly distributed spin vector polar angles $\theta_{1,2} \in [0, \pi)$ and azimuthal angles $\phi_{1,2} \in [0, 2\pi)$.
We set the initial GW frequency to 20\,Hz for $M\ge 100 M_\odot$ and to a linearly decreasing function of $M$ from 37.5 to 20\,Hz as $M$ increases from
40 to $100\, M_\odot$. We do this because the surrogate can at most yield waveforms of length $4300M$
\cite{Varma:2019csw} so the lighter-mass inspirals need to start from higher frequencies.
Figure~\ref{fig:nrsurr} shows the distributions of the unfaithfulness obtained for 
inclinations of $\iota=0, \pi/6$,$\pi/3$ and $\pi/2$.
We find that, for $\iota=0$, $98.8\%$ of the systems considered have unfaithfulness below $3\%$ and
$87.8\%$ below $1\%$, with a global distribution 
spanning the range $\bar{\mathcal{F}}_{\rm SM} \in [0.0027, 0.04]$ with median $0.0046$.

As previously observed, the situation worsens as the inclination increases, 
with $\bar{\mathcal{F}}_{\rm SM} \in [0.003, 0.05]$ for $\iota=\pi/6$,
$\bar{\mathcal{F}}_{\rm SM} \in [0.003, 0.1]$ for $\iota=\pi/3$ and $\bar{\mathcal{F}}_{\rm SM} \in [0.006, 0.14]$ for $\iota=\pi/2$.

For $\iota=\pi/3$, only $80\%$ ($21\%$) of the total mismatches are below the $3\%$ ($1\%$) threshold.
The degradation of the unfaithfulness is observed especially for asymmetric binaries
with large $\chi_p$ as can be discerned by comparing the middle and bottom panels of Fig.~\ref{fig:nrsurr}.

\subsection{Waveform evaluation timing}

We now test the computational efficiency of our EOB model, and compare it to other state of the art
precessing approximants for BBH and BNS coalescences, ${\tt SEOBNRv4PHM}$, ${\tt IMRPhenomXPHM}$ and ${\tt IMRPhenomPv2NRTidalv2}$. 
We choose one reference equal mass BBH binary, with $M = 60 \Msun$ and $\bm{\chi}_1 = (-0.6,0.1,0.2)$, $\bm{\chi}_2 = (0.1,-0.5,-0.3)$, 
and a list of initial frequencies $f_0 = \{10, 12.5, 15, 17.5, 20., 22.5, 25, 27.5, 30\}$ Hz.
For each initial frequency $f_0^i$ we calculate the average time (over 20 repetitions) needed 
to evolve the binary and produce the $h_+$ and $h_\times$ polarization. This process is then
repeated for a BNS configuration with $q=1, M=2.8 \Msun$ and same spins as the previous BBH system, and 
a choice of initial frequencies $f_0 = \{15, 20, 25, 30, 35\}$ Hz.
We performed this test on a Huawei MateBook 14 with AMD Ryzen 5 2500U processors and 8 Gb RAM.

The results are displayed in Fig.~\ref{fig:timing}. We find that, for BBH systems, \TEOB{} 
is approximately three to four times slower than ${\tt IMRPhenomXPHM}$ and about one order of
magnitude faster than ${\tt SEOBNRv4PHM}$. For BNS systems, instead, the FD model is about 
two times faster than its TD counterpart, and two times slower than the phenomenological $\tt IMRPhenomPv2NRTidalv2$.
We highlight that the main evaluation cost for both the TD and FD \TEOB{} models comes from
the twisting procedure itself, rather than from the solution of the two (PN and EOB) dynamics ODE
systems. 

\section{Parameter estimation}
\label{Sec:PE}

We demonstrate possible applications of our model by performing 
PE on real GW data.
We re-analyze the data of GW150914 and GW190412, and show that the posteriors
obtained are consistent with those presented in, e.g., Refs.~\cite{LIGOScientific:2018mvr, LIGOScientific:2020ibl}. Then, we analyze GW170817 ~\cite{TheLIGOScientific:2017qsa, Abbott:2018wiz,Abbott:2018exr} and compute the radius of a NS of mass $M \in [1.4, 2.1] \Msun$ using the fits of Ref.~\cite{Godzieba:2021vnz}.
All of our PE studies are performed with the \bajes{} pipeline \cite{Breschi:2021wzr}
and the  ${\tt dynesty}$ \cite{Speagle:2020} sampler.

\subsection{GW150914}
\label{Subsec:GW150914}
GW150914 \cite{Abbott:2016blz, LIGOScientific:2018mvr} was the first BBH event observed by the LIGO collaboration. 
For our study, we consider 8\,seconds of data
centered around the GPS time of the event. We employ 4096 livepoints, and analyze the frequencies between $20$ and $1024$ Hz. We fix the sampling rate to 4096 Hz, and sample the component masses enforcing that
the chirp mass lies in $\mathcal{M} \in [12.3, 45]M_\odot$, the mass ratio $q \in [1, 8]$, and the spin
magnitudes $|\chi_i| \in [0, 0.89]$ with $i =1,2$ with an isotropic prior for the tilt angles. We consider all modes up to $\ell=4$, and marginalize over the timeshift.
Finally, we employ 10 calibration nodes, and the PSD given in Ref.~\cite{LIGOScientific:2018mvr}.
Fig.~\ref{fig:GW150914} displays the posteriors we recovered from our analysis. We find that $\mathcal{M} = 31.7^{+2.0}_{-1.5} \Msun$, $q = 1.17^{+0.36}_{-0.16}$, $\chi_{\rm eff} = 0.04^{+0.09}_{-0.08}$ and $\chi_{\rm p} = 0.38^{+0.37}_{-0.29}$. 
Our results are consistent with the analyses presented in Refs.~\cite{Abbott:2016blz, TheLIGOScientific:2016wfe, LIGOScientific:2018mvr}, performed with other approximants, and with the PE conducted in Ref.~\cite{Breschi:2021wzr}, which employed the non-precessing version of $\TEOB$. 
The posteriors of $\chi_{\rm p}$ are consistent with the prior as GW150914 displays no evidence of precession. Notably, the introduction of additional spin components widens the credible intervals on the component masses with respect to the analysis of Ref.~\cite{Breschi:2021wzr}, obtained with the same approximant and similar settings.

\subsection{GW190412}
\label{Subsec:GW190412}
GW190412 was the first highly asymmetrical BBH event ($q\approx 3-4$), which was also 
one of the ``louder'' events of O3a with an SNR of $~19$ \cite{LIGOScientific:2020ibl, LIGOScientific:2020stg}.
The original LVC data analysis has yielded well constrained imprints of spin precession with $ 0.15 \lesssim \chi_\text{p}\lesssim 0.5$ and $\theta_1 = 0.80^{+0.52}_{-0.36}$ \cite{LIGOScientific:2020ibl}, support for $\chi_\text{eff}>0$ with 95\% credibility \cite{LIGOScientific:2020ibl}, and clear evidence of the subdominant modes carrying a significant portion of the signal SNR.
A number of following studies have further improved on the 
original analysis by investigating in more detail the effects of the higher modes \cite{Islam:2020reh} and of the chosen priors \cite{Colleoni:2020tgc} on the PE. 
The same works also carried out studies to understand the differences observed when 
different waveform models are employed to analyze the signal.
Although such detailed investigations lie beyond the scope of this paper, 
it is clear that the exceptional nature of GW190412 makes it very desirable to analyze with \TEOB.

We employ 4096 livepoints and analyze the frequencies between $20$ and $1024$ Hz with a fixed sampling rate of $4096$ Hz. 
We sample in the 
component masses, requiring that the chirp mass falls in $\mathcal{M} \in [8, 20]M_\odot$ and the mass ratio $q \in [1, 10]$. We sample in spin
magnitudes $|\chi_i| \in [0, 0.89]$ with $i =1,2$, enforcing an isotropic prior for the tilt angles. Once again, we consider all modes up to $\ell=4$, and marginalize over the timeshift.

Posteriors for the masses and spins are plotted in Fig.~\ref{fig:GW190412}. We compare the results obtained in our PE with the publicly available LVC posterior samples, obtained with the two independent models ${\tt SEOBNRv4PHM}$ \cite{Ossokine:2020kjp} and ${\tt IMRPhenomPv3HM}$ \cite{Khan:2019kot}. 
We find that $\TEOB$ gives estimates of GW190412 parameters that are overall consistent with those computed from the 
other two approximants. We obtain a slightly larger chirp mass, and overall wider $\chi_p$ and tighter $\chi_{\rm eff}$ posteriors.


\subsection{GW170817}
\label{Subsec:GW170817}

\begin{figure}
\includegraphics[width=0.5\textwidth]{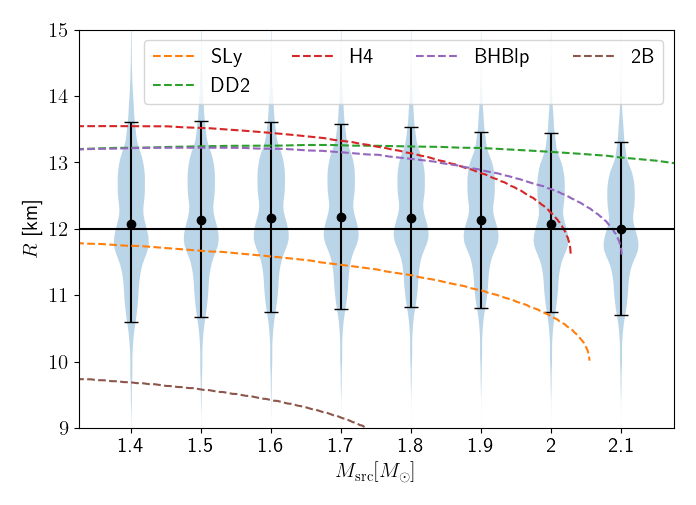}
\caption{\label{gw170817_radius} Marginalized posterior distributions of $R(M)$ computed via the tidal parameters and source masses obtained from our GW170817 analysis and the fits of Ref.~\cite{Godzieba:2021vnz}. Overlayed, we also display the $R(M)$ relations for a handful of well-known equations of state.}
\end{figure}

GW170817 was the first BNS inspiral/merger event. To date, it is still the loudest detected GW event with an SNR of 32 \cite{TheLIGOScientific:2017qsa}.
Though observations from millisecond pulsars yield at most dimensionless spins of $\chi \approx 0.5$ \cite{Hessels:2006ze} and the fastest observed
neutron star spin in electromagnetically observed binary pulsars is $\chi \lesssim 0.05$ \cite{Kramer:2009zza, Stovall:2018ouw},
the spins of the components of GW170817 are not well constrained \cite{TheLIGOScientific:2017qsa, LIGOScientific:2018mvr}.
For our re-analysis we employ 6000 livepoints and consider 128 seconds around the GPS time of the event, analyzing the frequencies between $20$ and $1024$ Hz to minimize waveform systematics \cite{Gamba:2020ljo}.
We sample the component masses imposing that $\mathcal{M} \in [1.1, 1.3] \Msun$ and $q \in [1, 3]$. The dimensionless spin magnitudes are sampled in the interval $[0, 0.89]$, with an isotropic prior for the tilt angles. We sample the dimensionless tidal deformabilities $\Lambda_1$, $\Lambda_2$ over a uniform prior $[5,5000]$.
Figure ~\ref{gw170817} displays the marginalized, two-dimensional posteriors for the detector masses $m_1$, $m_2$
of the neutron stars and the tidal parameters $\tilde\Lambda$ and $\delta\Lambda$, which parameterize the LO and NLO tidal corrections to the PN GW phase \cite{Wade:2014vqa}.
The masses are slightly bimodal. This effect is not unexpected, and has already been previously observed \cite{Abbott:2018wiz, LIGOScientific:2018mvr}.
Evidently, it is related to the modelling of spin precession: on the one hand, allowing spin magnitudes 
to vary in the large interval $[0, 0.89]$ increases the correlations between spins and mass ratio; on the other hand, precession effects can more easily fit features of the data which might be due to the noise.
This event too, much like {GW150914}, does not display evidence for precession or spinning components.
Indeed, we find that $\chi_p$ is consistent with its prior, and $\chi_{\rm eff} = 0.01^{+0.04}_{-0.02}$. Finally, we measure $\tilde\Lambda = 406^{+238}_{-150}$. This value is marginally larger than the one 
obtained with the {\tt IMRPhenomPv2NRTidal} model, consistently with Ref.~\cite{Gamba:2020wgg}, but slightly
smaller than the one obtained with the aligned spin model. By employing the fits of Ref.~\cite{Godzieba:2021vnz}, 
we map the source frame masses and the $\tilde\Lambda$ into posteriors for the radius $R$ of a NS with mass in $[1.4, 2.1] \Msun$ (See Fig. ~\ref{gw170817_radius}). Over this mass interval we find that the fits give values of $R$
which are weakly dependent on the mass of the neutron star, and $R \sim 12.0^{+1.5}_{-1.2}$ km.

\section{Conclusions}\label{Sec:conclusions}

In this work we have presented a new efficient multipolar EOB model for generic-spin binaries

The model presented builds on the previous work of Ref.~\cite{Akcay:2020qrj} and improves it by including a description for merger and ringdown, higher modes and time-evolving projected spin components. 
We also constructed an inspiral-merger frequency domain model for precessing BNS systems, which
can incorporate all the advancements added to the time domain approximant, and is -- to our knowledge -- 
the first multipolar, tidal and precessing frequency domain approximant for these systems.

We have investigated different realizations of radiation reaction included in the spin dynamics via the $\dot{\omega}(\omega)$ relation. 
By comparing hybrid EOB-PN expressions with a pure PN expression and the ``aligned'' EOB relation, we  
observed that the PN expression for $\dot{\omega}(\omega)$ employed in the previous work provides a satisfactory description of the EOB radiation reaction. We also presented a preliminary investigation of the importance of $m=0$ modes in asymmetrical binaries, showing that the contribution of the $h^T_{\ell, 0}$ modes to the total 
waveform polarizations is non-negligible close to merger.

We then validated \TEOB{} IMR BBH model using a total of $120$ SXS simulations, spanning a large portion of the precessing BBHs parameter space. We found that $96\%$ ($99\%$, $98\%$) of the total mismatches lie below the $3\%$ unfaithfulness threshold for $\iota=0$ ($\iota=\pi/3, \pi/2$). 
We also computed the mismatch between a subset of the same SXS simulations and the state-of-the-art phenomenological waveform approximant $\tt IMRPhenomXPHM$. We found good
consistency between the mismatches obtained with the two waveform models.
A similar comparison was performed against the NR surrogate
${\tt NRSur7dq4}$: when considering a large number of systems, which cover the surrogate's parameter space, we found that $98.8\%$ ($80\%$, $67\%$) of the systems considered have unfaithfulness below $3\%$ and
$87.8\%$ ($21\%$, $2\%$) below $1\%$ for $\iota=0$ ($\iota=\pi/3, \pi/2$). The worsening of the unfaithfulness for increasing inclination is expected, as 
for more edge-on binaries geometrical effects enhance the importance of higher modes and precession, which in turn deteriorate the EOB-NR agreement 
during the merger and ringdown phases of the coalescence.

Finally, we applied the model to the PE of three real LVC signals, namely GW150914, GW190412 and GW170817, and obtained results consistent with currently published analyses by relying on the $\tt bajes$ PE infrastructure and the ${\tt dynesty}$ sampler.
GW150914 and GW170817 display no evidence for precession, with $\chi_p$ closely following its prior distribution. 
Conversely, GW190412 is clearly found to be an asymmetrical, mildly precessing system with mass ratio $q \sim 3$
and $\chi_{\rm p} \sim 0.35 $, $\chi_{\rm eff} \sim 0.25$. Such values are compatible
with the ones recovered by the LVK collaboration employing the precessing waveform approximants
$\tt SEOBNRv4PHM$ and $\tt IMRPhenomPv3HM$. Marginal differences due to waveform systematics are nonetheless found 
in $\chi_p$ and $\chi_{\rm eff}$: we recover a tighter posterior distribution of the latter and a wider distribution of the former, as well as a slighty larger chirp mass. 
Critically, these studies demonstrate that $\tt TEOBResumS$ can be directly applied to 
PE, even of computationally challenging BNS systems, without the need for additional surrogates or reduced order models.

In spite of the satisfactory performance of the model for current detectors,
some work remains to be done in view of the continuously-increasing sensitivity of the instruments. 
In particular, the degradation of the performance of the model at high masses seen in Fig.~\ref{fig:sxs_short_eobnrmm} indicates
the need for an improved merger-ringdown description of the precessing waveform. 
This could come from a combination of three different yet complementary avenues: (i) an improved description of the ringdown of spin-aligned 
$(\ell, m) \neq (2,2)$ modes; (ii) a more accurate model for the evolution of the Euler angles $\alpha$, $\beta$ 
beyond the merger; (iii) an improved (analytical) fit for the remnant spin $\chi_f$ (see Fig.~\ref{fig:final_state}).

Regarding the first point, in particular, we mention that the Achille's heel of the aligned-spin \TEOB{} is the 
modeling of the $(\ell, m) = (2,1)$ mode, which is known to become inaccurate close to merger for large spins 
anti-aligned with the orbital angular momentum ($\chi^z_i < -0.8$)~\cite{Nagar:2020pcj}. This known issue is potentially even more important for precessing systems. 
That is because, clearly, the $\textit{twisted}$ modes are obtained as a superposition of spin-aligned multipoles. 
Hence, the $(2,1)$ mode directly affects also the $(2,2)$ and $(2,0)$ multipolar waveforms. However, once an appropriate 
solution for the issue is found for the spin-aligned case, this will immediately have a positive impact on the precessing model.  
Similarly, any improvement of the spin-aligned model (addition of analytical information, re-calibrations of the NR informed 
parameters, improved merger-ringdown) will \textit{immediately} be reflected on the precessing waveform, thanks to the modular nature
of our approximant.
 
With respect to the second point, instead, we observe that Ref.~\cite{Hamilton:2021pkf} recently proposed a phenomenological model to extend 
the Euler angles beyond merger, directly fit to NR simulations.
Since the spin evolution is independently evolved in our model, it is in principle straightforward to apply the model of Ref.~\cite{Hamilton:2021pkf} to our EOB waveform.\footnote{Notably, the same authors also highlight the need to move beyond the spin-aligned description of the co-precessing waveform.}

To summarize, the \TEOB \ model represents a new state-of-the-art, robust, faithful and efficient 
alternative to already existing waveform models, which we hope will prove useful to the gravitational 
wave community in the effort of interpreting GW data and understanding the nature of BNS and BBH
 systems in the years to come.

\begin{acknowledgments}
We thank Alessandro Nagar for helpful discussions and a careful reading of the manuscript. We thank Geraint Pratten for sharing with
us the list of 99 ``short'' {\tt SXS} simulations employed for the mismatch computations.
SA thanks Marta Colleoni and Jonathan Thompson for help with {\tt IMRPhenomXPHM}.
R.~G. acknowledges support from the Deutsche Forschungsgemeinschaft (DFG) under Grant No. 406116891 within the Research Training Group RTG 2522/1. 
S.~B. and S.~A. acknowledge support by the EU H2020 under ERC Starting Grant, no.~BinGraSp-714626.
S.~A. and J.~W. acknowledge support from the University College Dublin Ad Astra Fellowship.
The data analyses were performed on the supercomputer ARA at Jena. We acknowledge the computational resources provided
by Friedrich Schiller University Jena, supported in part by DFG grants
INST 275/334-1 FUGG and INST 275/363-1 FUGG and by EU H2020 ERC Starting Grant, no.~BinGraSp-714626.
Data postprocessing was performed on the Virgo ``Tullio'' server 
in Torino, supported by INFN.
\noindent
\TEOB{} is publicly developed and available at

\url{https://bitbucket.org/eob_ihes/teobresums/}.

\noindent
The precession implementation (branch) will be part of the next stable release {\rm v3}.

\bajes{} is publicly available at

\url{https://github.com/matteobreschi/bajes}

\noindent
This research has made use of data, software and/or web tools obtained 
from the Gravitational Wave Open Science Center (\url{https://www.gw-openscience.org}), 
a service of LIGO Laboratory, the LIGO Scientific Collaboration and the 
Virgo Collaboration. We have additionally employed the computational resources of
LIGO Scientific Collaboration DataGrid.
LIGO is funded by the U.S. National Science Foundation. 
Virgo is funded by the French Centre National de Recherche Scientifique (CNRS), 
the Italian Istituto Nazionale della Fisica Nucleare (INFN) and the 
Dutch Nationaal instituut voor subatomaire fysica (Nikhef), with contributions by Polish and Hungarian institutes.
\end{acknowledgments}

\appendix

\section{The analytic expression for $\alpha(0)$}\label{app:alphaIC}

\begin{figure}
\includegraphics[scale=.55]{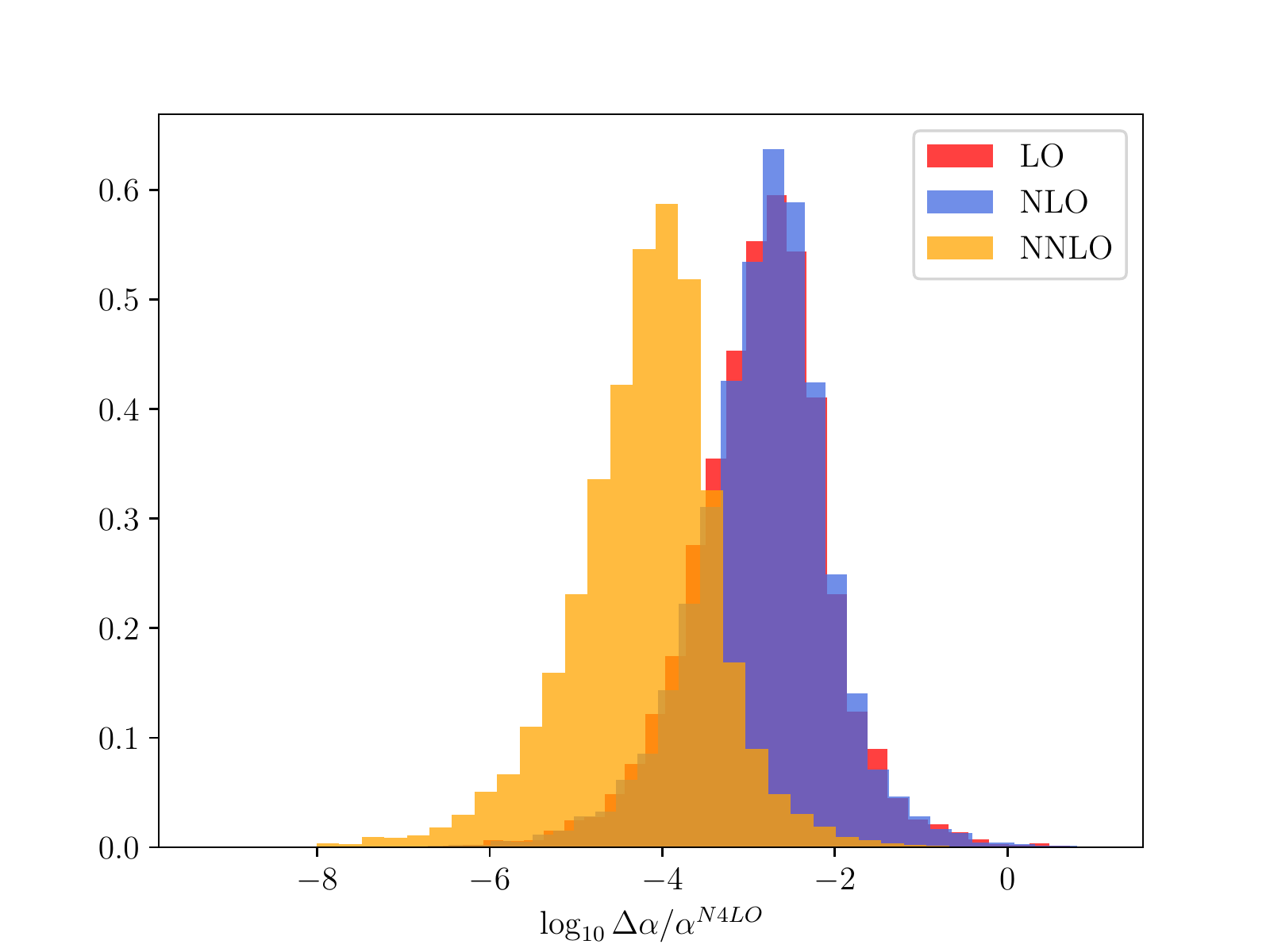}
\caption{\label{fig:alpha0_comparison}
Distributions of the relative differences between $\alpha(0)_{\rm N4LO}$ and $\alpha(0)_X$, with $\rm X= LO, NLO, NNLO$. The $9000$ points considered were sampled over the space
$q\in\ [0.1,1)$, $\theta_1,\theta_2\in[0,\pi)$, and
  $\phi_2\in[0,2\pi)$. 
  The median of the relative difference between the N4LO and LO or NLO distribution is $\Delta\alpha/\alpha \sim 10^{-3}$, about one 
  order of magnitude larger than the difference between the NNLO and N4LO expressions. Given the analytical 
  simplicity of the NLO expression, we decided to employ it for our current implementation for the $\alpha(0)$ computation.}
\end{figure}

At next-to-leading order (NLO), the $q\ge 1$ version of the orbital angular momentum and spin precession ODEs are given by Eqs.~(\ref{eq:Sidot_NLO}, \ref{eq:LNdot_NLO}) with the following initial conditions

\begin{align}
    \mathbf{S}_1(0)&=m_1^2(\chi_{1x,0},\chi_{1y,0},\chi_{1z,0}),\\
    \mathbf{S}_2(0)&=m_2^2(\chi_{2x,0},\chi_{2y,0},\chi_{2z,0}),
\end{align}

\begin{equation}
    \Lhat(0)=(0,0,1).
\end{equation}
Recall the definition of $\alpha$:
\begin{equation}
    \alpha=\arctan\left(\frac{\mathbf{L}_\text{Ny}}{\mathbf{L}_\text{Nx}}\right),
\end{equation}
which is initially undefined as $\LN$ points along the $z$ axis. 
However, as we explained in Sec.~\ref{Sec:model}, we mitigate this by using the initial torque instead as follows
\begin{equation}
  \alpha(0) = \arctan\left(\f{\Lhatdot{}_y(0)}{\Lhatdot{}_x(0)} \right) \label{eq:alpha0_new},
\end{equation}
where we employ the initial $x,y$ components of the $\Lhatdot$ ODE.
At NLO, this yields
\be
\alpha(0)=\arctan\left(-\frac{q (3+4q)\chi_{1x,0}+(4+3 q)\chi_{2x,0}}{q (3+4q)\chi_{1y,0}+(4+3 q)\chi_{2y,0}}\right). \label{eq:alpha0_NLO}
\ee

Eq.~\eqref{eq:alpha0_new} can be straightforwardly extended to higher orders.
Figure ~\ref{fig:alpha0_comparison} shows the relative difference between $\alpha(0)$ computed at N4LO and lower orders.

\section{Numerical relativity data}\label{app:nrdata}
\begin{figure*}
\includegraphics[scale=.33]{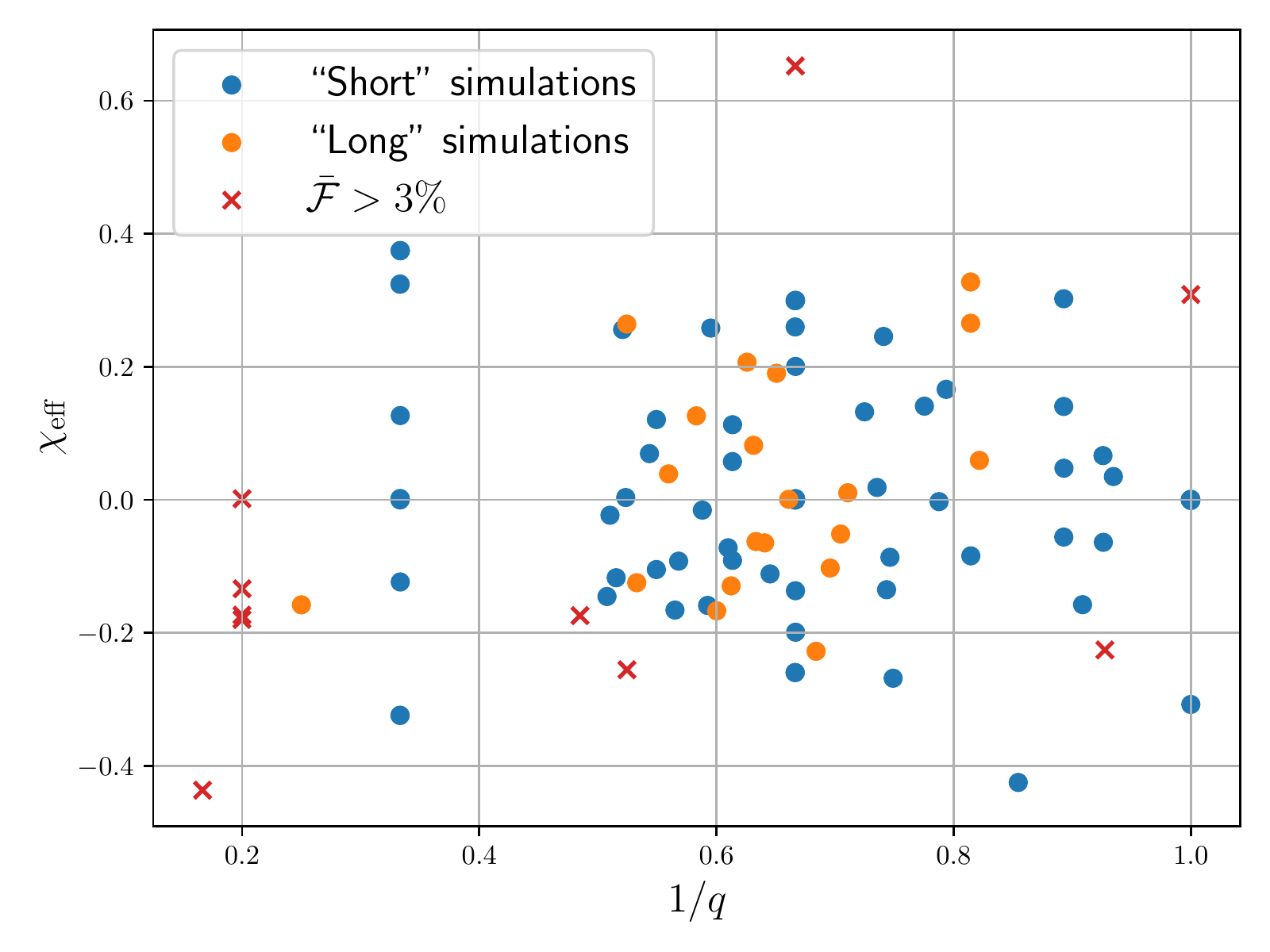}
\includegraphics[scale=.33]{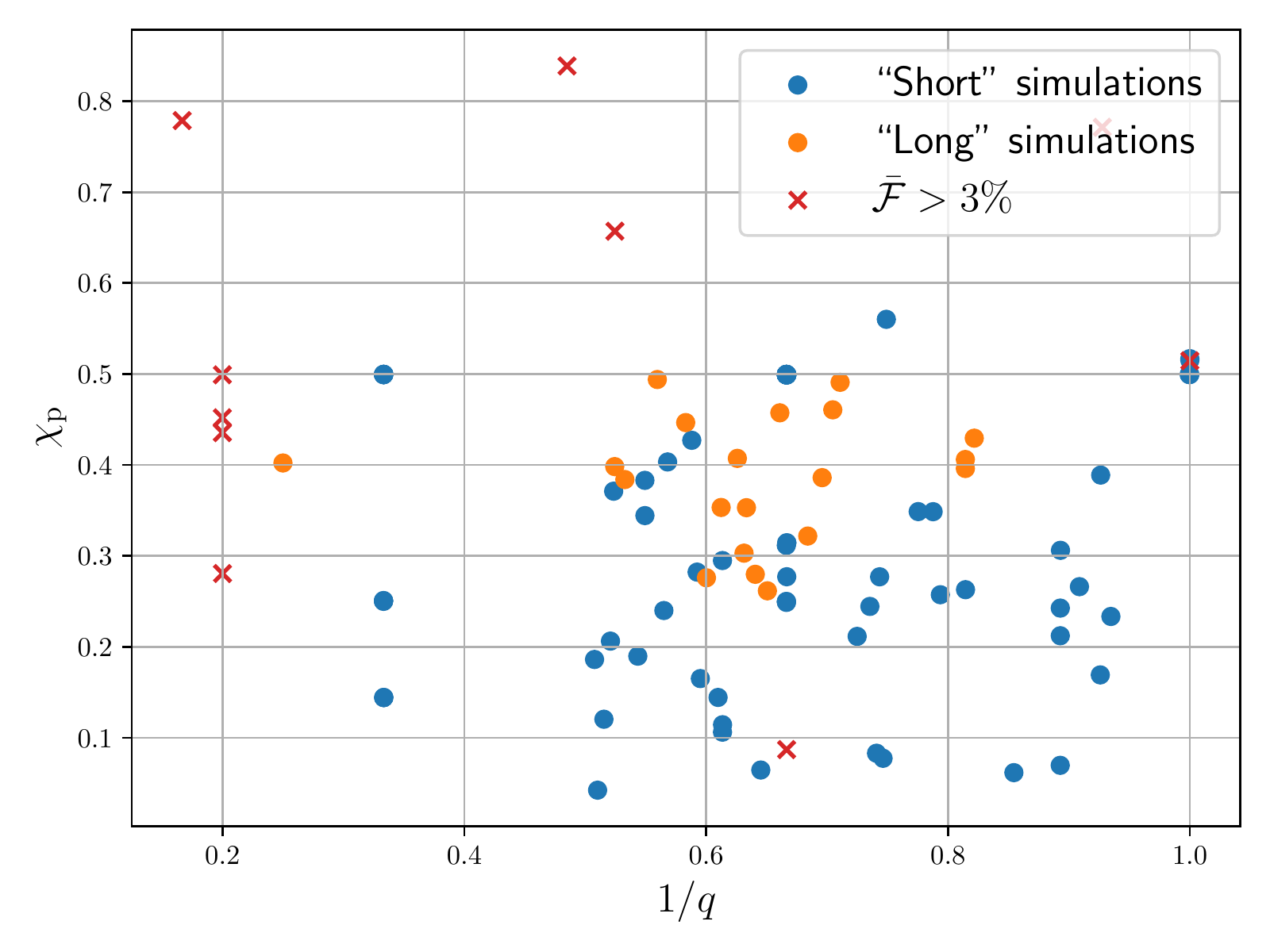}
\includegraphics[scale=.33]{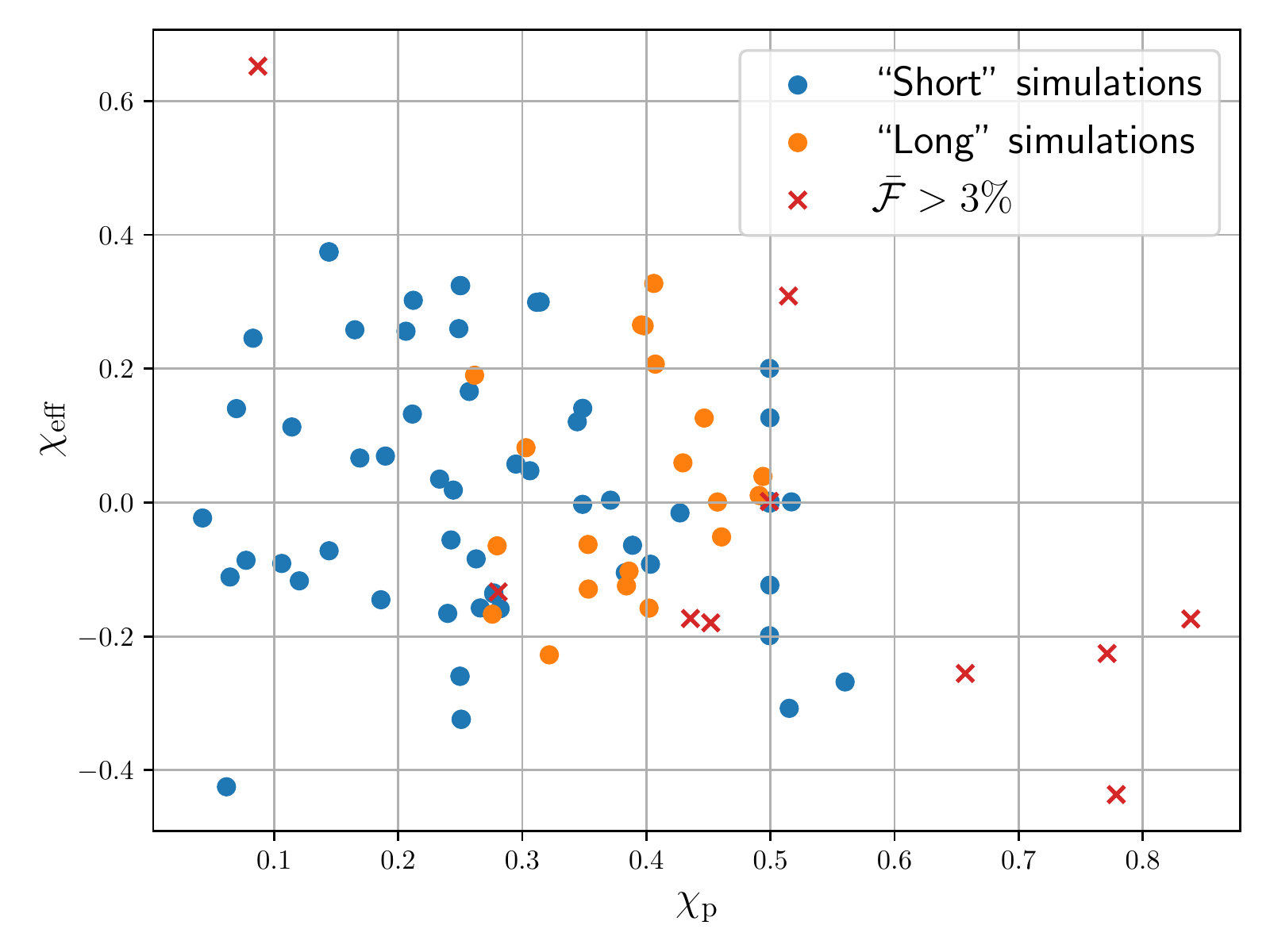}
\caption{\label{fig:nrdata} 
Properties of the two sets of NR data considered in Sec.~\ref{Sec:validation} \cite{Buchman:2012dw,Lovelace:2008tw,Pfeiffer:2007yz,Caudill:2006hw,Cook:2004kt} to validate our EOB model. Different colors denote the two simulations sets employed. Crosses are used to indicate configurations for which the maximum EOB/NR unfaithfulness 
$\bar{\mathcal{F}}_M = \max_{M, \iota} \bar{\mathcal{F}}$ surpasses the $3\%$ threshold.}
\end{figure*}

Figure~\ref{fig:nrdata} shows the properties of the two sets of NR data considered in this study \cite{Buchman:2012dw,Lovelace:2008tw,Pfeiffer:2007yz,Caudill:2006hw,Cook:2004kt}
in terms of inverse mass ratio $m_2/m_1= 1/q$, effective spin parameter $\chi_{\rm eff}$
and precessing spin parameter $\chi_{\rm p}$. 

For the two sets consided, we have $q \in [1, 6]$, $\chi_{p} \in [0.0424, 0.7787]$ and 
$\chi_{\rm eff} \in [-0.4364, 0.6522]$ (``short" simulations) and 
$q \in [1.217, 4]$, $\chi_p \in [0.2616, 0.4940]$, $\chi_{\rm eff} \in [-0.2276, 0.3275]$ (``long" simulations).

\input{paper20220609.bbl}

\end{document}

%% file: paper20220609.bbl
%

%% file: paper20220609.bbl
\begin{thebibliography}{128}%
\makeatletter
\providecommand \@ifxundefined [1]{%
 \@ifx{#1\undefined}
}%
\providecommand \@ifnum [1]{%
 \ifnum #1\expandafter \@firstoftwo
 \else \expandafter \@secondoftwo
 \fi
}%
\providecommand \@ifx [1]{%
 \ifx #1\expandafter \@firstoftwo
 \else \expandafter \@secondoftwo
 \fi
}%
\providecommand \natexlab [1]{#1}%
\providecommand \enquote  [1]{``#1''}%
\providecommand \bibnamefont  [1]{#1}%
\providecommand \bibfnamefont [1]{#1}%
\providecommand \citenamefont [1]{#1}%
\providecommand \href@noop [0]{\@secondoftwo}%
\providecommand \href [0]{\begingroup \@sanitize@url \@href}%
\providecommand \@href[1]{\@@startlink{#1}\@@href}%
\providecommand \@@href[1]{\endgroup#1\@@endlink}%
\providecommand \@sanitize@url [0]{\catcode `\\12\catcode `\$12\catcode
  `\&12\catcode `\#12\catcode `\^12\catcode `\_12\catcode `\%12\relax}%
\providecommand \@@startlink[1]{}%
\providecommand \@@endlink[0]{}%
\providecommand \url  [0]{\begingroup\@sanitize@url \@url }%
\providecommand \@url [1]{\endgroup\@href {#1}{\urlprefix }}%
\providecommand \urlprefix  [0]{URL }%
\providecommand \Eprint [0]{\href }%
\providecommand \doibase [0]{http://dx.doi.org/}%
\providecommand \selectlanguage [0]{\@gobble}%
\providecommand \bibinfo  [0]{\@secondoftwo}%
\providecommand \bibfield  [0]{\@secondoftwo}%
\providecommand \translation [1]{[#1]}%
\providecommand \BibitemOpen [0]{}%
\providecommand \bibitemStop [0]{}%
\providecommand \bibitemNoStop [0]{.\EOS\space}%
\providecommand \EOS [0]{\spacefactor3000\relax}%
\providecommand \BibitemShut  [1]{\csname bibitem#1\endcsname}%
\let\auto@bib@innerbib\@empty
\bibitem [{\citenamefont {Abbott}\ \emph
  {et~al.}(2016{\natexlab{a}})\citenamefont {Abbott} \emph
  {et~al.}}]{Abbott:2016blz}%
  \BibitemOpen
  \bibfield  {author} {\bibinfo {author} {\bibfnamefont {B.~P.}\ \bibnamefont
  {Abbott}} \emph {et~al.} (\bibinfo {collaboration} {Virgo, LIGO
  Scientific}),\ }\href {\doibase 10.1103/PhysRevLett.116.061102} {\bibfield
  {journal} {\bibinfo  {journal} {Phys. Rev. Lett.}\ }\textbf {\bibinfo
  {volume} {116}},\ \bibinfo {pages} {061102} (\bibinfo {year}
  {2016}{\natexlab{a}})},\ \Eprint {http://arxiv.org/abs/1602.03837}
  {arXiv:1602.03837 [gr-qc]} \BibitemShut {NoStop}%
\bibitem [{\citenamefont {Abbott}\ \emph
  {et~al.}(2019{\natexlab{a}})\citenamefont {Abbott} \emph
  {et~al.}}]{Abbott:2018wiz}%
  \BibitemOpen
  \bibfield  {author} {\bibinfo {author} {\bibfnamefont {B.~P.}\ \bibnamefont
  {Abbott}} \emph {et~al.} (\bibinfo {collaboration} {LIGO Scientific,
  Virgo}),\ }\href {\doibase 10.1103/PhysRevX.9.011001} {\bibfield  {journal}
  {\bibinfo  {journal} {Phys. Rev.}\ }\textbf {\bibinfo {volume} {X9}},\
  \bibinfo {pages} {011001} (\bibinfo {year} {2019}{\natexlab{a}})},\ \Eprint
  {http://arxiv.org/abs/1805.11579} {arXiv:1805.11579 [gr-qc]} \BibitemShut
  {NoStop}%
\bibitem [{\citenamefont {Abbott}\ \emph {et~al.}(2018)\citenamefont {Abbott}
  \emph {et~al.}}]{Abbott:2018exr}%
  \BibitemOpen
  \bibfield  {author} {\bibinfo {author} {\bibfnamefont {B.~P.}\ \bibnamefont
  {Abbott}} \emph {et~al.} (\bibinfo {collaboration} {LIGO Scientific,
  Virgo}),\ }\href {\doibase 10.1103/PhysRevLett.121.161101} {\bibfield
  {journal} {\bibinfo  {journal} {Phys. Rev. Lett.}\ }\textbf {\bibinfo
  {volume} {121}},\ \bibinfo {pages} {161101} (\bibinfo {year} {2018})},\
  \Eprint {http://arxiv.org/abs/1805.11581} {arXiv:1805.11581 [gr-qc]}
  \BibitemShut {NoStop}%
\bibitem [{\citenamefont {Abbott}\ \emph
  {et~al.}(2016{\natexlab{b}})\citenamefont {Abbott} \emph
  {et~al.}}]{Abbott:2016nmj}%
  \BibitemOpen
  \bibfield  {author} {\bibinfo {author} {\bibfnamefont {B.~P.}\ \bibnamefont
  {Abbott}} \emph {et~al.} (\bibinfo {collaboration} {Virgo, LIGO
  Scientific}),\ }\href {\doibase 10.1103/PhysRevLett.116.241103} {\bibfield
  {journal} {\bibinfo  {journal} {Phys. Rev. Lett.}\ }\textbf {\bibinfo
  {volume} {116}},\ \bibinfo {pages} {241103} (\bibinfo {year}
  {2016}{\natexlab{b}})},\ \Eprint {http://arxiv.org/abs/1606.04855}
  {arXiv:1606.04855 [gr-qc]} \BibitemShut {NoStop}%
\bibitem [{\citenamefont {Abbott}\ \emph
  {et~al.}(2017{\natexlab{a}})\citenamefont {Abbott} \emph
  {et~al.}}]{Abbott:2017gyy}%
  \BibitemOpen
  \bibfield  {author} {\bibinfo {author} {\bibfnamefont {B.~P.}\ \bibnamefont
  {Abbott}} \emph {et~al.} (\bibinfo {collaboration} {Virgo, LIGO
  Scientific}),\ }\href {\doibase 10.3847/2041-8213/aa9f0c} {\bibfield
  {journal} {\bibinfo  {journal} {Astrophys. J.}\ }\textbf {\bibinfo {volume}
  {851}},\ \bibinfo {pages} {L35} (\bibinfo {year} {2017}{\natexlab{a}})},\
  \Eprint {http://arxiv.org/abs/1711.05578} {arXiv:1711.05578 [astro-ph.HE]}
  \BibitemShut {NoStop}%
\bibitem [{\citenamefont {Abbott}\ \emph
  {et~al.}(2017{\natexlab{b}})\citenamefont {Abbott} \emph
  {et~al.}}]{Abbott:2017oio}%
  \BibitemOpen
  \bibfield  {author} {\bibinfo {author} {\bibfnamefont {B.~P.}\ \bibnamefont
  {Abbott}} \emph {et~al.} (\bibinfo {collaboration} {Virgo, LIGO
  Scientific}),\ }\href {\doibase 10.1103/PhysRevLett.119.141101} {\bibfield
  {journal} {\bibinfo  {journal} {Phys. Rev. Lett.}\ }\textbf {\bibinfo
  {volume} {119}},\ \bibinfo {pages} {141101} (\bibinfo {year}
  {2017}{\natexlab{b}})},\ \Eprint {http://arxiv.org/abs/1709.09660}
  {arXiv:1709.09660 [gr-qc]} \BibitemShut {NoStop}%
\bibitem [{\citenamefont {Abbott}\ \emph
  {et~al.}(2017{\natexlab{c}})\citenamefont {Abbott} \emph
  {et~al.}}]{Abbott:2017vtc}%
  \BibitemOpen
  \bibfield  {author} {\bibinfo {author} {\bibfnamefont {B.~P.}\ \bibnamefont
  {Abbott}} \emph {et~al.} (\bibinfo {collaboration} {VIRGO, LIGO
  Scientific}),\ }\href {\doibase 10.1103/PhysRevLett.118.221101} {\bibfield
  {journal} {\bibinfo  {journal} {Phys. Rev. Lett.}\ }\textbf {\bibinfo
  {volume} {118}},\ \bibinfo {pages} {221101} (\bibinfo {year}
  {2017}{\natexlab{c}})},\ \Eprint {http://arxiv.org/abs/1706.01812}
  {arXiv:1706.01812 [gr-qc]} \BibitemShut {NoStop}%
\bibitem [{\citenamefont {Abbott}\ \emph
  {et~al.}(2019{\natexlab{b}})\citenamefont {Abbott} \emph
  {et~al.}}]{LIGOScientific:2018mvr}%
  \BibitemOpen
  \bibfield  {author} {\bibinfo {author} {\bibfnamefont {B.~P.}\ \bibnamefont
  {Abbott}} \emph {et~al.} (\bibinfo {collaboration} {LIGO Scientific,
  Virgo}),\ }\href {\doibase 10.1103/PhysRevX.9.031040} {\bibfield  {journal}
  {\bibinfo  {journal} {Phys. Rev.}\ }\textbf {\bibinfo {volume} {X9}},\
  \bibinfo {pages} {031040} (\bibinfo {year} {2019}{\natexlab{b}})},\ \Eprint
  {http://arxiv.org/abs/1811.12907} {arXiv:1811.12907 [astro-ph.HE]}
  \BibitemShut {NoStop}%
\bibitem [{\citenamefont {Venumadhav}\ \emph {et~al.}(2019)\citenamefont
  {Venumadhav}, \citenamefont {Zackay}, \citenamefont {Roulet}, \citenamefont
  {Dai},\ and\ \citenamefont {Zaldarriaga}}]{Venumadhav:2019tad}%
  \BibitemOpen
  \bibfield  {author} {\bibinfo {author} {\bibfnamefont {T.}~\bibnamefont
  {Venumadhav}}, \bibinfo {author} {\bibfnamefont {B.}~\bibnamefont {Zackay}},
  \bibinfo {author} {\bibfnamefont {J.}~\bibnamefont {Roulet}}, \bibinfo
  {author} {\bibfnamefont {L.}~\bibnamefont {Dai}}, \ and\ \bibinfo {author}
  {\bibfnamefont {M.}~\bibnamefont {Zaldarriaga}},\ }\href {\doibase
  10.1103/PhysRevD.100.023011} {\bibfield  {journal} {\bibinfo  {journal}
  {Phys. Rev. D}\ }\textbf {\bibinfo {volume} {100}},\ \bibinfo {pages}
  {023011} (\bibinfo {year} {2019})},\ \Eprint
  {http://arxiv.org/abs/1902.10341} {arXiv:1902.10341 [astro-ph.IM]}
  \BibitemShut {NoStop}%
\bibitem [{\citenamefont {Venumadhav}\ \emph {et~al.}(2020)\citenamefont
  {Venumadhav}, \citenamefont {Zackay}, \citenamefont {Roulet}, \citenamefont
  {Dai},\ and\ \citenamefont {Zaldarriaga}}]{Venumadhav:2019lyq}%
  \BibitemOpen
  \bibfield  {author} {\bibinfo {author} {\bibfnamefont {T.}~\bibnamefont
  {Venumadhav}}, \bibinfo {author} {\bibfnamefont {B.}~\bibnamefont {Zackay}},
  \bibinfo {author} {\bibfnamefont {J.}~\bibnamefont {Roulet}}, \bibinfo
  {author} {\bibfnamefont {L.}~\bibnamefont {Dai}}, \ and\ \bibinfo {author}
  {\bibfnamefont {M.}~\bibnamefont {Zaldarriaga}},\ }\href {\doibase
  10.1103/PhysRevD.101.083030} {\bibfield  {journal} {\bibinfo  {journal}
  {Phys. Rev. D}\ }\textbf {\bibinfo {volume} {101}},\ \bibinfo {pages}
  {083030} (\bibinfo {year} {2020})},\ \Eprint
  {http://arxiv.org/abs/1904.07214} {arXiv:1904.07214 [astro-ph.HE]}
  \BibitemShut {NoStop}%
\bibitem [{\citenamefont {Nitz}\ \emph {et~al.}(2019)\citenamefont {Nitz},
  \citenamefont {Capano}, \citenamefont {Nielsen}, \citenamefont {Reyes},
  \citenamefont {White}, \citenamefont {Brown},\ and\ \citenamefont
  {Krishnan}}]{Nitz:2018imz}%
  \BibitemOpen
  \bibfield  {author} {\bibinfo {author} {\bibfnamefont {A.~H.}\ \bibnamefont
  {Nitz}}, \bibinfo {author} {\bibfnamefont {C.}~\bibnamefont {Capano}},
  \bibinfo {author} {\bibfnamefont {A.~B.}\ \bibnamefont {Nielsen}}, \bibinfo
  {author} {\bibfnamefont {S.}~\bibnamefont {Reyes}}, \bibinfo {author}
  {\bibfnamefont {R.}~\bibnamefont {White}}, \bibinfo {author} {\bibfnamefont
  {D.~A.}\ \bibnamefont {Brown}}, \ and\ \bibinfo {author} {\bibfnamefont
  {B.}~\bibnamefont {Krishnan}},\ }\href {\doibase 10.3847/1538-4357/ab0108}
  {\bibfield  {journal} {\bibinfo  {journal} {Astrophys. J.}\ }\textbf
  {\bibinfo {volume} {872}},\ \bibinfo {pages} {195} (\bibinfo {year}
  {2019})},\ \Eprint {http://arxiv.org/abs/1811.01921} {arXiv:1811.01921
  [gr-qc]} \BibitemShut {NoStop}%
\bibitem [{\citenamefont {Nitz}\ \emph {et~al.}(2020)\citenamefont {Nitz},
  \citenamefont {Dent}, \citenamefont {Davies}, \citenamefont {Kumar},
  \citenamefont {Capano}, \citenamefont {Harry}, \citenamefont {Mozzon},
  \citenamefont {Nuttall}, \citenamefont {Lundgren},\ and\ \citenamefont
  {T\'apai}}]{Nitz:2019hdf}%
  \BibitemOpen
  \bibfield  {author} {\bibinfo {author} {\bibfnamefont {A.~H.}\ \bibnamefont
  {Nitz}}, \bibinfo {author} {\bibfnamefont {T.}~\bibnamefont {Dent}}, \bibinfo
  {author} {\bibfnamefont {G.~S.}\ \bibnamefont {Davies}}, \bibinfo {author}
  {\bibfnamefont {S.}~\bibnamefont {Kumar}}, \bibinfo {author} {\bibfnamefont
  {C.~D.}\ \bibnamefont {Capano}}, \bibinfo {author} {\bibfnamefont
  {I.}~\bibnamefont {Harry}}, \bibinfo {author} {\bibfnamefont
  {S.}~\bibnamefont {Mozzon}}, \bibinfo {author} {\bibfnamefont
  {L.}~\bibnamefont {Nuttall}}, \bibinfo {author} {\bibfnamefont
  {A.}~\bibnamefont {Lundgren}}, \ and\ \bibinfo {author} {\bibfnamefont
  {M.}~\bibnamefont {T\'apai}},\ }\href {\doibase 10.3847/1538-4357/ab733f}
  {\bibfield  {journal} {\bibinfo  {journal} {Astrophys. J.}\ }\textbf
  {\bibinfo {volume} {891}},\ \bibinfo {pages} {123} (\bibinfo {year}
  {2020})},\ \Eprint {http://arxiv.org/abs/1910.05331} {arXiv:1910.05331
  [astro-ph.HE]} \BibitemShut {NoStop}%
\bibitem [{\citenamefont {Abbott}\ \emph
  {et~al.}(2020{\natexlab{a}})\citenamefont {Abbott} \emph
  {et~al.}}]{Abbott:2020khf}%
  \BibitemOpen
  \bibfield  {author} {\bibinfo {author} {\bibfnamefont {R.}~\bibnamefont
  {Abbott}} \emph {et~al.} (\bibinfo {collaboration} {LIGO Scientific,
  Virgo}),\ }\href {\doibase 10.3847/2041-8213/ab960f} {\bibfield  {journal}
  {\bibinfo  {journal} {Astrophys. J. Lett.}\ }\textbf {\bibinfo {volume}
  {896}},\ \bibinfo {pages} {L44} (\bibinfo {year} {2020}{\natexlab{a}})},\
  \Eprint {http://arxiv.org/abs/2006.12611} {arXiv:2006.12611 [astro-ph.HE]}
  \BibitemShut {NoStop}%
\bibitem [{\citenamefont {Abbott}\ \emph
  {et~al.}(2020{\natexlab{b}})\citenamefont {Abbott} \emph
  {et~al.}}]{LIGOScientific:2020stg}%
  \BibitemOpen
  \bibfield  {author} {\bibinfo {author} {\bibfnamefont {R.}~\bibnamefont
  {Abbott}} \emph {et~al.} (\bibinfo {collaboration} {LIGO Scientific,
  Virgo}),\ }\href {\doibase 10.1103/PhysRevD.102.043015} {\bibfield  {journal}
  {\bibinfo  {journal} {Phys. Rev. D}\ }\textbf {\bibinfo {volume} {102}},\
  \bibinfo {pages} {043015} (\bibinfo {year} {2020}{\natexlab{b}})},\ \Eprint
  {http://arxiv.org/abs/2004.08342} {arXiv:2004.08342 [astro-ph.HE]}
  \BibitemShut {NoStop}%
\bibitem [{\citenamefont {Abbott}\ \emph
  {et~al.}(2020{\natexlab{c}})\citenamefont {Abbott} \emph
  {et~al.}}]{Abbott:2020tfl}%
  \BibitemOpen
  \bibfield  {author} {\bibinfo {author} {\bibfnamefont {R.}~\bibnamefont
  {Abbott}} \emph {et~al.} (\bibinfo {collaboration} {LIGO Scientific,
  Virgo}),\ }\href {\doibase 10.1103/PhysRevLett.125.101102} {\bibfield
  {journal} {\bibinfo  {journal} {Phys. Rev. Lett.}\ }\textbf {\bibinfo
  {volume} {125}},\ \bibinfo {pages} {101102} (\bibinfo {year}
  {2020}{\natexlab{c}})},\ \Eprint {http://arxiv.org/abs/2009.01075}
  {arXiv:2009.01075 [gr-qc]} \BibitemShut {NoStop}%
\bibitem [{\citenamefont {Abbott}\ \emph
  {et~al.}(2017{\natexlab{d}})\citenamefont {Abbott} \emph
  {et~al.}}]{TheLIGOScientific:2017qsa}%
  \BibitemOpen
  \bibfield  {author} {\bibinfo {author} {\bibfnamefont {B.~P.}\ \bibnamefont
  {Abbott}} \emph {et~al.} (\bibinfo {collaboration} {Virgo, LIGO
  Scientific}),\ }\href {\doibase 10.1103/PhysRevLett.119.161101} {\bibfield
  {journal} {\bibinfo  {journal} {Phys. Rev. Lett.}\ }\textbf {\bibinfo
  {volume} {119}},\ \bibinfo {pages} {161101} (\bibinfo {year}
  {2017}{\natexlab{d}})},\ \Eprint {http://arxiv.org/abs/1710.05832}
  {arXiv:1710.05832 [gr-qc]} \BibitemShut {NoStop}%
\bibitem [{\citenamefont {Abbott}\ \emph
  {et~al.}(2020{\natexlab{d}})\citenamefont {Abbott} \emph
  {et~al.}}]{Abbott:2020uma}%
  \BibitemOpen
  \bibfield  {author} {\bibinfo {author} {\bibfnamefont {B.}~\bibnamefont
  {Abbott}} \emph {et~al.} (\bibinfo {collaboration} {LIGO Scientific,
  Virgo}),\ }\href {\doibase 10.3847/2041-8213/ab75f5} {\bibfield  {journal}
  {\bibinfo  {journal} {Astrophys. J. Lett.}\ }\textbf {\bibinfo {volume}
  {892}},\ \bibinfo {pages} {L3} (\bibinfo {year} {2020}{\natexlab{d}})},\
  \Eprint {http://arxiv.org/abs/2001.01761} {arXiv:2001.01761 [astro-ph.HE]}
  \BibitemShut {NoStop}%
\bibitem [{\citenamefont {Abbott}\ \emph
  {et~al.}(2021{\natexlab{a}})\citenamefont {Abbott} \emph
  {et~al.}}]{LIGOScientific:2021qlt}%
  \BibitemOpen
  \bibfield  {author} {\bibinfo {author} {\bibfnamefont {R.}~\bibnamefont
  {Abbott}} \emph {et~al.} (\bibinfo {collaboration} {LIGO Scientific, KAGRA,
  VIRGO}),\ }\href {\doibase 10.3847/2041-8213/ac082e} {\bibfield  {journal}
  {\bibinfo  {journal} {Astrophys. J. Lett.}\ }\textbf {\bibinfo {volume}
  {915}},\ \bibinfo {pages} {L5} (\bibinfo {year} {2021}{\natexlab{a}})},\
  \Eprint {http://arxiv.org/abs/2106.15163} {arXiv:2106.15163 [astro-ph.HE]}
  \BibitemShut {NoStop}%
\bibitem [{\citenamefont {Abbott}\ \emph
  {et~al.}(2020{\natexlab{e}})\citenamefont {Abbott} \emph
  {et~al.}}]{Abbott:2020mjq}%
  \BibitemOpen
  \bibfield  {author} {\bibinfo {author} {\bibfnamefont {R.}~\bibnamefont
  {Abbott}} \emph {et~al.} (\bibinfo {collaboration} {LIGO Scientific,
  Virgo}),\ }\href {\doibase 10.3847/2041-8213/aba493} {\bibfield  {journal}
  {\bibinfo  {journal} {Astrophys. J. Lett.}\ }\textbf {\bibinfo {volume}
  {900}},\ \bibinfo {pages} {L13} (\bibinfo {year} {2020}{\natexlab{e}})},\
  \Eprint {http://arxiv.org/abs/2009.01190} {arXiv:2009.01190 [astro-ph.HE]}
  \BibitemShut {NoStop}%
\bibitem [{\citenamefont {Bustillo}\ \emph {et~al.}(2021)\citenamefont
  {Bustillo}, \citenamefont {Sanchis-Gual}, \citenamefont {Torres-Forn\'e},
  \citenamefont {Font}, \citenamefont {Vajpeyi}, \citenamefont {Smith},
  \citenamefont {Herdeiro}, \citenamefont {Radu},\ and\ \citenamefont
  {Leong}}]{Bustillo:2020syj}%
  \BibitemOpen
  \bibfield  {author} {\bibinfo {author} {\bibfnamefont {J.~C.}\ \bibnamefont
  {Bustillo}}, \bibinfo {author} {\bibfnamefont {N.}~\bibnamefont
  {Sanchis-Gual}}, \bibinfo {author} {\bibfnamefont {A.}~\bibnamefont
  {Torres-Forn\'e}}, \bibinfo {author} {\bibfnamefont {J.~A.}\ \bibnamefont
  {Font}}, \bibinfo {author} {\bibfnamefont {A.}~\bibnamefont {Vajpeyi}},
  \bibinfo {author} {\bibfnamefont {R.}~\bibnamefont {Smith}}, \bibinfo
  {author} {\bibfnamefont {C.}~\bibnamefont {Herdeiro}}, \bibinfo {author}
  {\bibfnamefont {E.}~\bibnamefont {Radu}}, \ and\ \bibinfo {author}
  {\bibfnamefont {S.~H.~W.}\ \bibnamefont {Leong}},\ }\href {\doibase
  10.1103/PhysRevLett.126.081101} {\bibfield  {journal} {\bibinfo  {journal}
  {Phys. Rev. Lett.}\ }\textbf {\bibinfo {volume} {126}},\ \bibinfo {pages}
  {081101} (\bibinfo {year} {2021})},\ \Eprint
  {http://arxiv.org/abs/2009.05376} {arXiv:2009.05376 [gr-qc]} \BibitemShut
  {NoStop}%
\bibitem [{\citenamefont {Gayathri}\ \emph {et~al.}(2020)\citenamefont
  {Gayathri}, \citenamefont {Healy}, \citenamefont {Lange}, \citenamefont
  {O'Brien}, \citenamefont {Szczepanczyk}, \citenamefont {Bartos},
  \citenamefont {Campanelli}, \citenamefont {Klimenko}, \citenamefont
  {Lousto},\ and\ \citenamefont {O'Shaughnessy}}]{Gayathri:2020coq}%
  \BibitemOpen
  \bibfield  {author} {\bibinfo {author} {\bibfnamefont {V.}~\bibnamefont
  {Gayathri}}, \bibinfo {author} {\bibfnamefont {J.}~\bibnamefont {Healy}},
  \bibinfo {author} {\bibfnamefont {J.}~\bibnamefont {Lange}}, \bibinfo
  {author} {\bibfnamefont {B.}~\bibnamefont {O'Brien}}, \bibinfo {author}
  {\bibfnamefont {M.}~\bibnamefont {Szczepanczyk}}, \bibinfo {author}
  {\bibfnamefont {I.}~\bibnamefont {Bartos}}, \bibinfo {author} {\bibfnamefont
  {M.}~\bibnamefont {Campanelli}}, \bibinfo {author} {\bibfnamefont
  {S.}~\bibnamefont {Klimenko}}, \bibinfo {author} {\bibfnamefont
  {C.}~\bibnamefont {Lousto}}, \ and\ \bibinfo {author} {\bibfnamefont
  {R.}~\bibnamefont {O'Shaughnessy}},\ }\href@noop {} {\  (\bibinfo {year}
  {2020})},\ \Eprint {http://arxiv.org/abs/2009.05461} {arXiv:2009.05461
  [astro-ph.HE]} \BibitemShut {NoStop}%
\bibitem [{\citenamefont {Gamba}\ \emph
  {et~al.}(2021{\natexlab{a}})\citenamefont {Gamba}, \citenamefont {Breschi},
  \citenamefont {Carullo}, \citenamefont {Rettegno}, \citenamefont {Albanesi},
  \citenamefont {Bernuzzi},\ and\ \citenamefont {Nagar}}]{Gamba:2021gap}%
  \BibitemOpen
  \bibfield  {author} {\bibinfo {author} {\bibfnamefont {R.}~\bibnamefont
  {Gamba}}, \bibinfo {author} {\bibfnamefont {M.}~\bibnamefont {Breschi}},
  \bibinfo {author} {\bibfnamefont {G.}~\bibnamefont {Carullo}}, \bibinfo
  {author} {\bibfnamefont {P.}~\bibnamefont {Rettegno}}, \bibinfo {author}
  {\bibfnamefont {S.}~\bibnamefont {Albanesi}}, \bibinfo {author}
  {\bibfnamefont {S.}~\bibnamefont {Bernuzzi}}, \ and\ \bibinfo {author}
  {\bibfnamefont {A.}~\bibnamefont {Nagar}},\ }\href@noop {} {\bibfield
  {journal} {\bibinfo  {journal} {Submitted to Nature Astronomy}\ } (\bibinfo
  {year} {2021}{\natexlab{a}})},\ \Eprint {http://arxiv.org/abs/2106.05575}
  {arXiv:2106.05575 [gr-qc]} \BibitemShut {NoStop}%
\bibitem [{\citenamefont {Abbott}\ \emph
  {et~al.}(2021{\natexlab{b}})\citenamefont {Abbott} \emph
  {et~al.}}]{LIGOScientific:2020ibl}%
  \BibitemOpen
  \bibfield  {author} {\bibinfo {author} {\bibfnamefont {R.}~\bibnamefont
  {Abbott}} \emph {et~al.} (\bibinfo {collaboration} {LIGO Scientific,
  Virgo}),\ }\href {\doibase 10.1103/PhysRevX.11.021053} {\bibfield  {journal}
  {\bibinfo  {journal} {Phys. Rev. X}\ }\textbf {\bibinfo {volume} {11}},\
  \bibinfo {pages} {021053} (\bibinfo {year} {2021}{\natexlab{b}})},\ \Eprint
  {http://arxiv.org/abs/2010.14527} {arXiv:2010.14527 [gr-qc]} \BibitemShut
  {NoStop}%
\bibitem [{\citenamefont {Abbott}\ \emph
  {et~al.}(2021{\natexlab{c}})\citenamefont {Abbott} \emph
  {et~al.}}]{LIGOScientific:2020kqk}%
  \BibitemOpen
  \bibfield  {author} {\bibinfo {author} {\bibfnamefont {R.}~\bibnamefont
  {Abbott}} \emph {et~al.} (\bibinfo {collaboration} {LIGO Scientific,
  Virgo}),\ }\href {\doibase 10.3847/2041-8213/abe949} {\bibfield  {journal}
  {\bibinfo  {journal} {Astrophys. J. Lett.}\ }\textbf {\bibinfo {volume}
  {913}},\ \bibinfo {pages} {L7} (\bibinfo {year} {2021}{\natexlab{c}})},\
  \Eprint {http://arxiv.org/abs/2010.14533} {arXiv:2010.14533 [astro-ph.HE]}
  \BibitemShut {NoStop}%
\bibitem [{\citenamefont {Apostolatos}\ \emph {et~al.}(1994)\citenamefont
  {Apostolatos}, \citenamefont {Cutler}, \citenamefont {Sussman},\ and\
  \citenamefont {Thorne}}]{Apostolatos:1994mx}%
  \BibitemOpen
  \bibfield  {author} {\bibinfo {author} {\bibfnamefont {T.~A.}\ \bibnamefont
  {Apostolatos}}, \bibinfo {author} {\bibfnamefont {C.}~\bibnamefont {Cutler}},
  \bibinfo {author} {\bibfnamefont {G.~J.}\ \bibnamefont {Sussman}}, \ and\
  \bibinfo {author} {\bibfnamefont {K.~S.}\ \bibnamefont {Thorne}},\ }\href
  {\doibase 10.1103/PhysRevD.49.6274} {\bibfield  {journal} {\bibinfo
  {journal} {Phys. Rev.}\ }\textbf {\bibinfo {volume} {D49}},\ \bibinfo {pages}
  {6274} (\bibinfo {year} {1994})}\BibitemShut {NoStop}%
\bibitem [{\citenamefont {Buonanno}\ and\ \citenamefont
  {Damour}(1999)}]{Buonanno:1998gg}%
  \BibitemOpen
  \bibfield  {author} {\bibinfo {author} {\bibfnamefont {A.}~\bibnamefont
  {Buonanno}}\ and\ \bibinfo {author} {\bibfnamefont {T.}~\bibnamefont
  {Damour}},\ }\href {\doibase 10.1103/PhysRevD.59.084006} {\bibfield
  {journal} {\bibinfo  {journal} {Phys. Rev.}\ }\textbf {\bibinfo {volume}
  {D59}},\ \bibinfo {pages} {084006} (\bibinfo {year} {1999})},\ \Eprint
  {http://arxiv.org/abs/gr-qc/9811091} {arXiv:gr-qc/9811091} \BibitemShut
  {NoStop}%
\bibitem [{\citenamefont {Buonanno}\ and\ \citenamefont
  {Damour}(2000)}]{Buonanno:2000ef}%
  \BibitemOpen
  \bibfield  {author} {\bibinfo {author} {\bibfnamefont {A.}~\bibnamefont
  {Buonanno}}\ and\ \bibinfo {author} {\bibfnamefont {T.}~\bibnamefont
  {Damour}},\ }\href {\doibase 10.1103/PhysRevD.62.064015} {\bibfield
  {journal} {\bibinfo  {journal} {Phys. Rev.}\ }\textbf {\bibinfo {volume}
  {D62}},\ \bibinfo {pages} {064015} (\bibinfo {year} {2000})},\ \Eprint
  {http://arxiv.org/abs/gr-qc/0001013} {arXiv:gr-qc/0001013} \BibitemShut
  {NoStop}%
\bibitem [{\citenamefont {Damour}\ \emph {et~al.}(2000)\citenamefont {Damour},
  \citenamefont {Jaranowski},\ and\ \citenamefont {Schaefer}}]{Damour:2000we}%
  \BibitemOpen
  \bibfield  {author} {\bibinfo {author} {\bibfnamefont {T.}~\bibnamefont
  {Damour}}, \bibinfo {author} {\bibfnamefont {P.}~\bibnamefont {Jaranowski}},
  \ and\ \bibinfo {author} {\bibfnamefont {G.}~\bibnamefont {Schaefer}},\
  }\href {\doibase 10.1103/PhysRevD.62.084011} {\bibfield  {journal} {\bibinfo
  {journal} {Phys. Rev.}\ }\textbf {\bibinfo {volume} {D62}},\ \bibinfo {pages}
  {084011} (\bibinfo {year} {2000})},\ \Eprint
  {http://arxiv.org/abs/gr-qc/0005034} {arXiv:gr-qc/0005034 [gr-qc]}
  \BibitemShut {NoStop}%
\bibitem [{\citenamefont {Damour}(2001)}]{Damour:2001tu}%
  \BibitemOpen
  \bibfield  {author} {\bibinfo {author} {\bibfnamefont {T.}~\bibnamefont
  {Damour}},\ }\href {\doibase 10.1103/PhysRevD.64.124013} {\bibfield
  {journal} {\bibinfo  {journal} {Phys. Rev.}\ }\textbf {\bibinfo {volume}
  {D64}},\ \bibinfo {pages} {124013} (\bibinfo {year} {2001})},\ \Eprint
  {http://arxiv.org/abs/gr-qc/0103018} {arXiv:gr-qc/0103018} \BibitemShut
  {NoStop}%
\bibitem [{\citenamefont {Buonanno}\ \emph {et~al.}(2006)\citenamefont
  {Buonanno}, \citenamefont {Chen},\ and\ \citenamefont
  {Damour}}]{Buonanno:2005xu}%
  \BibitemOpen
  \bibfield  {author} {\bibinfo {author} {\bibfnamefont {A.}~\bibnamefont
  {Buonanno}}, \bibinfo {author} {\bibfnamefont {Y.}~\bibnamefont {Chen}}, \
  and\ \bibinfo {author} {\bibfnamefont {T.}~\bibnamefont {Damour}},\ }\href
  {\doibase 10.1103/PhysRevD.74.104005} {\bibfield  {journal} {\bibinfo
  {journal} {Phys. Rev.}\ }\textbf {\bibinfo {volume} {D74}},\ \bibinfo {pages}
  {104005} (\bibinfo {year} {2006})},\ \Eprint
  {http://arxiv.org/abs/gr-qc/0508067} {arXiv:gr-qc/0508067} \BibitemShut
  {NoStop}%
\bibitem [{\citenamefont {Damour}\ \emph {et~al.}(2015)\citenamefont {Damour},
  \citenamefont {Jaranowski},\ and\ \citenamefont {Schäfer}}]{Damour:2015isa}%
  \BibitemOpen
  \bibfield  {author} {\bibinfo {author} {\bibfnamefont {T.}~\bibnamefont
  {Damour}}, \bibinfo {author} {\bibfnamefont {P.}~\bibnamefont {Jaranowski}},
  \ and\ \bibinfo {author} {\bibfnamefont {G.}~\bibnamefont {Schäfer}},\
  }\href {\doibase 10.1103/PhysRevD.91.084024} {\bibfield  {journal} {\bibinfo
  {journal} {Phys. Rev.}\ }\textbf {\bibinfo {volume} {D91}},\ \bibinfo {pages}
  {084024} (\bibinfo {year} {2015})},\ \Eprint
  {http://arxiv.org/abs/1502.07245} {arXiv:1502.07245 [gr-qc]} \BibitemShut
  {NoStop}%
\bibitem [{\citenamefont {Damour}\ and\ \citenamefont
  {Nagar}(2010)}]{Damour:2009wj}%
  \BibitemOpen
  \bibfield  {author} {\bibinfo {author} {\bibfnamefont {T.}~\bibnamefont
  {Damour}}\ and\ \bibinfo {author} {\bibfnamefont {A.}~\bibnamefont {Nagar}},\
  }\href {\doibase 10.1103/PhysRevD.81.084016} {\bibfield  {journal} {\bibinfo
  {journal} {Phys. Rev.}\ }\textbf {\bibinfo {volume} {D81}},\ \bibinfo {pages}
  {084016} (\bibinfo {year} {2010})},\ \Eprint {http://arxiv.org/abs/0911.5041}
  {arXiv:0911.5041 [gr-qc]} \BibitemShut {NoStop}%
\bibitem [{\citenamefont {Damour}\ \emph {et~al.}(2012)\citenamefont {Damour},
  \citenamefont {Nagar},\ and\ \citenamefont {Villain}}]{Damour:2012yf}%
  \BibitemOpen
  \bibfield  {author} {\bibinfo {author} {\bibfnamefont {T.}~\bibnamefont
  {Damour}}, \bibinfo {author} {\bibfnamefont {A.}~\bibnamefont {Nagar}}, \
  and\ \bibinfo {author} {\bibfnamefont {L.}~\bibnamefont {Villain}},\ }\href
  {\doibase 10.1103/PhysRevD.85.123007} {\bibfield  {journal} {\bibinfo
  {journal} {Phys.Rev.}\ }\textbf {\bibinfo {volume} {D85}},\ \bibinfo {pages}
  {123007} (\bibinfo {year} {2012})},\ \Eprint {http://arxiv.org/abs/1203.4352}
  {arXiv:1203.4352 [gr-qc]} \BibitemShut {NoStop}%
\bibitem [{\citenamefont {Bernuzzi}\ \emph {et~al.}(2012)\citenamefont
  {Bernuzzi}, \citenamefont {Nagar}, \citenamefont {Thierfelder},\ and\
  \citenamefont {Br{\"u}gmann}}]{Bernuzzi:2012ci}%
  \BibitemOpen
  \bibfield  {author} {\bibinfo {author} {\bibfnamefont {S.}~\bibnamefont
  {Bernuzzi}}, \bibinfo {author} {\bibfnamefont {A.}~\bibnamefont {Nagar}},
  \bibinfo {author} {\bibfnamefont {M.}~\bibnamefont {Thierfelder}}, \ and\
  \bibinfo {author} {\bibfnamefont {B.}~\bibnamefont {Br{\"u}gmann}},\ }\href
  {\doibase 10.1103/PhysRevD.86.044030} {\bibfield  {journal} {\bibinfo
  {journal} {Phys.Rev.}\ }\textbf {\bibinfo {volume} {D86}},\ \bibinfo {pages}
  {044030} (\bibinfo {year} {2012})},\ \Eprint {http://arxiv.org/abs/1205.3403}
  {arXiv:1205.3403 [gr-qc]} \BibitemShut {NoStop}%
\bibitem [{\citenamefont {Boh{\'e}}\ \emph {et~al.}(2017)\citenamefont
  {Boh{\'e}} \emph {et~al.}}]{Bohe:2016gbl}%
  \BibitemOpen
  \bibfield  {author} {\bibinfo {author} {\bibfnamefont {A.}~\bibnamefont
  {Boh{\'e}}} \emph {et~al.},\ }\href {\doibase 10.1103/PhysRevD.95.044028}
  {\bibfield  {journal} {\bibinfo  {journal} {Phys. Rev.}\ }\textbf {\bibinfo
  {volume} {D95}},\ \bibinfo {pages} {044028} (\bibinfo {year} {2017})},\
  \Eprint {http://arxiv.org/abs/1611.03703} {arXiv:1611.03703 [gr-qc]}
  \BibitemShut {NoStop}%
\bibitem [{\citenamefont {Babak}\ \emph {et~al.}(2017)\citenamefont {Babak},
  \citenamefont {Taracchini},\ and\ \citenamefont {Buonanno}}]{Babak:2016tgq}%
  \BibitemOpen
  \bibfield  {author} {\bibinfo {author} {\bibfnamefont {S.}~\bibnamefont
  {Babak}}, \bibinfo {author} {\bibfnamefont {A.}~\bibnamefont {Taracchini}}, \
  and\ \bibinfo {author} {\bibfnamefont {A.}~\bibnamefont {Buonanno}},\ }\href
  {\doibase 10.1103/PhysRevD.95.024010} {\bibfield  {journal} {\bibinfo
  {journal} {Phys. Rev.}\ }\textbf {\bibinfo {volume} {D95}},\ \bibinfo {pages}
  {024010} (\bibinfo {year} {2017})},\ \Eprint
  {http://arxiv.org/abs/1607.05661} {arXiv:1607.05661 [gr-qc]} \BibitemShut
  {NoStop}%
\bibitem [{\citenamefont {Cotesta}\ \emph {et~al.}(2018)\citenamefont
  {Cotesta}, \citenamefont {Buonanno}, \citenamefont {Boh\'e}, \citenamefont
  {Taracchini}, \citenamefont {Hinder},\ and\ \citenamefont
  {Ossokine}}]{Cotesta:2018fcv}%
  \BibitemOpen
  \bibfield  {author} {\bibinfo {author} {\bibfnamefont {R.}~\bibnamefont
  {Cotesta}}, \bibinfo {author} {\bibfnamefont {A.}~\bibnamefont {Buonanno}},
  \bibinfo {author} {\bibfnamefont {A.}~\bibnamefont {Boh\'e}}, \bibinfo
  {author} {\bibfnamefont {A.}~\bibnamefont {Taracchini}}, \bibinfo {author}
  {\bibfnamefont {I.}~\bibnamefont {Hinder}}, \ and\ \bibinfo {author}
  {\bibfnamefont {S.}~\bibnamefont {Ossokine}},\ }\href {\doibase
  10.1103/PhysRevD.98.084028} {\bibfield  {journal} {\bibinfo  {journal} {Phys.
  Rev.}\ }\textbf {\bibinfo {volume} {D98}},\ \bibinfo {pages} {084028}
  (\bibinfo {year} {2018})},\ \Eprint {http://arxiv.org/abs/1803.10701}
  {arXiv:1803.10701 [gr-qc]} \BibitemShut {NoStop}%
\bibitem [{\citenamefont {Hinderer}\ \emph {et~al.}(2016)\citenamefont
  {Hinderer} \emph {et~al.}}]{Hinderer:2016eia}%
  \BibitemOpen
  \bibfield  {author} {\bibinfo {author} {\bibfnamefont {T.}~\bibnamefont
  {Hinderer}} \emph {et~al.},\ }\href {\doibase 10.1103/PhysRevLett.116.181101}
  {\bibfield  {journal} {\bibinfo  {journal} {Phys. Rev. Lett.}\ }\textbf
  {\bibinfo {volume} {116}},\ \bibinfo {pages} {181101} (\bibinfo {year}
  {2016})},\ \Eprint {http://arxiv.org/abs/1602.00599} {arXiv:1602.00599
  [gr-qc]} \BibitemShut {NoStop}%
\bibitem [{\citenamefont {Steinhoff}\ \emph {et~al.}(2016)\citenamefont
  {Steinhoff}, \citenamefont {Hinderer}, \citenamefont {Buonanno},\ and\
  \citenamefont {Taracchini}}]{Steinhoff:2016rfi}%
  \BibitemOpen
  \bibfield  {author} {\bibinfo {author} {\bibfnamefont {J.}~\bibnamefont
  {Steinhoff}}, \bibinfo {author} {\bibfnamefont {T.}~\bibnamefont {Hinderer}},
  \bibinfo {author} {\bibfnamefont {A.}~\bibnamefont {Buonanno}}, \ and\
  \bibinfo {author} {\bibfnamefont {A.}~\bibnamefont {Taracchini}},\ }\href
  {\doibase 10.1103/PhysRevD.94.104028} {\bibfield  {journal} {\bibinfo
  {journal} {Phys. Rev.}\ }\textbf {\bibinfo {volume} {D94}},\ \bibinfo {pages}
  {104028} (\bibinfo {year} {2016})},\ \Eprint
  {http://arxiv.org/abs/1608.01907} {arXiv:1608.01907 [gr-qc]} \BibitemShut
  {NoStop}%
\bibitem [{\citenamefont {Lackey}\ \emph {et~al.}(2019)\citenamefont {Lackey},
  \citenamefont {Pürrer}, \citenamefont {Taracchini},\ and\ \citenamefont
  {Marsat}}]{Lackey:2018zvw}%
  \BibitemOpen
  \bibfield  {author} {\bibinfo {author} {\bibfnamefont {B.~D.}\ \bibnamefont
  {Lackey}}, \bibinfo {author} {\bibfnamefont {M.}~\bibnamefont {Pürrer}},
  \bibinfo {author} {\bibfnamefont {A.}~\bibnamefont {Taracchini}}, \ and\
  \bibinfo {author} {\bibfnamefont {S.}~\bibnamefont {Marsat}},\ }\href
  {\doibase 10.1103/PhysRevD.100.024002} {\bibfield  {journal} {\bibinfo
  {journal} {Phys. Rev. D}\ }\textbf {\bibinfo {volume} {100}},\ \bibinfo
  {pages} {024002} (\bibinfo {year} {2019})},\ \Eprint
  {http://arxiv.org/abs/1812.08643} {arXiv:1812.08643 [gr-qc]} \BibitemShut
  {NoStop}%
\bibitem [{\citenamefont {Matas}\ \emph {et~al.}(2020)\citenamefont {Matas}
  \emph {et~al.}}]{Matas:2020wab}%
  \BibitemOpen
  \bibfield  {author} {\bibinfo {author} {\bibfnamefont {A.}~\bibnamefont
  {Matas}} \emph {et~al.},\ }\href {\doibase 10.1103/PhysRevD.102.043023}
  {\bibfield  {journal} {\bibinfo  {journal} {Phys. Rev. D}\ }\textbf {\bibinfo
  {volume} {102}},\ \bibinfo {pages} {043023} (\bibinfo {year} {2020})},\
  \Eprint {http://arxiv.org/abs/2004.10001} {arXiv:2004.10001 [gr-qc]}
  \BibitemShut {NoStop}%
\bibitem [{\citenamefont {Ossokine}\ \emph {et~al.}(2020)\citenamefont
  {Ossokine} \emph {et~al.}}]{Ossokine:2020kjp}%
  \BibitemOpen
  \bibfield  {author} {\bibinfo {author} {\bibfnamefont {S.}~\bibnamefont
  {Ossokine}} \emph {et~al.},\ }\href {\doibase 10.1103/PhysRevD.102.044055}
  {\bibfield  {journal} {\bibinfo  {journal} {Phys. Rev. D}\ }\textbf {\bibinfo
  {volume} {102}},\ \bibinfo {pages} {044055} (\bibinfo {year} {2020})},\
  \Eprint {http://arxiv.org/abs/2004.09442} {arXiv:2004.09442 [gr-qc]}
  \BibitemShut {NoStop}%
\bibitem [{\citenamefont {Damour}\ and\ \citenamefont
  {Nagar}(2014{\natexlab{a}})}]{Damour:2014yha}%
  \BibitemOpen
  \bibfield  {author} {\bibinfo {author} {\bibfnamefont {T.}~\bibnamefont
  {Damour}}\ and\ \bibinfo {author} {\bibfnamefont {A.}~\bibnamefont {Nagar}},\
  }\href {\doibase 10.1103/PhysRevD.90.024054} {\bibfield  {journal} {\bibinfo
  {journal} {Phys.Rev.}\ }\textbf {\bibinfo {volume} {D90}},\ \bibinfo {pages}
  {024054} (\bibinfo {year} {2014}{\natexlab{a}})},\ \Eprint
  {http://arxiv.org/abs/1406.0401} {arXiv:1406.0401 [gr-qc]} \BibitemShut
  {NoStop}%
\bibitem [{\citenamefont {Bernuzzi}\ \emph {et~al.}(2015)\citenamefont
  {Bernuzzi}, \citenamefont {Nagar}, \citenamefont {Dietrich},\ and\
  \citenamefont {Damour}}]{Bernuzzi:2014owa}%
  \BibitemOpen
  \bibfield  {author} {\bibinfo {author} {\bibfnamefont {S.}~\bibnamefont
  {Bernuzzi}}, \bibinfo {author} {\bibfnamefont {A.}~\bibnamefont {Nagar}},
  \bibinfo {author} {\bibfnamefont {T.}~\bibnamefont {Dietrich}}, \ and\
  \bibinfo {author} {\bibfnamefont {T.}~\bibnamefont {Damour}},\ }\href
  {\doibase 10.1103/PhysRevLett.114.161103} {\bibfield  {journal} {\bibinfo
  {journal} {Phys.Rev.Lett.}\ }\textbf {\bibinfo {volume} {114}},\ \bibinfo
  {pages} {161103} (\bibinfo {year} {2015})},\ \Eprint
  {http://arxiv.org/abs/1412.4553} {arXiv:1412.4553 [gr-qc]} \BibitemShut
  {NoStop}%
\bibitem [{\citenamefont {Nagar}\ \emph {et~al.}(2017)\citenamefont {Nagar},
  \citenamefont {Riemenschneider},\ and\ \citenamefont
  {Pratten}}]{Nagar:2017jdw}%
  \BibitemOpen
  \bibfield  {author} {\bibinfo {author} {\bibfnamefont {A.}~\bibnamefont
  {Nagar}}, \bibinfo {author} {\bibfnamefont {G.}~\bibnamefont
  {Riemenschneider}}, \ and\ \bibinfo {author} {\bibfnamefont {G.}~\bibnamefont
  {Pratten}},\ }\href {\doibase 10.1103/PhysRevD.96.084045} {\bibfield
  {journal} {\bibinfo  {journal} {Phys. Rev.}\ }\textbf {\bibinfo {volume}
  {D96}},\ \bibinfo {pages} {084045} (\bibinfo {year} {2017})},\ \Eprint
  {http://arxiv.org/abs/1703.06814} {arXiv:1703.06814 [gr-qc]} \BibitemShut
  {NoStop}%
\bibitem [{\citenamefont {Nagar}\ \emph {et~al.}(2018)\citenamefont {Nagar}
  \emph {et~al.}}]{Nagar:2018zoe}%
  \BibitemOpen
  \bibfield  {author} {\bibinfo {author} {\bibfnamefont {A.}~\bibnamefont
  {Nagar}} \emph {et~al.},\ }\href {\doibase 10.1103/PhysRevD.98.104052}
  {\bibfield  {journal} {\bibinfo  {journal} {Phys. Rev.}\ }\textbf {\bibinfo
  {volume} {D98}},\ \bibinfo {pages} {104052} (\bibinfo {year} {2018})},\
  \Eprint {http://arxiv.org/abs/1806.01772} {arXiv:1806.01772 [gr-qc]}
  \BibitemShut {NoStop}%
\bibitem [{\citenamefont {Akcay}\ \emph {et~al.}(2019)\citenamefont {Akcay},
  \citenamefont {Bernuzzi}, \citenamefont {Messina}, \citenamefont {Nagar},
  \citenamefont {Ortiz},\ and\ \citenamefont {Rettegno}}]{Akcay:2018yyh}%
  \BibitemOpen
  \bibfield  {author} {\bibinfo {author} {\bibfnamefont {S.}~\bibnamefont
  {Akcay}}, \bibinfo {author} {\bibfnamefont {S.}~\bibnamefont {Bernuzzi}},
  \bibinfo {author} {\bibfnamefont {F.}~\bibnamefont {Messina}}, \bibinfo
  {author} {\bibfnamefont {A.}~\bibnamefont {Nagar}}, \bibinfo {author}
  {\bibfnamefont {N.}~\bibnamefont {Ortiz}}, \ and\ \bibinfo {author}
  {\bibfnamefont {P.}~\bibnamefont {Rettegno}},\ }\href {\doibase
  10.1103/PhysRevD.99.044051} {\bibfield  {journal} {\bibinfo  {journal} {Phys.
  Rev.}\ }\textbf {\bibinfo {volume} {D99}},\ \bibinfo {pages} {044051}
  (\bibinfo {year} {2019})},\ \Eprint {http://arxiv.org/abs/1812.02744}
  {arXiv:1812.02744 [gr-qc]} \BibitemShut {NoStop}%
\bibitem [{\citenamefont {Nagar}\ \emph {et~al.}(2020)\citenamefont {Nagar},
  \citenamefont {Riemenschneider}, \citenamefont {Pratten}, \citenamefont
  {Rettegno},\ and\ \citenamefont {Messina}}]{Nagar:2020pcj}%
  \BibitemOpen
  \bibfield  {author} {\bibinfo {author} {\bibfnamefont {A.}~\bibnamefont
  {Nagar}}, \bibinfo {author} {\bibfnamefont {G.}~\bibnamefont
  {Riemenschneider}}, \bibinfo {author} {\bibfnamefont {G.}~\bibnamefont
  {Pratten}}, \bibinfo {author} {\bibfnamefont {P.}~\bibnamefont {Rettegno}}, \
  and\ \bibinfo {author} {\bibfnamefont {F.}~\bibnamefont {Messina}},\ }\href
  {\doibase 10.1103/PhysRevD.102.024077} {\bibfield  {journal} {\bibinfo
  {journal} {Phys. Rev. D}\ }\textbf {\bibinfo {volume} {102}},\ \bibinfo
  {pages} {024077} (\bibinfo {year} {2020})},\ \Eprint
  {http://arxiv.org/abs/2001.09082} {arXiv:2001.09082 [gr-qc]} \BibitemShut
  {NoStop}%
\bibitem [{\citenamefont {Rettegno}\ \emph {et~al.}(2019)\citenamefont
  {Rettegno}, \citenamefont {Martinetti}, \citenamefont {Nagar}, \citenamefont
  {Bini}, \citenamefont {Riemenschneider},\ and\ \citenamefont
  {Damour}}]{Rettegno:2019tzh}%
  \BibitemOpen
  \bibfield  {author} {\bibinfo {author} {\bibfnamefont {P.}~\bibnamefont
  {Rettegno}}, \bibinfo {author} {\bibfnamefont {F.}~\bibnamefont
  {Martinetti}}, \bibinfo {author} {\bibfnamefont {A.}~\bibnamefont {Nagar}},
  \bibinfo {author} {\bibfnamefont {D.}~\bibnamefont {Bini}}, \bibinfo {author}
  {\bibfnamefont {G.}~\bibnamefont {Riemenschneider}}, \ and\ \bibinfo {author}
  {\bibfnamefont {T.}~\bibnamefont {Damour}},\ }\href@noop {} {\  (\bibinfo
  {year} {2019})},\ \Eprint {http://arxiv.org/abs/1911.10818} {arXiv:1911.10818
  [gr-qc]} \BibitemShut {NoStop}%
\bibitem [{\citenamefont {Chiaramello}\ and\ \citenamefont
  {Nagar}(2020)}]{Chiaramello:2020ehz}%
  \BibitemOpen
  \bibfield  {author} {\bibinfo {author} {\bibfnamefont {D.}~\bibnamefont
  {Chiaramello}}\ and\ \bibinfo {author} {\bibfnamefont {A.}~\bibnamefont
  {Nagar}},\ }\href {\doibase 10.1103/PhysRevD.101.101501} {\bibfield
  {journal} {\bibinfo  {journal} {Phys. Rev. D}\ }\textbf {\bibinfo {volume}
  {101}},\ \bibinfo {pages} {101501} (\bibinfo {year} {2020})},\ \Eprint
  {http://arxiv.org/abs/2001.11736} {arXiv:2001.11736 [gr-qc]} \BibitemShut
  {NoStop}%
\bibitem [{\citenamefont {Nagar}\ \emph
  {et~al.}(2021{\natexlab{a}})\citenamefont {Nagar}, \citenamefont {Rettegno},
  \citenamefont {Gamba},\ and\ \citenamefont {Bernuzzi}}]{Nagar:2020xsk}%
  \BibitemOpen
  \bibfield  {author} {\bibinfo {author} {\bibfnamefont {A.}~\bibnamefont
  {Nagar}}, \bibinfo {author} {\bibfnamefont {P.}~\bibnamefont {Rettegno}},
  \bibinfo {author} {\bibfnamefont {R.}~\bibnamefont {Gamba}}, \ and\ \bibinfo
  {author} {\bibfnamefont {S.}~\bibnamefont {Bernuzzi}},\ }\href {\doibase
  10.1103/PhysRevD.103.064013} {\bibfield  {journal} {\bibinfo  {journal}
  {Phys. Rev. D}\ }\textbf {\bibinfo {volume} {103}},\ \bibinfo {pages}
  {064013} (\bibinfo {year} {2021}{\natexlab{a}})},\ \Eprint
  {http://arxiv.org/abs/2009.12857} {arXiv:2009.12857 [gr-qc]} \BibitemShut
  {NoStop}%
\bibitem [{\citenamefont {Albanesi}\ \emph {et~al.}(2021)\citenamefont
  {Albanesi}, \citenamefont {Nagar},\ and\ \citenamefont
  {Bernuzzi}}]{Albanesi:2021rby}%
  \BibitemOpen
  \bibfield  {author} {\bibinfo {author} {\bibfnamefont {S.}~\bibnamefont
  {Albanesi}}, \bibinfo {author} {\bibfnamefont {A.}~\bibnamefont {Nagar}}, \
  and\ \bibinfo {author} {\bibfnamefont {S.}~\bibnamefont {Bernuzzi}},\ }\href
  {\doibase 10.1103/PhysRevD.104.024067} {\bibfield  {journal} {\bibinfo
  {journal} {Phys. Rev. D}\ }\textbf {\bibinfo {volume} {104}},\ \bibinfo
  {pages} {024067} (\bibinfo {year} {2021})},\ \Eprint
  {http://arxiv.org/abs/2104.10559} {arXiv:2104.10559 [gr-qc]} \BibitemShut
  {NoStop}%
\bibitem [{\citenamefont {Nagar}\ \emph
  {et~al.}(2021{\natexlab{b}})\citenamefont {Nagar}, \citenamefont {Bonino},\
  and\ \citenamefont {Rettegno}}]{Nagar:2021gss}%
  \BibitemOpen
  \bibfield  {author} {\bibinfo {author} {\bibfnamefont {A.}~\bibnamefont
  {Nagar}}, \bibinfo {author} {\bibfnamefont {A.}~\bibnamefont {Bonino}}, \
  and\ \bibinfo {author} {\bibfnamefont {P.}~\bibnamefont {Rettegno}},\ }\href
  {\doibase 10.1103/PhysRevD.103.104021} {\bibfield  {journal} {\bibinfo
  {journal} {Phys. Rev. D}\ }\textbf {\bibinfo {volume} {103}},\ \bibinfo
  {pages} {104021} (\bibinfo {year} {2021}{\natexlab{b}})},\ \Eprint
  {http://arxiv.org/abs/2101.08624} {arXiv:2101.08624 [gr-qc]} \BibitemShut
  {NoStop}%
\bibitem [{\citenamefont {Nagar}\ and\ \citenamefont
  {Rettegno}(2021)}]{Nagar:2021xnh}%
  \BibitemOpen
  \bibfield  {author} {\bibinfo {author} {\bibfnamefont {A.}~\bibnamefont
  {Nagar}}\ and\ \bibinfo {author} {\bibfnamefont {P.}~\bibnamefont
  {Rettegno}},\ }\href@noop {} {\  (\bibinfo {year} {2021})},\ \Eprint
  {http://arxiv.org/abs/2108.02043} {arXiv:2108.02043 [gr-qc]} \BibitemShut
  {NoStop}%
\bibitem [{\citenamefont {Ajith}\ \emph {et~al.}(2007)\citenamefont {Ajith},
  \citenamefont {Babak}, \citenamefont {Chen}, \citenamefont {Hewitson},
  \citenamefont {Krishnan} \emph {et~al.}}]{Ajith:2007qp}%
  \BibitemOpen
  \bibfield  {author} {\bibinfo {author} {\bibfnamefont {P.}~\bibnamefont
  {Ajith}}, \bibinfo {author} {\bibfnamefont {S.}~\bibnamefont {Babak}},
  \bibinfo {author} {\bibfnamefont {Y.}~\bibnamefont {Chen}}, \bibinfo {author}
  {\bibfnamefont {M.}~\bibnamefont {Hewitson}}, \bibinfo {author}
  {\bibfnamefont {B.}~\bibnamefont {Krishnan}},  \emph {et~al.},\ }\href
  {\doibase 10.1088/0264-9381/24/19/S31} {\bibfield  {journal} {\bibinfo
  {journal} {Class.Quant.Grav.}\ }\textbf {\bibinfo {volume} {24}},\ \bibinfo
  {pages} {S689} (\bibinfo {year} {2007})},\ \Eprint
  {http://arxiv.org/abs/0704.3764} {arXiv:0704.3764 [gr-qc]} \BibitemShut
  {NoStop}%
\bibitem [{\citenamefont {Ajith}\ \emph {et~al.}(2008)\citenamefont {Ajith},
  \citenamefont {Babak}, \citenamefont {Chen}, \citenamefont {Hewitson},
  \citenamefont {Krishnan} \emph {et~al.}}]{Ajith:2007kx}%
  \BibitemOpen
  \bibfield  {author} {\bibinfo {author} {\bibfnamefont {P.}~\bibnamefont
  {Ajith}}, \bibinfo {author} {\bibfnamefont {S.}~\bibnamefont {Babak}},
  \bibinfo {author} {\bibfnamefont {Y.}~\bibnamefont {Chen}}, \bibinfo {author}
  {\bibfnamefont {M.}~\bibnamefont {Hewitson}}, \bibinfo {author}
  {\bibfnamefont {B.}~\bibnamefont {Krishnan}},  \emph {et~al.},\ }\href
  {\doibase 10.1103/PhysRevD.79.129901, 10.1103/PhysRevD.77.104017} {\bibfield
  {journal} {\bibinfo  {journal} {Phys.Rev.}\ }\textbf {\bibinfo {volume}
  {D77}},\ \bibinfo {pages} {104017} (\bibinfo {year} {2008})},\ \Eprint
  {http://arxiv.org/abs/0710.2335} {arXiv:0710.2335 [gr-qc]} \BibitemShut
  {NoStop}%
\bibitem [{\citenamefont {Ajith}\ \emph {et~al.}(2011)\citenamefont {Ajith},
  \citenamefont {Hannam}, \citenamefont {Husa}, \citenamefont {Chen},
  \citenamefont {Br{\"u}gmann} \emph {et~al.}}]{Ajith:2009bn}%
  \BibitemOpen
  \bibfield  {author} {\bibinfo {author} {\bibfnamefont {P.}~\bibnamefont
  {Ajith}}, \bibinfo {author} {\bibfnamefont {M.}~\bibnamefont {Hannam}},
  \bibinfo {author} {\bibfnamefont {S.}~\bibnamefont {Husa}}, \bibinfo {author}
  {\bibfnamefont {Y.}~\bibnamefont {Chen}}, \bibinfo {author} {\bibfnamefont
  {B.}~\bibnamefont {Br{\"u}gmann}},  \emph {et~al.},\ }\href {\doibase
  10.1103/PhysRevLett.106.241101} {\bibfield  {journal} {\bibinfo  {journal}
  {Phys.Rev.Lett.}\ }\textbf {\bibinfo {volume} {106}},\ \bibinfo {pages}
  {241101} (\bibinfo {year} {2011})},\ \Eprint {http://arxiv.org/abs/0909.2867}
  {arXiv:0909.2867 [gr-qc]} \BibitemShut {NoStop}%
\bibitem [{\citenamefont {Santamaria}\ \emph {et~al.}(2010)\citenamefont
  {Santamaria}, \citenamefont {Ohme}, \citenamefont {Ajith}, \citenamefont
  {Br{\"u}gmann}, \citenamefont {Dorband} \emph {et~al.}}]{Santamaria:2010yb}%
  \BibitemOpen
  \bibfield  {author} {\bibinfo {author} {\bibfnamefont {L.}~\bibnamefont
  {Santamaria}}, \bibinfo {author} {\bibfnamefont {F.}~\bibnamefont {Ohme}},
  \bibinfo {author} {\bibfnamefont {P.}~\bibnamefont {Ajith}}, \bibinfo
  {author} {\bibfnamefont {B.}~\bibnamefont {Br{\"u}gmann}}, \bibinfo {author}
  {\bibfnamefont {N.}~\bibnamefont {Dorband}},  \emph {et~al.},\ }\href
  {\doibase 10.1103/PhysRevD.82.064016} {\bibfield  {journal} {\bibinfo
  {journal} {Phys.Rev.}\ }\textbf {\bibinfo {volume} {D82}},\ \bibinfo {pages}
  {064016} (\bibinfo {year} {2010})},\ \Eprint {http://arxiv.org/abs/1005.3306}
  {arXiv:1005.3306 [gr-qc]} \BibitemShut {NoStop}%
\bibitem [{\citenamefont {Husa}\ \emph {et~al.}(2016)\citenamefont {Husa},
  \citenamefont {Khan}, \citenamefont {Hannam}, \citenamefont {P{\"u}rrer},
  \citenamefont {Ohme}, \citenamefont {Jim{\'e}nez~Forteza},\ and\
  \citenamefont {Boh{\'e}}}]{Husa:2015iqa}%
  \BibitemOpen
  \bibfield  {author} {\bibinfo {author} {\bibfnamefont {S.}~\bibnamefont
  {Husa}}, \bibinfo {author} {\bibfnamefont {S.}~\bibnamefont {Khan}}, \bibinfo
  {author} {\bibfnamefont {M.}~\bibnamefont {Hannam}}, \bibinfo {author}
  {\bibfnamefont {M.}~\bibnamefont {P{\"u}rrer}}, \bibinfo {author}
  {\bibfnamefont {F.}~\bibnamefont {Ohme}}, \bibinfo {author} {\bibfnamefont
  {X.}~\bibnamefont {Jim{\'e}nez~Forteza}}, \ and\ \bibinfo {author}
  {\bibfnamefont {A.}~\bibnamefont {Boh{\'e}}},\ }\href {\doibase
  10.1103/PhysRevD.93.044006} {\bibfield  {journal} {\bibinfo  {journal} {Phys.
  Rev.}\ }\textbf {\bibinfo {volume} {D93}},\ \bibinfo {pages} {044006}
  (\bibinfo {year} {2016})},\ \Eprint {http://arxiv.org/abs/1508.07250}
  {arXiv:1508.07250 [gr-qc]} \BibitemShut {NoStop}%
\bibitem [{\citenamefont {Khan}\ \emph {et~al.}(2016)\citenamefont {Khan},
  \citenamefont {Husa}, \citenamefont {Hannam}, \citenamefont {Ohme},
  \citenamefont {P{\"u}rrer}, \citenamefont {Jim{\'e}nez~Forteza},\ and\
  \citenamefont {Boh{\'e}}}]{Khan:2015jqa}%
  \BibitemOpen
  \bibfield  {author} {\bibinfo {author} {\bibfnamefont {S.}~\bibnamefont
  {Khan}}, \bibinfo {author} {\bibfnamefont {S.}~\bibnamefont {Husa}}, \bibinfo
  {author} {\bibfnamefont {M.}~\bibnamefont {Hannam}}, \bibinfo {author}
  {\bibfnamefont {F.}~\bibnamefont {Ohme}}, \bibinfo {author} {\bibfnamefont
  {M.}~\bibnamefont {P{\"u}rrer}}, \bibinfo {author} {\bibfnamefont
  {X.}~\bibnamefont {Jim{\'e}nez~Forteza}}, \ and\ \bibinfo {author}
  {\bibfnamefont {A.}~\bibnamefont {Boh{\'e}}},\ }\href {\doibase
  10.1103/PhysRevD.93.044007} {\bibfield  {journal} {\bibinfo  {journal} {Phys.
  Rev.}\ }\textbf {\bibinfo {volume} {D93}},\ \bibinfo {pages} {044007}
  (\bibinfo {year} {2016})},\ \Eprint {http://arxiv.org/abs/1508.07253}
  {arXiv:1508.07253 [gr-qc]} \BibitemShut {NoStop}%
\bibitem [{\citenamefont {Pratten}\ \emph {et~al.}(2020)\citenamefont
  {Pratten}, \citenamefont {Husa}, \citenamefont {Garcia-Quiros}, \citenamefont
  {Colleoni}, \citenamefont {Ramos-Buades}, \citenamefont {Estelles},\ and\
  \citenamefont {Jaume}}]{Pratten:2020fqn}%
  \BibitemOpen
  \bibfield  {author} {\bibinfo {author} {\bibfnamefont {G.}~\bibnamefont
  {Pratten}}, \bibinfo {author} {\bibfnamefont {S.}~\bibnamefont {Husa}},
  \bibinfo {author} {\bibfnamefont {C.}~\bibnamefont {Garcia-Quiros}}, \bibinfo
  {author} {\bibfnamefont {M.}~\bibnamefont {Colleoni}}, \bibinfo {author}
  {\bibfnamefont {A.}~\bibnamefont {Ramos-Buades}}, \bibinfo {author}
  {\bibfnamefont {H.}~\bibnamefont {Estelles}}, \ and\ \bibinfo {author}
  {\bibfnamefont {R.}~\bibnamefont {Jaume}},\ }\href {\doibase
  10.1103/PhysRevD.102.064001} {\bibfield  {journal} {\bibinfo  {journal}
  {Phys. Rev. D}\ }\textbf {\bibinfo {volume} {102}},\ \bibinfo {pages}
  {064001} (\bibinfo {year} {2020})},\ \Eprint
  {http://arxiv.org/abs/2001.11412} {arXiv:2001.11412 [gr-qc]} \BibitemShut
  {NoStop}%
\bibitem [{\citenamefont {Estell\'es}\ \emph
  {et~al.}(2021{\natexlab{a}})\citenamefont {Estell\'es}, \citenamefont
  {Ramos-Buades}, \citenamefont {Husa}, \citenamefont {Garc\'\i{}a-Quir\'os},
  \citenamefont {Colleoni}, \citenamefont {Haegel},\ and\ \citenamefont
  {Jaume}}]{Estelles:2020osj}%
  \BibitemOpen
  \bibfield  {author} {\bibinfo {author} {\bibfnamefont {H.}~\bibnamefont
  {Estell\'es}}, \bibinfo {author} {\bibfnamefont {A.}~\bibnamefont
  {Ramos-Buades}}, \bibinfo {author} {\bibfnamefont {S.}~\bibnamefont {Husa}},
  \bibinfo {author} {\bibfnamefont {C.}~\bibnamefont {Garc\'\i{}a-Quir\'os}},
  \bibinfo {author} {\bibfnamefont {M.}~\bibnamefont {Colleoni}}, \bibinfo
  {author} {\bibfnamefont {L.}~\bibnamefont {Haegel}}, \ and\ \bibinfo {author}
  {\bibfnamefont {R.}~\bibnamefont {Jaume}},\ }\href {\doibase
  10.1103/PhysRevD.103.124060} {\bibfield  {journal} {\bibinfo  {journal}
  {Phys. Rev. D}\ }\textbf {\bibinfo {volume} {103}},\ \bibinfo {pages}
  {124060} (\bibinfo {year} {2021}{\natexlab{a}})},\ \Eprint
  {http://arxiv.org/abs/2004.08302} {arXiv:2004.08302 [gr-qc]} \BibitemShut
  {NoStop}%
\bibitem [{\citenamefont {Estell\'es}\ \emph {et~al.}(2020)\citenamefont
  {Estell\'es}, \citenamefont {Husa}, \citenamefont {Colleoni}, \citenamefont
  {Keitel}, \citenamefont {Mateu-Lucena}, \citenamefont {Garc\'\i{}a-Quir\'os},
  \citenamefont {Ramos-Buades},\ and\ \citenamefont
  {Borchers}}]{Estelles:2020twz}%
  \BibitemOpen
  \bibfield  {author} {\bibinfo {author} {\bibfnamefont {H.}~\bibnamefont
  {Estell\'es}}, \bibinfo {author} {\bibfnamefont {S.}~\bibnamefont {Husa}},
  \bibinfo {author} {\bibfnamefont {M.}~\bibnamefont {Colleoni}}, \bibinfo
  {author} {\bibfnamefont {D.}~\bibnamefont {Keitel}}, \bibinfo {author}
  {\bibfnamefont {M.}~\bibnamefont {Mateu-Lucena}}, \bibinfo {author}
  {\bibfnamefont {C.}~\bibnamefont {Garc\'\i{}a-Quir\'os}}, \bibinfo {author}
  {\bibfnamefont {A.}~\bibnamefont {Ramos-Buades}}, \ and\ \bibinfo {author}
  {\bibfnamefont {A.}~\bibnamefont {Borchers}},\ }\href@noop {} {\  (\bibinfo
  {year} {2020})},\ \Eprint {http://arxiv.org/abs/2012.11923} {arXiv:2012.11923
  [gr-qc]} \BibitemShut {NoStop}%
\bibitem [{\citenamefont {Estell\'es}\ \emph
  {et~al.}(2021{\natexlab{b}})\citenamefont {Estell\'es}, \citenamefont
  {Colleoni}, \citenamefont {Garc\'\i{}a-Quir\'os}, \citenamefont {Husa},
  \citenamefont {Keitel}, \citenamefont {Mateu-Lucena}, \citenamefont
  {Planas},\ and\ \citenamefont {Ramos-Buades}}]{Estelles:2021gvs}%
  \BibitemOpen
  \bibfield  {author} {\bibinfo {author} {\bibfnamefont {H.}~\bibnamefont
  {Estell\'es}}, \bibinfo {author} {\bibfnamefont {M.}~\bibnamefont
  {Colleoni}}, \bibinfo {author} {\bibfnamefont {C.}~\bibnamefont
  {Garc\'\i{}a-Quir\'os}}, \bibinfo {author} {\bibfnamefont {S.}~\bibnamefont
  {Husa}}, \bibinfo {author} {\bibfnamefont {D.}~\bibnamefont {Keitel}},
  \bibinfo {author} {\bibfnamefont {M.}~\bibnamefont {Mateu-Lucena}}, \bibinfo
  {author} {\bibfnamefont {M.~d.~L.}\ \bibnamefont {Planas}}, \ and\ \bibinfo
  {author} {\bibfnamefont {A.}~\bibnamefont {Ramos-Buades}},\ }\href@noop {} {\
   (\bibinfo {year} {2021}{\natexlab{b}})},\ \Eprint
  {http://arxiv.org/abs/2105.05872} {arXiv:2105.05872 [gr-qc]} \BibitemShut
  {NoStop}%
\bibitem [{\citenamefont {Hamilton}\ \emph {et~al.}(2021)\citenamefont
  {Hamilton}, \citenamefont {London}, \citenamefont {Thompson}, \citenamefont
  {Fauchon-Jones}, \citenamefont {Hannam}, \citenamefont {Kalaghatgi},
  \citenamefont {Khan}, \citenamefont {Pannarale},\ and\ \citenamefont
  {Vano-Vinuales}}]{Hamilton:2021pkf}%
  \BibitemOpen
  \bibfield  {author} {\bibinfo {author} {\bibfnamefont {E.}~\bibnamefont
  {Hamilton}}, \bibinfo {author} {\bibfnamefont {L.}~\bibnamefont {London}},
  \bibinfo {author} {\bibfnamefont {J.~E.}\ \bibnamefont {Thompson}}, \bibinfo
  {author} {\bibfnamefont {E.}~\bibnamefont {Fauchon-Jones}}, \bibinfo {author}
  {\bibfnamefont {M.}~\bibnamefont {Hannam}}, \bibinfo {author} {\bibfnamefont
  {C.}~\bibnamefont {Kalaghatgi}}, \bibinfo {author} {\bibfnamefont
  {S.}~\bibnamefont {Khan}}, \bibinfo {author} {\bibfnamefont {F.}~\bibnamefont
  {Pannarale}}, \ and\ \bibinfo {author} {\bibfnamefont {A.}~\bibnamefont
  {Vano-Vinuales}},\ }\href@noop {} {\  (\bibinfo {year} {2021})},\ \Eprint
  {http://arxiv.org/abs/2107.08876} {arXiv:2107.08876 [gr-qc]} \BibitemShut
  {NoStop}%
\bibitem [{\citenamefont {London}\ \emph {et~al.}(2018)\citenamefont {London},
  \citenamefont {Khan}, \citenamefont {Fauchon-Jones}, \citenamefont {Forteza},
  \citenamefont {Hannam}, \citenamefont {Husa}, \citenamefont {Kalaghatgi},
  \citenamefont {Ohme},\ and\ \citenamefont {Pannarale}}]{London:2017bcn}%
  \BibitemOpen
  \bibfield  {author} {\bibinfo {author} {\bibfnamefont {L.}~\bibnamefont
  {London}}, \bibinfo {author} {\bibfnamefont {S.}~\bibnamefont {Khan}},
  \bibinfo {author} {\bibfnamefont {E.}~\bibnamefont {Fauchon-Jones}}, \bibinfo
  {author} {\bibfnamefont {X.~J.}\ \bibnamefont {Forteza}}, \bibinfo {author}
  {\bibfnamefont {M.}~\bibnamefont {Hannam}}, \bibinfo {author} {\bibfnamefont
  {S.}~\bibnamefont {Husa}}, \bibinfo {author} {\bibfnamefont {C.}~\bibnamefont
  {Kalaghatgi}}, \bibinfo {author} {\bibfnamefont {F.}~\bibnamefont {Ohme}}, \
  and\ \bibinfo {author} {\bibfnamefont {F.}~\bibnamefont {Pannarale}},\ }\href
  {\doibase 10.1103/PhysRevLett.120.161102} {\bibfield  {journal} {\bibinfo
  {journal} {Phys. Rev. Lett.}\ }\textbf {\bibinfo {volume} {120}},\ \bibinfo
  {pages} {161102} (\bibinfo {year} {2018})},\ \Eprint
  {http://arxiv.org/abs/1708.00404} {arXiv:1708.00404 [gr-qc]} \BibitemShut
  {NoStop}%
\bibitem [{\citenamefont {Garc\'\i{}a-Quir\'os}\ \emph
  {et~al.}(2020)\citenamefont {Garc\'\i{}a-Quir\'os}, \citenamefont {Colleoni},
  \citenamefont {Husa}, \citenamefont {Estell\'es}, \citenamefont {Pratten},
  \citenamefont {Ramos-Buades}, \citenamefont {Mateu-Lucena},\ and\
  \citenamefont {Jaume}}]{Garcia-Quiros:2020qpx}%
  \BibitemOpen
  \bibfield  {author} {\bibinfo {author} {\bibfnamefont {C.}~\bibnamefont
  {Garc\'\i{}a-Quir\'os}}, \bibinfo {author} {\bibfnamefont {M.}~\bibnamefont
  {Colleoni}}, \bibinfo {author} {\bibfnamefont {S.}~\bibnamefont {Husa}},
  \bibinfo {author} {\bibfnamefont {H.}~\bibnamefont {Estell\'es}}, \bibinfo
  {author} {\bibfnamefont {G.}~\bibnamefont {Pratten}}, \bibinfo {author}
  {\bibfnamefont {A.}~\bibnamefont {Ramos-Buades}}, \bibinfo {author}
  {\bibfnamefont {M.}~\bibnamefont {Mateu-Lucena}}, \ and\ \bibinfo {author}
  {\bibfnamefont {R.}~\bibnamefont {Jaume}},\ }\href {\doibase
  10.1103/PhysRevD.102.064002} {\bibfield  {journal} {\bibinfo  {journal}
  {Phys. Rev. D}\ }\textbf {\bibinfo {volume} {102}},\ \bibinfo {pages}
  {064002} (\bibinfo {year} {2020})},\ \Eprint
  {http://arxiv.org/abs/2001.10914} {arXiv:2001.10914 [gr-qc]} \BibitemShut
  {NoStop}%
\bibitem [{\citenamefont {Khan}\ \emph
  {et~al.}(2019{\natexlab{a}})\citenamefont {Khan}, \citenamefont {Ohme},
  \citenamefont {Chatziioannou},\ and\ \citenamefont {Hannam}}]{Khan:2019kot}%
  \BibitemOpen
  \bibfield  {author} {\bibinfo {author} {\bibfnamefont {S.}~\bibnamefont
  {Khan}}, \bibinfo {author} {\bibfnamefont {F.}~\bibnamefont {Ohme}}, \bibinfo
  {author} {\bibfnamefont {K.}~\bibnamefont {Chatziioannou}}, \ and\ \bibinfo
  {author} {\bibfnamefont {M.}~\bibnamefont {Hannam}},\ }\href@noop {} {\
  (\bibinfo {year} {2019}{\natexlab{a}})},\ \Eprint
  {http://arxiv.org/abs/1911.06050} {arXiv:1911.06050 [gr-qc]} \BibitemShut
  {NoStop}%
\bibitem [{\citenamefont {Hannam}\ \emph {et~al.}(2014)\citenamefont {Hannam},
  \citenamefont {Schmidt}, \citenamefont {Boh{\'e}}, \citenamefont {Haegel},
  \citenamefont {Husa}, \citenamefont {Ohme}, \citenamefont {Pratten},\ and\
  \citenamefont {P{\"u}rrer}}]{Hannam:2013oca}%
  \BibitemOpen
  \bibfield  {author} {\bibinfo {author} {\bibfnamefont {M.}~\bibnamefont
  {Hannam}}, \bibinfo {author} {\bibfnamefont {P.}~\bibnamefont {Schmidt}},
  \bibinfo {author} {\bibfnamefont {A.}~\bibnamefont {Boh{\'e}}}, \bibinfo
  {author} {\bibfnamefont {L.}~\bibnamefont {Haegel}}, \bibinfo {author}
  {\bibfnamefont {S.}~\bibnamefont {Husa}}, \bibinfo {author} {\bibfnamefont
  {F.}~\bibnamefont {Ohme}}, \bibinfo {author} {\bibfnamefont {G.}~\bibnamefont
  {Pratten}}, \ and\ \bibinfo {author} {\bibfnamefont {M.}~\bibnamefont
  {P{\"u}rrer}},\ }\href {\doibase 10.1103/PhysRevLett.113.151101} {\bibfield
  {journal} {\bibinfo  {journal} {Phys. Rev. Lett.}\ }\textbf {\bibinfo
  {volume} {113}},\ \bibinfo {pages} {151101} (\bibinfo {year} {2014})},\
  \Eprint {http://arxiv.org/abs/1308.3271} {arXiv:1308.3271 [gr-qc]}
  \BibitemShut {NoStop}%
\bibitem [{\citenamefont {Schmidt}\ \emph {et~al.}(2015)\citenamefont
  {Schmidt}, \citenamefont {Ohme},\ and\ \citenamefont
  {Hannam}}]{Schmidt:2014iyl}%
  \BibitemOpen
  \bibfield  {author} {\bibinfo {author} {\bibfnamefont {P.}~\bibnamefont
  {Schmidt}}, \bibinfo {author} {\bibfnamefont {F.}~\bibnamefont {Ohme}}, \
  and\ \bibinfo {author} {\bibfnamefont {M.}~\bibnamefont {Hannam}},\ }\href
  {\doibase 10.1103/PhysRevD.91.024043} {\bibfield  {journal} {\bibinfo
  {journal} {Phys. Rev.}\ }\textbf {\bibinfo {volume} {D91}},\ \bibinfo {pages}
  {024043} (\bibinfo {year} {2015})},\ \Eprint {http://arxiv.org/abs/1408.1810}
  {arXiv:1408.1810 [gr-qc]} \BibitemShut {NoStop}%
\bibitem [{\citenamefont {Khan}\ \emph
  {et~al.}(2019{\natexlab{b}})\citenamefont {Khan}, \citenamefont
  {Chatziioannou}, \citenamefont {Hannam},\ and\ \citenamefont
  {Ohme}}]{Khan:2018fmp}%
  \BibitemOpen
  \bibfield  {author} {\bibinfo {author} {\bibfnamefont {S.}~\bibnamefont
  {Khan}}, \bibinfo {author} {\bibfnamefont {K.}~\bibnamefont {Chatziioannou}},
  \bibinfo {author} {\bibfnamefont {M.}~\bibnamefont {Hannam}}, \ and\ \bibinfo
  {author} {\bibfnamefont {F.}~\bibnamefont {Ohme}},\ }\href {\doibase
  10.1103/PhysRevD.100.024059} {\bibfield  {journal} {\bibinfo  {journal}
  {Phys. Rev.}\ }\textbf {\bibinfo {volume} {D100}},\ \bibinfo {pages} {024059}
  (\bibinfo {year} {2019}{\natexlab{b}})},\ \Eprint
  {http://arxiv.org/abs/1809.10113} {arXiv:1809.10113 [gr-qc]} \BibitemShut
  {NoStop}%
\bibitem [{\citenamefont {Pratten}\ \emph {et~al.}(2021)\citenamefont {Pratten}
  \emph {et~al.}}]{Pratten:2020ceb}%
  \BibitemOpen
  \bibfield  {author} {\bibinfo {author} {\bibfnamefont {G.}~\bibnamefont
  {Pratten}} \emph {et~al.},\ }\href {\doibase 10.1103/PhysRevD.103.104056}
  {\bibfield  {journal} {\bibinfo  {journal} {Phys. Rev. D}\ }\textbf {\bibinfo
  {volume} {103}},\ \bibinfo {pages} {104056} (\bibinfo {year} {2021})},\
  \Eprint {http://arxiv.org/abs/2004.06503} {arXiv:2004.06503 [gr-qc]}
  \BibitemShut {NoStop}%
\bibitem [{\citenamefont {Dietrich}\ \emph {et~al.}(2017)\citenamefont
  {Dietrich}, \citenamefont {Bernuzzi},\ and\ \citenamefont
  {Tichy}}]{Dietrich:2017aum}%
  \BibitemOpen
  \bibfield  {author} {\bibinfo {author} {\bibfnamefont {T.}~\bibnamefont
  {Dietrich}}, \bibinfo {author} {\bibfnamefont {S.}~\bibnamefont {Bernuzzi}},
  \ and\ \bibinfo {author} {\bibfnamefont {W.}~\bibnamefont {Tichy}},\ }\href
  {\doibase 10.1103/PhysRevD.96.121501} {\bibfield  {journal} {\bibinfo
  {journal} {Phys. Rev.}\ }\textbf {\bibinfo {volume} {D96}},\ \bibinfo {pages}
  {121501} (\bibinfo {year} {2017})},\ \Eprint
  {http://arxiv.org/abs/1706.02969} {arXiv:1706.02969 [gr-qc]} \BibitemShut
  {NoStop}%
\bibitem [{\citenamefont {Dietrich}\ \emph {et~al.}(2019)\citenamefont
  {Dietrich}, \citenamefont {Samajdar}, \citenamefont {Khan}, \citenamefont
  {Johnson-McDaniel}, \citenamefont {Dudi},\ and\ \citenamefont
  {Tichy}}]{Dietrich:2019kaq}%
  \BibitemOpen
  \bibfield  {author} {\bibinfo {author} {\bibfnamefont {T.}~\bibnamefont
  {Dietrich}}, \bibinfo {author} {\bibfnamefont {A.}~\bibnamefont {Samajdar}},
  \bibinfo {author} {\bibfnamefont {S.}~\bibnamefont {Khan}}, \bibinfo {author}
  {\bibfnamefont {N.~K.}\ \bibnamefont {Johnson-McDaniel}}, \bibinfo {author}
  {\bibfnamefont {R.}~\bibnamefont {Dudi}}, \ and\ \bibinfo {author}
  {\bibfnamefont {W.}~\bibnamefont {Tichy}},\ }\href {\doibase
  10.1103/PhysRevD.100.044003} {\bibfield  {journal} {\bibinfo  {journal}
  {Phys. Rev.}\ }\textbf {\bibinfo {volume} {D100}},\ \bibinfo {pages} {044003}
  (\bibinfo {year} {2019})},\ \Eprint {http://arxiv.org/abs/1905.06011}
  {arXiv:1905.06011 [gr-qc]} \BibitemShut {NoStop}%
\bibitem [{\citenamefont {Blackman}\ \emph {et~al.}(2014)\citenamefont
  {Blackman}, \citenamefont {Szilagyi}, \citenamefont {Galley},\ and\
  \citenamefont {Tiglio}}]{Blackman:2014maa}%
  \BibitemOpen
  \bibfield  {author} {\bibinfo {author} {\bibfnamefont {J.}~\bibnamefont
  {Blackman}}, \bibinfo {author} {\bibfnamefont {B.}~\bibnamefont {Szilagyi}},
  \bibinfo {author} {\bibfnamefont {C.~R.}\ \bibnamefont {Galley}}, \ and\
  \bibinfo {author} {\bibfnamefont {M.}~\bibnamefont {Tiglio}},\ }\href
  {\doibase 10.1103/PhysRevLett.113.021101} {\bibfield  {journal} {\bibinfo
  {journal} {Phys. Rev. Lett.}\ }\textbf {\bibinfo {volume} {113}},\ \bibinfo
  {pages} {021101} (\bibinfo {year} {2014})},\ \Eprint
  {http://arxiv.org/abs/1401.7038} {arXiv:1401.7038 [gr-qc]} \BibitemShut
  {NoStop}%
\bibitem [{\citenamefont {Blackman}\ \emph {et~al.}(2017)\citenamefont
  {Blackman}, \citenamefont {Field}, \citenamefont {Scheel}, \citenamefont
  {Galley}, \citenamefont {Hemberger}, \citenamefont {Schmidt},\ and\
  \citenamefont {Smith}}]{Blackman:2017dfb}%
  \BibitemOpen
  \bibfield  {author} {\bibinfo {author} {\bibfnamefont {J.}~\bibnamefont
  {Blackman}}, \bibinfo {author} {\bibfnamefont {S.~E.}\ \bibnamefont {Field}},
  \bibinfo {author} {\bibfnamefont {M.~A.}\ \bibnamefont {Scheel}}, \bibinfo
  {author} {\bibfnamefont {C.~R.}\ \bibnamefont {Galley}}, \bibinfo {author}
  {\bibfnamefont {D.~A.}\ \bibnamefont {Hemberger}}, \bibinfo {author}
  {\bibfnamefont {P.}~\bibnamefont {Schmidt}}, \ and\ \bibinfo {author}
  {\bibfnamefont {R.}~\bibnamefont {Smith}},\ }\href {\doibase
  10.1103/PhysRevD.95.104023} {\bibfield  {journal} {\bibinfo  {journal} {Phys.
  Rev.}\ }\textbf {\bibinfo {volume} {D95}},\ \bibinfo {pages} {104023}
  (\bibinfo {year} {2017})},\ \Eprint {http://arxiv.org/abs/1701.00550}
  {arXiv:1701.00550 [gr-qc]} \BibitemShut {NoStop}%
\bibitem [{\citenamefont {Varma}\ \emph
  {et~al.}(2019{\natexlab{a}})\citenamefont {Varma}, \citenamefont {Field},
  \citenamefont {Scheel}, \citenamefont {Blackman}, \citenamefont {Kidder},\
  and\ \citenamefont {Pfeiffer}}]{Varma:2018mmi}%
  \BibitemOpen
  \bibfield  {author} {\bibinfo {author} {\bibfnamefont {V.}~\bibnamefont
  {Varma}}, \bibinfo {author} {\bibfnamefont {S.~E.}\ \bibnamefont {Field}},
  \bibinfo {author} {\bibfnamefont {M.~A.}\ \bibnamefont {Scheel}}, \bibinfo
  {author} {\bibfnamefont {J.}~\bibnamefont {Blackman}}, \bibinfo {author}
  {\bibfnamefont {L.~E.}\ \bibnamefont {Kidder}}, \ and\ \bibinfo {author}
  {\bibfnamefont {H.~P.}\ \bibnamefont {Pfeiffer}},\ }\href {\doibase
  10.1103/PhysRevD.99.064045} {\bibfield  {journal} {\bibinfo  {journal} {Phys.
  Rev.}\ }\textbf {\bibinfo {volume} {D99}},\ \bibinfo {pages} {064045}
  (\bibinfo {year} {2019}{\natexlab{a}})},\ \Eprint
  {http://arxiv.org/abs/1812.07865} {arXiv:1812.07865 [gr-qc]} \BibitemShut
  {NoStop}%
\bibitem [{\citenamefont {Varma}\ \emph
  {et~al.}(2019{\natexlab{b}})\citenamefont {Varma}, \citenamefont {Field},
  \citenamefont {Scheel}, \citenamefont {Blackman}, \citenamefont {Gerosa},
  \citenamefont {Stein}, \citenamefont {Kidder},\ and\ \citenamefont
  {Pfeiffer}}]{Varma:2019csw}%
  \BibitemOpen
  \bibfield  {author} {\bibinfo {author} {\bibfnamefont {V.}~\bibnamefont
  {Varma}}, \bibinfo {author} {\bibfnamefont {S.~E.}\ \bibnamefont {Field}},
  \bibinfo {author} {\bibfnamefont {M.~A.}\ \bibnamefont {Scheel}}, \bibinfo
  {author} {\bibfnamefont {J.}~\bibnamefont {Blackman}}, \bibinfo {author}
  {\bibfnamefont {D.}~\bibnamefont {Gerosa}}, \bibinfo {author} {\bibfnamefont
  {L.~C.}\ \bibnamefont {Stein}}, \bibinfo {author} {\bibfnamefont {L.~E.}\
  \bibnamefont {Kidder}}, \ and\ \bibinfo {author} {\bibfnamefont {H.~P.}\
  \bibnamefont {Pfeiffer}},\ }\href {\doibase 10.1103/PhysRevResearch.1.033015}
  {\bibfield  {journal} {\bibinfo  {journal} {Phys. Rev. Research.}\ }\textbf
  {\bibinfo {volume} {1}},\ \bibinfo {pages} {033015} (\bibinfo {year}
  {2019}{\natexlab{b}})},\ \Eprint {http://arxiv.org/abs/1905.09300}
  {arXiv:1905.09300 [gr-qc]} \BibitemShut {NoStop}%
\bibitem [{\citenamefont {Williams}\ \emph {et~al.}(2019)\citenamefont
  {Williams}, \citenamefont {Heng}, \citenamefont {Gair}, \citenamefont
  {Clark},\ and\ \citenamefont {Khamesra}}]{Williams:2019vub}%
  \BibitemOpen
  \bibfield  {author} {\bibinfo {author} {\bibfnamefont {D.}~\bibnamefont
  {Williams}}, \bibinfo {author} {\bibfnamefont {I.~S.}\ \bibnamefont {Heng}},
  \bibinfo {author} {\bibfnamefont {J.}~\bibnamefont {Gair}}, \bibinfo {author}
  {\bibfnamefont {J.~A.}\ \bibnamefont {Clark}}, \ and\ \bibinfo {author}
  {\bibfnamefont {B.}~\bibnamefont {Khamesra}},\ }\href@noop {} {\  (\bibinfo
  {year} {2019})},\ \Eprint {http://arxiv.org/abs/1903.09204} {arXiv:1903.09204
  [gr-qc]} \BibitemShut {NoStop}%
\bibitem [{\citenamefont {Akcay}\ \emph {et~al.}(2021)\citenamefont {Akcay},
  \citenamefont {Gamba},\ and\ \citenamefont {Bernuzzi}}]{Akcay:2020qrj}%
  \BibitemOpen
  \bibfield  {author} {\bibinfo {author} {\bibfnamefont {S.}~\bibnamefont
  {Akcay}}, \bibinfo {author} {\bibfnamefont {R.}~\bibnamefont {Gamba}}, \ and\
  \bibinfo {author} {\bibfnamefont {S.}~\bibnamefont {Bernuzzi}},\ }\href
  {\doibase 10.1103/PhysRevD.103.024014} {\bibfield  {journal} {\bibinfo
  {journal} {Phys. Rev. D}\ }\textbf {\bibinfo {volume} {103}},\ \bibinfo
  {pages} {024014} (\bibinfo {year} {2021})},\ \Eprint
  {http://arxiv.org/abs/2005.05338} {arXiv:2005.05338 [gr-qc]} \BibitemShut
  {NoStop}%
\bibitem [{\citenamefont {Gamba}\ \emph
  {et~al.}(2021{\natexlab{b}})\citenamefont {Gamba}, \citenamefont {Breschi},
  \citenamefont {Bernuzzi}, \citenamefont {Agathos},\ and\ \citenamefont
  {Nagar}}]{Gamba:2020wgg}%
  \BibitemOpen
  \bibfield  {author} {\bibinfo {author} {\bibfnamefont {R.}~\bibnamefont
  {Gamba}}, \bibinfo {author} {\bibfnamefont {M.}~\bibnamefont {Breschi}},
  \bibinfo {author} {\bibfnamefont {S.}~\bibnamefont {Bernuzzi}}, \bibinfo
  {author} {\bibfnamefont {M.}~\bibnamefont {Agathos}}, \ and\ \bibinfo
  {author} {\bibfnamefont {A.}~\bibnamefont {Nagar}},\ }\href {\doibase
  10.1103/PhysRevD.103.124015} {\bibfield  {journal} {\bibinfo  {journal}
  {Phys. Rev. D}\ }\textbf {\bibinfo {volume} {103}},\ \bibinfo {pages}
  {124015} (\bibinfo {year} {2021}{\natexlab{b}})},\ \Eprint
  {http://arxiv.org/abs/2009.08467} {arXiv:2009.08467 [gr-qc]} \BibitemShut
  {NoStop}%
\bibitem [{SXS()}]{SXS:catalog}%
  \BibitemOpen
  \href@noop {} {\enquote {\bibinfo {title} {{SXS Gravitational Waveform
  Database}},}\ }\bibinfo {howpublished}
  {\url{https://data.black-holes.org/waveforms/index.html}}\BibitemShut
  {NoStop}%
\bibitem [{\citenamefont {Ajith}(2011)}]{Ajith:2011ec}%
  \BibitemOpen
  \bibfield  {author} {\bibinfo {author} {\bibfnamefont {P.}~\bibnamefont
  {Ajith}},\ }\href {\doibase 10.1103/PhysRevD.84.084037} {\bibfield  {journal}
  {\bibinfo  {journal} {Phys.Rev.}\ }\textbf {\bibinfo {volume} {D84}},\
  \bibinfo {pages} {084037} (\bibinfo {year} {2011})},\ \Eprint
  {http://arxiv.org/abs/1107.1267} {arXiv:1107.1267 [gr-qc]} \BibitemShut
  {NoStop}%
\bibitem [{\citenamefont {Racine}(2008)}]{Racine:2008qv}%
  \BibitemOpen
  \bibfield  {author} {\bibinfo {author} {\bibfnamefont {E.}~\bibnamefont
  {Racine}},\ }\href {\doibase 10.1103/PhysRevD.78.044021} {\bibfield
  {journal} {\bibinfo  {journal} {Phys. Rev.}\ }\textbf {\bibinfo {volume}
  {D78}},\ \bibinfo {pages} {044021} (\bibinfo {year} {2008})},\ \Eprint
  {http://arxiv.org/abs/0803.1820} {arXiv:0803.1820 [gr-qc]} \BibitemShut
  {NoStop}%
\bibitem [{\citenamefont {Schmidt}\ \emph {et~al.}(2011)\citenamefont
  {Schmidt}, \citenamefont {Hannam}, \citenamefont {Husa},\ and\ \citenamefont
  {Ajith}}]{Schmidt:2010it}%
  \BibitemOpen
  \bibfield  {author} {\bibinfo {author} {\bibfnamefont {P.}~\bibnamefont
  {Schmidt}}, \bibinfo {author} {\bibfnamefont {M.}~\bibnamefont {Hannam}},
  \bibinfo {author} {\bibfnamefont {S.}~\bibnamefont {Husa}}, \ and\ \bibinfo
  {author} {\bibfnamefont {P.}~\bibnamefont {Ajith}},\ }\href {\doibase
  10.1103/PhysRevD.84.024046} {\bibfield  {journal} {\bibinfo  {journal} {Phys.
  Rev.}\ }\textbf {\bibinfo {volume} {D84}},\ \bibinfo {pages} {024046}
  (\bibinfo {year} {2011})},\ \Eprint {http://arxiv.org/abs/1012.2879}
  {arXiv:1012.2879 [gr-qc]} \BibitemShut {NoStop}%
\bibitem [{\citenamefont {Schmidt}\ \emph {et~al.}(2012)\citenamefont
  {Schmidt}, \citenamefont {Hannam},\ and\ \citenamefont
  {Husa}}]{Schmidt:2012rh}%
  \BibitemOpen
  \bibfield  {author} {\bibinfo {author} {\bibfnamefont {P.}~\bibnamefont
  {Schmidt}}, \bibinfo {author} {\bibfnamefont {M.}~\bibnamefont {Hannam}}, \
  and\ \bibinfo {author} {\bibfnamefont {S.}~\bibnamefont {Husa}},\ }\href
  {\doibase 10.1103/PhysRevD.86.104063} {\bibfield  {journal} {\bibinfo
  {journal} {Phys. Rev.}\ }\textbf {\bibinfo {volume} {D86}},\ \bibinfo {pages}
  {104063} (\bibinfo {year} {2012})},\ \Eprint {http://arxiv.org/abs/1207.3088}
  {arXiv:1207.3088 [gr-qc]} \BibitemShut {NoStop}%
\bibitem [{\citenamefont {Boyle}\ \emph {et~al.}(2011)\citenamefont {Boyle},
  \citenamefont {Owen},\ and\ \citenamefont {Pfeiffer}}]{Boyle:2011gg}%
  \BibitemOpen
  \bibfield  {author} {\bibinfo {author} {\bibfnamefont {M.}~\bibnamefont
  {Boyle}}, \bibinfo {author} {\bibfnamefont {R.}~\bibnamefont {Owen}}, \ and\
  \bibinfo {author} {\bibfnamefont {H.~P.}\ \bibnamefont {Pfeiffer}},\ }\href
  {\doibase 10.1103/PhysRevD.84.124011} {\bibfield  {journal} {\bibinfo
  {journal} {Phys. Rev.}\ }\textbf {\bibinfo {volume} {D84}},\ \bibinfo {pages}
  {124011} (\bibinfo {year} {2011})},\ \Eprint {http://arxiv.org/abs/1110.2965}
  {arXiv:1110.2965 [gr-qc]} \BibitemShut {NoStop}%
\bibitem [{\citenamefont {O'Shaughnessy}\ \emph {et~al.}(2011)\citenamefont
  {O'Shaughnessy}, \citenamefont {Vaishnav}, \citenamefont {Healy},
  \citenamefont {Meeks},\ and\ \citenamefont
  {Shoemaker}}]{OShaughnessy:2011pmr}%
  \BibitemOpen
  \bibfield  {author} {\bibinfo {author} {\bibfnamefont {R.}~\bibnamefont
  {O'Shaughnessy}}, \bibinfo {author} {\bibfnamefont {B.}~\bibnamefont
  {Vaishnav}}, \bibinfo {author} {\bibfnamefont {J.}~\bibnamefont {Healy}},
  \bibinfo {author} {\bibfnamefont {Z.}~\bibnamefont {Meeks}}, \ and\ \bibinfo
  {author} {\bibfnamefont {D.}~\bibnamefont {Shoemaker}},\ }\href {\doibase
  10.1103/PhysRevD.84.124002} {\bibfield  {journal} {\bibinfo  {journal} {Phys.
  Rev.}\ }\textbf {\bibinfo {volume} {D84}},\ \bibinfo {pages} {124002}
  (\bibinfo {year} {2011})},\ \Eprint {http://arxiv.org/abs/1109.5224}
  {arXiv:1109.5224 [gr-qc]} \BibitemShut {NoStop}%
\bibitem [{\citenamefont {Pan}\ \emph {et~al.}(2014{\natexlab{a}})\citenamefont
  {Pan}, \citenamefont {Buonanno}, \citenamefont {Taracchini}, \citenamefont
  {Kidder}, \citenamefont {Mroue} \emph {et~al.}}]{Pan:2013rra}%
  \BibitemOpen
  \bibfield  {author} {\bibinfo {author} {\bibfnamefont {Y.}~\bibnamefont
  {Pan}}, \bibinfo {author} {\bibfnamefont {A.}~\bibnamefont {Buonanno}},
  \bibinfo {author} {\bibfnamefont {A.}~\bibnamefont {Taracchini}}, \bibinfo
  {author} {\bibfnamefont {L.~E.}\ \bibnamefont {Kidder}}, \bibinfo {author}
  {\bibfnamefont {A.~H.}\ \bibnamefont {Mroue}},  \emph {et~al.},\ }\href
  {\doibase 10.1103/PhysRevD.89.084006} {\bibfield  {journal} {\bibinfo
  {journal} {Phys.Rev.}\ }\textbf {\bibinfo {volume} {D89}},\ \bibinfo {pages}
  {084006} (\bibinfo {year} {2014}{\natexlab{a}})},\ \Eprint
  {http://arxiv.org/abs/1307.6232} {arXiv:1307.6232 [gr-qc]} \BibitemShut
  {NoStop}%
\bibitem [{\citenamefont {Nagar}\ \emph
  {et~al.}(2019{\natexlab{a}})\citenamefont {Nagar}, \citenamefont {Pratten},
  \citenamefont {Riemenschneider},\ and\ \citenamefont
  {Gamba}}]{Nagar:2019wds}%
  \BibitemOpen
  \bibfield  {author} {\bibinfo {author} {\bibfnamefont {A.}~\bibnamefont
  {Nagar}}, \bibinfo {author} {\bibfnamefont {G.}~\bibnamefont {Pratten}},
  \bibinfo {author} {\bibfnamefont {G.}~\bibnamefont {Riemenschneider}}, \ and\
  \bibinfo {author} {\bibfnamefont {R.}~\bibnamefont {Gamba}},\ }\href@noop {}
  {\  (\bibinfo {year} {2019}{\natexlab{a}})},\ \Eprint
  {http://arxiv.org/abs/1904.09550} {arXiv:1904.09550 [gr-qc]} \BibitemShut
  {NoStop}%
\bibitem [{\citenamefont {Damour}\ and\ \citenamefont
  {Nagar}(2014{\natexlab{b}})}]{Damour:2014sva}%
  \BibitemOpen
  \bibfield  {author} {\bibinfo {author} {\bibfnamefont {T.}~\bibnamefont
  {Damour}}\ and\ \bibinfo {author} {\bibfnamefont {A.}~\bibnamefont {Nagar}},\
  }\href {\doibase 10.1103/PhysRevD.90.044018} {\bibfield  {journal} {\bibinfo
  {journal} {Phys.Rev.}\ }\textbf {\bibinfo {volume} {D90}},\ \bibinfo {pages}
  {044018} (\bibinfo {year} {2014}{\natexlab{b}})},\ \Eprint
  {http://arxiv.org/abs/1406.6913} {arXiv:1406.6913 [gr-qc]} \BibitemShut
  {NoStop}%
\bibitem [{\citenamefont {Damour}\ \emph {et~al.}(2008)\citenamefont {Damour},
  \citenamefont {Jaranowski},\ and\ \citenamefont
  {Sch{\"a}fer}}]{Damour:2008qf}%
  \BibitemOpen
  \bibfield  {author} {\bibinfo {author} {\bibfnamefont {T.}~\bibnamefont
  {Damour}}, \bibinfo {author} {\bibfnamefont {P.}~\bibnamefont {Jaranowski}},
  \ and\ \bibinfo {author} {\bibfnamefont {G.}~\bibnamefont {Sch{\"a}fer}},\
  }\href {\doibase 10.1103/PhysRevD.78.024009} {\bibfield  {journal} {\bibinfo
  {journal} {Phys.Rev.}\ }\textbf {\bibinfo {volume} {D78}},\ \bibinfo {pages}
  {024009} (\bibinfo {year} {2008})},\ \Eprint {http://arxiv.org/abs/0803.0915}
  {arXiv:0803.0915 [gr-qc]} \BibitemShut {NoStop}%
\bibitem [{\citenamefont {Nagar}\ \emph
  {et~al.}(2019{\natexlab{b}})\citenamefont {Nagar}, \citenamefont {Messina},
  \citenamefont {Rettegno}, \citenamefont {Bini}, \citenamefont {Damour},
  \citenamefont {Geralico}, \citenamefont {Akcay},\ and\ \citenamefont
  {Bernuzzi}}]{Nagar:2018plt}%
  \BibitemOpen
  \bibfield  {author} {\bibinfo {author} {\bibfnamefont {A.}~\bibnamefont
  {Nagar}}, \bibinfo {author} {\bibfnamefont {F.}~\bibnamefont {Messina}},
  \bibinfo {author} {\bibfnamefont {P.}~\bibnamefont {Rettegno}}, \bibinfo
  {author} {\bibfnamefont {D.}~\bibnamefont {Bini}}, \bibinfo {author}
  {\bibfnamefont {T.}~\bibnamefont {Damour}}, \bibinfo {author} {\bibfnamefont
  {A.}~\bibnamefont {Geralico}}, \bibinfo {author} {\bibfnamefont
  {S.}~\bibnamefont {Akcay}}, \ and\ \bibinfo {author} {\bibfnamefont
  {S.}~\bibnamefont {Bernuzzi}},\ }\href {\doibase 10.1103/PhysRevD.99.044007}
  {\bibfield  {journal} {\bibinfo  {journal} {Phys. Rev.}\ }\textbf {\bibinfo
  {volume} {D99}},\ \bibinfo {pages} {044007} (\bibinfo {year}
  {2019}{\natexlab{b}})},\ \Eprint {http://arxiv.org/abs/1812.07923}
  {arXiv:1812.07923 [gr-qc]} \BibitemShut {NoStop}%
\bibitem [{\citenamefont {Bini}\ \emph {et~al.}(2012)\citenamefont {Bini},
  \citenamefont {Damour},\ and\ \citenamefont {Faye}}]{Bini:2012gu}%
  \BibitemOpen
  \bibfield  {author} {\bibinfo {author} {\bibfnamefont {D.}~\bibnamefont
  {Bini}}, \bibinfo {author} {\bibfnamefont {T.}~\bibnamefont {Damour}}, \ and\
  \bibinfo {author} {\bibfnamefont {G.}~\bibnamefont {Faye}},\ }\href {\doibase
  10.1103/PhysRevD.85.124034} {\bibfield  {journal} {\bibinfo  {journal}
  {Phys.Rev.}\ }\textbf {\bibinfo {volume} {D85}},\ \bibinfo {pages} {124034}
  (\bibinfo {year} {2012})},\ \Eprint {http://arxiv.org/abs/1202.3565}
  {arXiv:1202.3565 [gr-qc]} \BibitemShut {NoStop}%
\bibitem [{\citenamefont {Bini}\ and\ \citenamefont
  {Damour}(2014)}]{Bini:2014zxa}%
  \BibitemOpen
  \bibfield  {author} {\bibinfo {author} {\bibfnamefont {D.}~\bibnamefont
  {Bini}}\ and\ \bibinfo {author} {\bibfnamefont {T.}~\bibnamefont {Damour}},\
  }\href {\doibase 10.1103/PhysRevD.90.124037} {\bibfield  {journal} {\bibinfo
  {journal} {Phys.Rev.}\ }\textbf {\bibinfo {volume} {D90}},\ \bibinfo {pages}
  {124037} (\bibinfo {year} {2014})},\ \Eprint {http://arxiv.org/abs/1409.6933}
  {arXiv:1409.6933 [gr-qc]} \BibitemShut {NoStop}%
\bibitem [{\citenamefont {Nagar}(2011)}]{Nagar:2011fx}%
  \BibitemOpen
  \bibfield  {author} {\bibinfo {author} {\bibfnamefont {A.}~\bibnamefont
  {Nagar}},\ }\href {\doibase 10.1103/PhysRevD.84.084028} {\bibfield  {journal}
  {\bibinfo  {journal} {Phys.Rev.}\ }\textbf {\bibinfo {volume} {D84}},\
  \bibinfo {pages} {084028} (\bibinfo {year} {2011})},\ \Eprint
  {http://arxiv.org/abs/1106.4349} {arXiv:1106.4349 [gr-qc]} \BibitemShut
  {NoStop}%
\bibitem [{\citenamefont {Buonanno}\ \emph {et~al.}(2003)\citenamefont
  {Buonanno}, \citenamefont {Chen},\ and\ \citenamefont
  {Vallisneri}}]{Buonanno:2002fy}%
  \BibitemOpen
  \bibfield  {author} {\bibinfo {author} {\bibfnamefont {A.}~\bibnamefont
  {Buonanno}}, \bibinfo {author} {\bibfnamefont {Y.-b.}\ \bibnamefont {Chen}},
  \ and\ \bibinfo {author} {\bibfnamefont {M.}~\bibnamefont {Vallisneri}},\
  }\href {\doibase 10.1103/PhysRevD.67.104025, 10.1103/PhysRevD.74.029904}
  {\bibfield  {journal} {\bibinfo  {journal} {Phys. Rev.}\ }\textbf {\bibinfo
  {volume} {D67}},\ \bibinfo {pages} {104025} (\bibinfo {year} {2003})},\
  \bibinfo {note} {[Erratum: Phys. Rev.D74,029904(2006)]},\ \Eprint
  {http://arxiv.org/abs/gr-qc/0211087} {arXiv:gr-qc/0211087 [gr-qc]}
  \BibitemShut {NoStop}%
\bibitem [{\citenamefont {Buonanno}\ \emph {et~al.}(2009)\citenamefont
  {Buonanno}, \citenamefont {Iyer}, \citenamefont {Ochsner}, \citenamefont
  {Pan},\ and\ \citenamefont {Sathyaprakash}}]{Buonanno:2009zt}%
  \BibitemOpen
  \bibfield  {author} {\bibinfo {author} {\bibfnamefont {A.}~\bibnamefont
  {Buonanno}}, \bibinfo {author} {\bibfnamefont {B.}~\bibnamefont {Iyer}},
  \bibinfo {author} {\bibfnamefont {E.}~\bibnamefont {Ochsner}}, \bibinfo
  {author} {\bibfnamefont {Y.}~\bibnamefont {Pan}}, \ and\ \bibinfo {author}
  {\bibfnamefont {B.}~\bibnamefont {Sathyaprakash}},\ }\href {\doibase
  10.1103/PhysRevD.80.084043} {\bibfield  {journal} {\bibinfo  {journal}
  {Phys.Rev.}\ }\textbf {\bibinfo {volume} {D80}},\ \bibinfo {pages} {084043}
  (\bibinfo {year} {2009})},\ \Eprint {http://arxiv.org/abs/0907.0700}
  {arXiv:0907.0700 [gr-qc]} \BibitemShut {NoStop}%
\bibitem [{\citenamefont {Chatziioannou}\ \emph {et~al.}(2013)\citenamefont
  {Chatziioannou}, \citenamefont {Klein}, \citenamefont {Yunes},\ and\
  \citenamefont {Cornish}}]{Chatziioannou:2013dza}%
  \BibitemOpen
  \bibfield  {author} {\bibinfo {author} {\bibfnamefont {K.}~\bibnamefont
  {Chatziioannou}}, \bibinfo {author} {\bibfnamefont {A.}~\bibnamefont
  {Klein}}, \bibinfo {author} {\bibfnamefont {N.}~\bibnamefont {Yunes}}, \ and\
  \bibinfo {author} {\bibfnamefont {N.}~\bibnamefont {Cornish}},\ }\href
  {\doibase 10.1103/PhysRevD.88.063011} {\bibfield  {journal} {\bibinfo
  {journal} {Phys. Rev.}\ }\textbf {\bibinfo {volume} {D88}},\ \bibinfo {pages}
  {063011} (\bibinfo {year} {2013})},\ \Eprint {http://arxiv.org/abs/1307.4418}
  {arXiv:1307.4418 [gr-qc]} \BibitemShut {NoStop}%
\bibitem [{\citenamefont {Pan}\ \emph {et~al.}(2014{\natexlab{b}})\citenamefont
  {Pan}, \citenamefont {Buonanno}, \citenamefont {Taracchini}, \citenamefont
  {Boyle}, \citenamefont {Kidder} \emph {et~al.}}]{Pan:2013tva}%
  \BibitemOpen
  \bibfield  {author} {\bibinfo {author} {\bibfnamefont {Y.}~\bibnamefont
  {Pan}}, \bibinfo {author} {\bibfnamefont {A.}~\bibnamefont {Buonanno}},
  \bibinfo {author} {\bibfnamefont {A.}~\bibnamefont {Taracchini}}, \bibinfo
  {author} {\bibfnamefont {M.}~\bibnamefont {Boyle}}, \bibinfo {author}
  {\bibfnamefont {L.~E.}\ \bibnamefont {Kidder}},  \emph {et~al.},\ }\href
  {\doibase 10.1103/PhysRevD.89.061501} {\bibfield  {journal} {\bibinfo
  {journal} {Phys.Rev.}\ }\textbf {\bibinfo {volume} {D89}},\ \bibinfo {pages}
  {061501} (\bibinfo {year} {2014}{\natexlab{b}})},\ \Eprint
  {http://arxiv.org/abs/1311.2565} {arXiv:1311.2565 [gr-qc]} \BibitemShut
  {NoStop}%
\bibitem [{\citenamefont {Nagar}\ and\ \citenamefont
  {Rettegno}(2019)}]{Nagar:2018gnk}%
  \BibitemOpen
  \bibfield  {author} {\bibinfo {author} {\bibfnamefont {A.}~\bibnamefont
  {Nagar}}\ and\ \bibinfo {author} {\bibfnamefont {P.}~\bibnamefont
  {Rettegno}},\ }\href {\doibase 10.1103/PhysRevD.99.021501} {\bibfield
  {journal} {\bibinfo  {journal} {Phys. Rev.}\ }\textbf {\bibinfo {volume}
  {D99}},\ \bibinfo {pages} {021501} (\bibinfo {year} {2019})},\ \Eprint
  {http://arxiv.org/abs/1805.03891} {arXiv:1805.03891 [gr-qc]} \BibitemShut
  {NoStop}%
\bibitem [{\citenamefont {Jim\'enez-Forteza}\ \emph {et~al.}(2017)\citenamefont
  {Jim\'enez-Forteza}, \citenamefont {Keitel}, \citenamefont {Husa},
  \citenamefont {Hannam}, \citenamefont {Khan},\ and\ \citenamefont
  {P{\"u}rrer}}]{Jimenez-Forteza:2016oae}%
  \BibitemOpen
  \bibfield  {author} {\bibinfo {author} {\bibfnamefont {X.}~\bibnamefont
  {Jim\'enez-Forteza}}, \bibinfo {author} {\bibfnamefont {D.}~\bibnamefont
  {Keitel}}, \bibinfo {author} {\bibfnamefont {S.}~\bibnamefont {Husa}},
  \bibinfo {author} {\bibfnamefont {M.}~\bibnamefont {Hannam}}, \bibinfo
  {author} {\bibfnamefont {S.}~\bibnamefont {Khan}}, \ and\ \bibinfo {author}
  {\bibfnamefont {M.}~\bibnamefont {P{\"u}rrer}},\ }\href {\doibase
  10.1103/PhysRevD.95.064024} {\bibfield  {journal} {\bibinfo  {journal} {Phys.
  Rev.}\ }\textbf {\bibinfo {volume} {D95}},\ \bibinfo {pages} {064024}
  (\bibinfo {year} {2017})},\ \Eprint {http://arxiv.org/abs/1611.00332}
  {arXiv:1611.00332 [gr-qc]} \BibitemShut {NoStop}%
\bibitem [{\citenamefont {Varma}\ \emph
  {et~al.}(2019{\natexlab{c}})\citenamefont {Varma}, \citenamefont {Gerosa},
  \citenamefont {Stein}, \citenamefont {H\'ebert},\ and\ \citenamefont
  {Zhang}}]{Varma:2018aht}%
  \BibitemOpen
  \bibfield  {author} {\bibinfo {author} {\bibfnamefont {V.}~\bibnamefont
  {Varma}}, \bibinfo {author} {\bibfnamefont {D.}~\bibnamefont {Gerosa}},
  \bibinfo {author} {\bibfnamefont {L.~C.}\ \bibnamefont {Stein}}, \bibinfo
  {author} {\bibfnamefont {F.}~\bibnamefont {H\'ebert}}, \ and\ \bibinfo
  {author} {\bibfnamefont {H.}~\bibnamefont {Zhang}},\ }\href {\doibase
  10.1103/PhysRevLett.122.011101} {\bibfield  {journal} {\bibinfo  {journal}
  {Phys. Rev. Lett.}\ }\textbf {\bibinfo {volume} {122}},\ \bibinfo {pages}
  {011101} (\bibinfo {year} {2019}{\natexlab{c}})},\ \Eprint
  {http://arxiv.org/abs/1809.09125} {arXiv:1809.09125 [gr-qc]} \BibitemShut
  {NoStop}%
\bibitem [{\citenamefont {O'Shaughnessy}\ \emph {et~al.}(2013)\citenamefont
  {O'Shaughnessy}, \citenamefont {London}, \citenamefont {Healy},\ and\
  \citenamefont {Shoemaker}}]{OShaughnessy:2012iol}%
  \BibitemOpen
  \bibfield  {author} {\bibinfo {author} {\bibfnamefont {R.}~\bibnamefont
  {O'Shaughnessy}}, \bibinfo {author} {\bibfnamefont {L.}~\bibnamefont
  {London}}, \bibinfo {author} {\bibfnamefont {J.}~\bibnamefont {Healy}}, \
  and\ \bibinfo {author} {\bibfnamefont {D.}~\bibnamefont {Shoemaker}},\ }\href
  {\doibase 10.1103/PhysRevD.87.044038} {\bibfield  {journal} {\bibinfo
  {journal} {Phys. Rev. D}\ }\textbf {\bibinfo {volume} {87}},\ \bibinfo
  {pages} {044038} (\bibinfo {year} {2013})},\ \Eprint
  {http://arxiv.org/abs/1209.3712} {arXiv:1209.3712 [gr-qc]} \BibitemShut
  {NoStop}%
\bibitem [{\citenamefont {Berti}\ \emph {et~al.}(2006)\citenamefont {Berti},
  \citenamefont {Cardoso},\ and\ \citenamefont {Will}}]{Berti:2005ys}%
  \BibitemOpen
  \bibfield  {author} {\bibinfo {author} {\bibfnamefont {E.}~\bibnamefont
  {Berti}}, \bibinfo {author} {\bibfnamefont {V.}~\bibnamefont {Cardoso}}, \
  and\ \bibinfo {author} {\bibfnamefont {C.~M.}\ \bibnamefont {Will}},\ }\href
  {\doibase 10.1103/PhysRevD.73.064030} {\bibfield  {journal} {\bibinfo
  {journal} {Phys. Rev.}\ }\textbf {\bibinfo {volume} {D73}},\ \bibinfo {pages}
  {064030} (\bibinfo {year} {2006})},\ \Eprint
  {http://arxiv.org/abs/gr-qc/0512160} {arXiv:gr-qc/0512160} \BibitemShut
  {NoStop}%
\bibitem [{\citenamefont {Ramos-Buades}\ \emph {et~al.}(2020)\citenamefont
  {Ramos-Buades}, \citenamefont {Schmidt}, \citenamefont {Pratten},\ and\
  \citenamefont {Husa}}]{Ramos-Buades:2020noq}%
  \BibitemOpen
  \bibfield  {author} {\bibinfo {author} {\bibfnamefont {A.}~\bibnamefont
  {Ramos-Buades}}, \bibinfo {author} {\bibfnamefont {P.}~\bibnamefont
  {Schmidt}}, \bibinfo {author} {\bibfnamefont {G.}~\bibnamefont {Pratten}}, \
  and\ \bibinfo {author} {\bibfnamefont {S.}~\bibnamefont {Husa}},\ }\href
  {\doibase 10.1103/PhysRevD.101.103014} {\bibfield  {journal} {\bibinfo
  {journal} {Phys. Rev. D}\ }\textbf {\bibinfo {volume} {101}},\ \bibinfo
  {pages} {103014} (\bibinfo {year} {2020})},\ \Eprint
  {http://arxiv.org/abs/2001.10936} {arXiv:2001.10936 [gr-qc]} \BibitemShut
  {NoStop}%
\bibitem [{\citenamefont {Gamba}\ \emph
  {et~al.}(2021{\natexlab{c}})\citenamefont {Gamba}, \citenamefont {Bernuzzi},\
  and\ \citenamefont {Nagar}}]{Gamba:2020ljo}%
  \BibitemOpen
  \bibfield  {author} {\bibinfo {author} {\bibfnamefont {R.}~\bibnamefont
  {Gamba}}, \bibinfo {author} {\bibfnamefont {S.}~\bibnamefont {Bernuzzi}}, \
  and\ \bibinfo {author} {\bibfnamefont {A.}~\bibnamefont {Nagar}},\ }\href
  {\doibase 10.1103/PhysRevD.104.084058} {\bibfield  {journal} {\bibinfo
  {journal} {Phys. Rev. D}\ }\textbf {\bibinfo {volume} {104}},\ \bibinfo
  {pages} {084058} (\bibinfo {year} {2021}{\natexlab{c}})},\ \Eprint
  {http://arxiv.org/abs/2012.00027} {arXiv:2012.00027 [gr-qc]} \BibitemShut
  {NoStop}%
\bibitem [{\citenamefont {Zackay}\ \emph {et~al.}(2018)\citenamefont {Zackay},
  \citenamefont {Dai},\ and\ \citenamefont {Venumadhav}}]{Zackay:2018qdy}%
  \BibitemOpen
  \bibfield  {author} {\bibinfo {author} {\bibfnamefont {B.}~\bibnamefont
  {Zackay}}, \bibinfo {author} {\bibfnamefont {L.}~\bibnamefont {Dai}}, \ and\
  \bibinfo {author} {\bibfnamefont {T.}~\bibnamefont {Venumadhav}},\
  }\href@noop {} {\  (\bibinfo {year} {2018})},\ \Eprint
  {http://arxiv.org/abs/1806.08792} {arXiv:1806.08792 [astro-ph.IM]}
  \BibitemShut {NoStop}%
\bibitem [{\citenamefont {Vinciguerra}\ \emph {et~al.}(2017)\citenamefont
  {Vinciguerra}, \citenamefont {Veitch},\ and\ \citenamefont
  {Mandel}}]{Vinciguerra:2017ngf}%
  \BibitemOpen
  \bibfield  {author} {\bibinfo {author} {\bibfnamefont {S.}~\bibnamefont
  {Vinciguerra}}, \bibinfo {author} {\bibfnamefont {J.}~\bibnamefont {Veitch}},
  \ and\ \bibinfo {author} {\bibfnamefont {I.}~\bibnamefont {Mandel}},\ }\href
  {\doibase 10.1088/1361-6382/aa6d44} {\bibfield  {journal} {\bibinfo
  {journal} {Class. Quant. Grav.}\ }\textbf {\bibinfo {volume} {34}},\ \bibinfo
  {pages} {115006} (\bibinfo {year} {2017})},\ \Eprint
  {http://arxiv.org/abs/1703.02062} {arXiv:1703.02062 [gr-qc]} \BibitemShut
  {NoStop}%
\bibitem [{\citenamefont {Harry}\ \emph {et~al.}(2016)\citenamefont {Harry},
  \citenamefont {Privitera}, \citenamefont {Bohé},\ and\ \citenamefont
  {Buonanno}}]{Harry:2016ijz}%
  \BibitemOpen
  \bibfield  {author} {\bibinfo {author} {\bibfnamefont {I.}~\bibnamefont
  {Harry}}, \bibinfo {author} {\bibfnamefont {S.}~\bibnamefont {Privitera}},
  \bibinfo {author} {\bibfnamefont {A.}~\bibnamefont {Bohé}}, \ and\ \bibinfo
  {author} {\bibfnamefont {A.}~\bibnamefont {Buonanno}},\ }\href {\doibase
  10.1103/PhysRevD.94.024012} {\bibfield  {journal} {\bibinfo  {journal} {Phys.
  Rev.}\ }\textbf {\bibinfo {volume} {D94}},\ \bibinfo {pages} {024012}
  (\bibinfo {year} {2016})},\ \Eprint {http://arxiv.org/abs/1603.02444}
  {arXiv:1603.02444 [gr-qc]} \BibitemShut {NoStop}%
\bibitem [{sxs()}]{sxs_tools}%
  \BibitemOpen
  \href@noop {} {\enquote {\bibinfo {title} {{SXS collaboration catalog
  tools}},}\ }\bibinfo {howpublished}
  {\url{https://github.com/sxs-collaboration/catalog_tools}}\BibitemShut
  {NoStop}%
\bibitem [{\citenamefont {Schmidt}\ \emph {et~al.}(2017)\citenamefont
  {Schmidt}, \citenamefont {Harry},\ and\ \citenamefont
  {Pfeiffer}}]{Schmidt:2017btt}%
  \BibitemOpen
  \bibfield  {author} {\bibinfo {author} {\bibfnamefont {P.}~\bibnamefont
  {Schmidt}}, \bibinfo {author} {\bibfnamefont {I.~W.}\ \bibnamefont {Harry}},
  \ and\ \bibinfo {author} {\bibfnamefont {H.~P.}\ \bibnamefont {Pfeiffer}},\
  }\href@noop {} {\  (\bibinfo {year} {2017})},\ \Eprint
  {http://arxiv.org/abs/1703.01076} {arXiv:1703.01076 [gr-qc]} \BibitemShut
  {NoStop}%
\bibitem [{aLI()}]{aLIGODesign_PSD}%
  \BibitemOpen
  \href@noop {} {\enquote {\bibinfo {title} {{Updated Advanced LIGO sensitivity
  design curve}},}\ }\bibinfo {howpublished}
  {\url{https://dcc.ligo.org/LIGO-T1800044/public}}\BibitemShut {NoStop}%
\bibitem [{\citenamefont {Breschi}\ \emph {et~al.}(2021)\citenamefont
  {Breschi}, \citenamefont {Gamba},\ and\ \citenamefont
  {Bernuzzi}}]{Breschi:2021wzr}%
  \BibitemOpen
  \bibfield  {author} {\bibinfo {author} {\bibfnamefont {M.}~\bibnamefont
  {Breschi}}, \bibinfo {author} {\bibfnamefont {R.}~\bibnamefont {Gamba}}, \
  and\ \bibinfo {author} {\bibfnamefont {S.}~\bibnamefont {Bernuzzi}},\ }\href
  {\doibase 10.1103/PhysRevD.104.042001} {\bibfield  {journal} {\bibinfo
  {journal} {Phys. Rev. D}\ }\textbf {\bibinfo {volume} {104}},\ \bibinfo
  {pages} {042001} (\bibinfo {year} {2021})},\ \Eprint
  {http://arxiv.org/abs/2102.00017} {arXiv:2102.00017 [gr-qc]} \BibitemShut
  {NoStop}%
\bibitem [{\citenamefont {Godzieba}\ and\ \citenamefont
  {Radice}(2021)}]{Godzieba:2021vnz}%
  \BibitemOpen
  \bibfield  {author} {\bibinfo {author} {\bibfnamefont {D.~A.}\ \bibnamefont
  {Godzieba}}\ and\ \bibinfo {author} {\bibfnamefont {D.}~\bibnamefont
  {Radice}},\ }\href {\doibase 10.3390/universe7100368} {\bibfield  {journal}
  {\bibinfo  {journal} {Universe}\ }\textbf {\bibinfo {volume} {7}},\ \bibinfo
  {pages} {368} (\bibinfo {year} {2021})},\ \Eprint
  {http://arxiv.org/abs/2109.01159} {arXiv:2109.01159 [astro-ph.HE]}
  \BibitemShut {NoStop}%
\bibitem [{\citenamefont {Speagle}(2020)}]{Speagle:2020}%
  \BibitemOpen
  \bibfield  {author} {\bibinfo {author} {\bibfnamefont {J.~S.}\ \bibnamefont
  {Speagle}},\ }\href {\doibase 10.1093/mnras/staa278} {\bibfield  {journal}
  {\bibinfo  {journal} {Monthly Notices of the Royal Astronomical Society}\
  }\textbf {\bibinfo {volume} {493}},\ \bibinfo {pages} {3132?3158} (\bibinfo
  {year} {2020})}\BibitemShut {NoStop}%
\bibitem [{\citenamefont {Abbott}\ \emph
  {et~al.}(2016{\natexlab{c}})\citenamefont {Abbott} \emph
  {et~al.}}]{TheLIGOScientific:2016wfe}%
  \BibitemOpen
  \bibfield  {author} {\bibinfo {author} {\bibfnamefont {B.~P.}\ \bibnamefont
  {Abbott}} \emph {et~al.} (\bibinfo {collaboration} {Virgo, LIGO
  Scientific}),\ }\href {\doibase 10.1103/PhysRevLett.116.241102} {\bibfield
  {journal} {\bibinfo  {journal} {Phys. Rev. Lett.}\ }\textbf {\bibinfo
  {volume} {116}},\ \bibinfo {pages} {241102} (\bibinfo {year}
  {2016}{\natexlab{c}})},\ \Eprint {http://arxiv.org/abs/1602.03840}
  {arXiv:1602.03840 [gr-qc]} \BibitemShut {NoStop}%
\bibitem [{\citenamefont {Islam}\ \emph {et~al.}(2021)\citenamefont {Islam},
  \citenamefont {Field}, \citenamefont {Haster},\ and\ \citenamefont
  {Smith}}]{Islam:2020reh}%
  \BibitemOpen
  \bibfield  {author} {\bibinfo {author} {\bibfnamefont {T.}~\bibnamefont
  {Islam}}, \bibinfo {author} {\bibfnamefont {S.~E.}\ \bibnamefont {Field}},
  \bibinfo {author} {\bibfnamefont {C.-J.}\ \bibnamefont {Haster}}, \ and\
  \bibinfo {author} {\bibfnamefont {R.}~\bibnamefont {Smith}},\ }\href
  {\doibase 10.1103/PhysRevD.103.104027} {\bibfield  {journal} {\bibinfo
  {journal} {Phys. Rev. D}\ }\textbf {\bibinfo {volume} {103}},\ \bibinfo
  {pages} {104027} (\bibinfo {year} {2021})},\ \Eprint
  {http://arxiv.org/abs/2010.04848} {arXiv:2010.04848 [gr-qc]} \BibitemShut
  {NoStop}%
\bibitem [{\citenamefont {Colleoni}\ \emph {et~al.}(2021)\citenamefont
  {Colleoni}, \citenamefont {Mateu-Lucena}, \citenamefont {Estell\'es},
  \citenamefont {Garc\'\i{}a-Quir\'os}, \citenamefont {Keitel}, \citenamefont
  {Pratten}, \citenamefont {Ramos-Buades},\ and\ \citenamefont
  {Husa}}]{Colleoni:2020tgc}%
  \BibitemOpen
  \bibfield  {author} {\bibinfo {author} {\bibfnamefont {M.}~\bibnamefont
  {Colleoni}}, \bibinfo {author} {\bibfnamefont {M.}~\bibnamefont
  {Mateu-Lucena}}, \bibinfo {author} {\bibfnamefont {H.}~\bibnamefont
  {Estell\'es}}, \bibinfo {author} {\bibfnamefont {C.}~\bibnamefont
  {Garc\'\i{}a-Quir\'os}}, \bibinfo {author} {\bibfnamefont {D.}~\bibnamefont
  {Keitel}}, \bibinfo {author} {\bibfnamefont {G.}~\bibnamefont {Pratten}},
  \bibinfo {author} {\bibfnamefont {A.}~\bibnamefont {Ramos-Buades}}, \ and\
  \bibinfo {author} {\bibfnamefont {S.}~\bibnamefont {Husa}},\ }\href {\doibase
  10.1103/PhysRevD.103.024029} {\bibfield  {journal} {\bibinfo  {journal}
  {Phys. Rev. D}\ }\textbf {\bibinfo {volume} {103}},\ \bibinfo {pages}
  {024029} (\bibinfo {year} {2021})},\ \Eprint
  {http://arxiv.org/abs/2010.05830} {arXiv:2010.05830 [gr-qc]} \BibitemShut
  {NoStop}%
\bibitem [{\citenamefont {Hessels}\ \emph {et~al.}(2006)\citenamefont
  {Hessels}, \citenamefont {Ransom}, \citenamefont {Stairs}, \citenamefont
  {Freire}, \citenamefont {Kaspi},\ and\ \citenamefont
  {Camilo}}]{Hessels:2006ze}%
  \BibitemOpen
  \bibfield  {author} {\bibinfo {author} {\bibfnamefont {J.~W.~T.}\
  \bibnamefont {Hessels}}, \bibinfo {author} {\bibfnamefont {S.~M.}\
  \bibnamefont {Ransom}}, \bibinfo {author} {\bibfnamefont {I.~H.}\
  \bibnamefont {Stairs}}, \bibinfo {author} {\bibfnamefont {P.~C.~C.}\
  \bibnamefont {Freire}}, \bibinfo {author} {\bibfnamefont {V.~M.}\
  \bibnamefont {Kaspi}}, \ and\ \bibinfo {author} {\bibfnamefont
  {F.}~\bibnamefont {Camilo}},\ }\href {\doibase 10.1126/science.1123430}
  {\bibfield  {journal} {\bibinfo  {journal} {Science}\ }\textbf {\bibinfo
  {volume} {311}},\ \bibinfo {pages} {1901} (\bibinfo {year} {2006})},\ \Eprint
  {http://arxiv.org/abs/astro-ph/0601337} {arXiv:astro-ph/0601337 [astro-ph]}
  \BibitemShut {NoStop}%
\bibitem [{\citenamefont {Kramer}\ and\ \citenamefont
  {Wex}(2009)}]{Kramer:2009zza}%
  \BibitemOpen
  \bibfield  {author} {\bibinfo {author} {\bibfnamefont {M.}~\bibnamefont
  {Kramer}}\ and\ \bibinfo {author} {\bibfnamefont {N.}~\bibnamefont {Wex}},\
  }\href {\doibase 10.1088/0264-9381/26/7/073001} {\bibfield  {journal}
  {\bibinfo  {journal} {Class. Quant. Grav.}\ }\textbf {\bibinfo {volume}
  {26}},\ \bibinfo {pages} {073001} (\bibinfo {year} {2009})}\BibitemShut
  {NoStop}%
\bibitem [{\citenamefont {Stovall}\ \emph {et~al.}(2018)\citenamefont {Stovall}
  \emph {et~al.}}]{Stovall:2018ouw}%
  \BibitemOpen
  \bibfield  {author} {\bibinfo {author} {\bibfnamefont {K.}~\bibnamefont
  {Stovall}} \emph {et~al.},\ }\href {\doibase 10.3847/2041-8213/aaad06}
  {\bibfield  {journal} {\bibinfo  {journal} {Astrophys. J.}\ }\textbf
  {\bibinfo {volume} {854}},\ \bibinfo {pages} {L22} (\bibinfo {year}
  {2018})},\ \Eprint {http://arxiv.org/abs/1802.01707} {arXiv:1802.01707
  [astro-ph.HE]} \BibitemShut {NoStop}%
\bibitem [{\citenamefont {Wade}\ \emph {et~al.}(2014)\citenamefont {Wade},
  \citenamefont {Creighton}, \citenamefont {Ochsner}, \citenamefont {Lackey},
  \citenamefont {Farr}, \citenamefont {Littenberg},\ and\ \citenamefont
  {Raymond}}]{Wade:2014vqa}%
  \BibitemOpen
  \bibfield  {author} {\bibinfo {author} {\bibfnamefont {L.}~\bibnamefont
  {Wade}}, \bibinfo {author} {\bibfnamefont {J.~D.~E.}\ \bibnamefont
  {Creighton}}, \bibinfo {author} {\bibfnamefont {E.}~\bibnamefont {Ochsner}},
  \bibinfo {author} {\bibfnamefont {B.~D.}\ \bibnamefont {Lackey}}, \bibinfo
  {author} {\bibfnamefont {B.~F.}\ \bibnamefont {Farr}}, \bibinfo {author}
  {\bibfnamefont {T.~B.}\ \bibnamefont {Littenberg}}, \ and\ \bibinfo {author}
  {\bibfnamefont {V.}~\bibnamefont {Raymond}},\ }\href {\doibase
  10.1103/PhysRevD.89.103012} {\bibfield  {journal} {\bibinfo  {journal} {Phys.
  Rev.}\ }\textbf {\bibinfo {volume} {D89}},\ \bibinfo {pages} {103012}
  (\bibinfo {year} {2014})},\ \Eprint {http://arxiv.org/abs/1402.5156}
  {arXiv:1402.5156 [gr-qc]} \BibitemShut {NoStop}%
\bibitem [{\citenamefont {Buchman}\ \emph {et~al.}(2012)\citenamefont
  {Buchman}, \citenamefont {Pfeiffer}, \citenamefont {Scheel},\ and\
  \citenamefont {Szilagyi}}]{Buchman:2012dw}%
  \BibitemOpen
  \bibfield  {author} {\bibinfo {author} {\bibfnamefont {L.~T.}\ \bibnamefont
  {Buchman}}, \bibinfo {author} {\bibfnamefont {H.~P.}\ \bibnamefont
  {Pfeiffer}}, \bibinfo {author} {\bibfnamefont {M.~A.}\ \bibnamefont
  {Scheel}}, \ and\ \bibinfo {author} {\bibfnamefont {B.}~\bibnamefont
  {Szilagyi}},\ }\href {\doibase 10.1103/PhysRevD.86.084033} {\bibfield
  {journal} {\bibinfo  {journal} {Phys. Rev.}\ }\textbf {\bibinfo {volume}
  {D86}},\ \bibinfo {pages} {084033} (\bibinfo {year} {2012})},\ \Eprint
  {http://arxiv.org/abs/1206.3015} {arXiv:1206.3015 [gr-qc]} \BibitemShut
  {NoStop}%
\bibitem [{\citenamefont {Lovelace}\ \emph {et~al.}(2008)\citenamefont
  {Lovelace}, \citenamefont {Owen}, \citenamefont {Pfeiffer},\ and\
  \citenamefont {Chu}}]{Lovelace:2008tw}%
  \BibitemOpen
  \bibfield  {author} {\bibinfo {author} {\bibfnamefont {G.}~\bibnamefont
  {Lovelace}}, \bibinfo {author} {\bibfnamefont {R.}~\bibnamefont {Owen}},
  \bibinfo {author} {\bibfnamefont {H.~P.}\ \bibnamefont {Pfeiffer}}, \ and\
  \bibinfo {author} {\bibfnamefont {T.}~\bibnamefont {Chu}},\ }\href {\doibase
  10.1103/PhysRevD.78.084017} {\bibfield  {journal} {\bibinfo  {journal} {Phys.
  Rev. D}\ }\textbf {\bibinfo {volume} {78}},\ \bibinfo {pages} {084017}
  (\bibinfo {year} {2008})},\ \Eprint {http://arxiv.org/abs/0805.4192}
  {arXiv:0805.4192 [gr-qc]} \BibitemShut {NoStop}%
\bibitem [{\citenamefont {Pfeiffer}\ \emph {et~al.}(2007)\citenamefont
  {Pfeiffer}, \citenamefont {Brown}, \citenamefont {Kidder}, \citenamefont
  {Lindblom}, \citenamefont {Lovelace},\ and\ \citenamefont
  {Scheel}}]{Pfeiffer:2007yz}%
  \BibitemOpen
  \bibfield  {author} {\bibinfo {author} {\bibfnamefont {H.~P.}\ \bibnamefont
  {Pfeiffer}}, \bibinfo {author} {\bibfnamefont {D.~A.}\ \bibnamefont {Brown}},
  \bibinfo {author} {\bibfnamefont {L.~E.}\ \bibnamefont {Kidder}}, \bibinfo
  {author} {\bibfnamefont {L.}~\bibnamefont {Lindblom}}, \bibinfo {author}
  {\bibfnamefont {G.}~\bibnamefont {Lovelace}}, \ and\ \bibinfo {author}
  {\bibfnamefont {M.~A.}\ \bibnamefont {Scheel}},\ }\bibfield  {booktitle}
  {\emph {\bibinfo {booktitle} {{New frontiers in numerical relativity.
  Proceedings, International Meeting, NFNR 2006, Potsdam, Germany, July 17-21,
  2006}}},\ }\href {\doibase 10.1088/0264-9381/24/12/S06} {\bibfield  {journal}
  {\bibinfo  {journal} {Class. Quant. Grav.}\ }\textbf {\bibinfo {volume}
  {24}},\ \bibinfo {pages} {S59} (\bibinfo {year} {2007})},\ \Eprint
  {http://arxiv.org/abs/gr-qc/0702106} {arXiv:gr-qc/0702106 [gr-qc]}
  \BibitemShut {NoStop}%
\bibitem [{\citenamefont {Caudill}\ \emph {et~al.}(2006)\citenamefont
  {Caudill}, \citenamefont {Cook}, \citenamefont {Grigsby},\ and\ \citenamefont
  {Pfeiffer}}]{Caudill:2006hw}%
  \BibitemOpen
  \bibfield  {author} {\bibinfo {author} {\bibfnamefont {M.}~\bibnamefont
  {Caudill}}, \bibinfo {author} {\bibfnamefont {G.~B.}\ \bibnamefont {Cook}},
  \bibinfo {author} {\bibfnamefont {J.~D.}\ \bibnamefont {Grigsby}}, \ and\
  \bibinfo {author} {\bibfnamefont {H.~P.}\ \bibnamefont {Pfeiffer}},\ }\href
  {\doibase 10.1103/PhysRevD.74.064011} {\bibfield  {journal} {\bibinfo
  {journal} {Phys. Rev. D}\ }\textbf {\bibinfo {volume} {74}},\ \bibinfo
  {pages} {064011} (\bibinfo {year} {2006})},\ \Eprint
  {http://arxiv.org/abs/gr-qc/0605053} {arXiv:gr-qc/0605053} \BibitemShut
  {NoStop}%
\bibitem [{\citenamefont {Cook}\ and\ \citenamefont
  {Pfeiffer}(2004)}]{Cook:2004kt}%
  \BibitemOpen
  \bibfield  {author} {\bibinfo {author} {\bibfnamefont {G.~B.}\ \bibnamefont
  {Cook}}\ and\ \bibinfo {author} {\bibfnamefont {H.~P.}\ \bibnamefont
  {Pfeiffer}},\ }\href {\doibase 10.1103/PhysRevD.70.104016} {\bibfield
  {journal} {\bibinfo  {journal} {Phys. Rev.}\ }\textbf {\bibinfo {volume}
  {D70}},\ \bibinfo {pages} {104016} (\bibinfo {year} {2004})},\ \Eprint
  {http://arxiv.org/abs/gr-qc/0407078} {arXiv:gr-qc/0407078 [gr-qc]}
  \BibitemShut {NoStop}%
\end{thebibliography}
